\documentclass[letterpaper,twocolumn,10pt]{article}
\usepackage{usenix2019_v3}

\usepackage[english]{babel}
\usepackage{blindtext}

\usepackage{ifthen}
\usepackage{xcolor}
\usepackage{blindtext}
\usepackage{algorithm}
\usepackage{graphicx}
\usepackage{amsmath}
\usepackage[noend]{algpseudocode}
\usepackage{subfig}
\usepackage{xspace}
\usepackage{amsthm}
\usepackage{balance}

\usepackage{graphicx}
\usepackage{footnote}
\usepackage{comment}
\usepackage{kantlipsum}
\makesavenoteenv{tabular}
\makesavenoteenv{table}
\usepackage{listings}
\usepackage{colortbl}
\usepackage{multicol}
\usepackage[skins]{tcolorbox}
\usepackage[framemethod=tikz]{mdframed}

\makeatletter
\def\thanks#1{\protected@xdef\@thanks{\@thanks
        \protect\footnotetext{#1}}}
\makeatother

\newcommand{\name}{{\sc{MLCC}}\xspace}
\newcommand{\namebf}{\bfseries{\scshape{MLCC}}\xspace}
\newcommand{\nameshort}{{\sc{ML}}\xspace}
\newcommand{\nameshortbf}{\bfseries{\scshape{ML}}\xspace}

\newcommand{\peak}{{phase}\xspace}
\newcommand{\peaks}{{phases}\xspace}

\makeatletter
\let\OldStatex\Statex
\renewcommand{\Statex}[1][3]{%
  \setlength\@tempdima{\algorithmicindent}%
  \OldStatex\hskip\dimexpr#1\@tempdima\relax}
\makeatother
\algnewcommand{\LeftComment}[1]{\OldStatex \(\triangleright\) #1}
\algnewcommand{\LineComment}[1]{\OldStatex \(\triangleright\) #1}

\newcommand{\new}[1]{{\color{black} #1}}

\newcommand{\para}[1]{{\textbf{{#1}}}}

\algnewcommand{\IOComment}[1]{\OldStatex \(\triangleright\) #1}
\algnewcommand{\firstLeftComment}[1]{\OldStatex \(\indent\triangleright\) #1}
\algnewcommand{\secondLeftComment}[1]{\OldStatex \(\indent\indent\triangleright\) #1}
\algnewcommand{\thirdLeftComment}[1]{\OldStatex \(\indent\indent\indent\triangleright\) #1}
\algrenewcommand\algorithmicwhile{\textbf{With probability}}
\definecolor{LightCyan}{rgb}{0.88,1,1}
\definecolor{celadon}{rgb}{0.67, 0.88, 0.69}

\newcommand{\exclude}[1]{}
\newcommand{\showComments}{yes}
\newcommand{\note}[2]{
    \ifthenelse{\equal{\showComments}{yes}}{\textcolor{#1}{#2}}{}
}

\newcommand{\bytes}{bytes\_ratio}

\newcommand{\f}{{\color{cyan}{\mathcal{F}(\bytes)\xspace}}} 
\newcommand{\fblack}{\mathcal{F}(\bytes)\xspace}
\newcommand*\req[1]{%
\scalebox{0.9}{\begin{tikzpicture}[baseline=-3pt]
  \node[draw=black,circle,inner sep=0.5pt, fill=white, thick] {\textcolor{black}{\textsf{\textbf{#1}}}};
\end{tikzpicture}}}
\newcommand*\reqsq[1]{%
\scalebox{0.9}{\begin{tikzpicture}[baseline=-3pt]
  \node[draw=black,rectangle,inner sep=0.5pt, fill=white, thick, minimum size=3mm] {\textcolor{black}{\textsf{\textbf{#1}}}};
\end{tikzpicture}}}

\usepackage{titlesec}
\titlespacing*{\paragraph}{0pt}{0.1ex plus 1ex minus .2ex}{1em}

\everypar{\looseness=-1}
\microtypecontext{spacing=nonfrench}

\titlespacing*{\section}{0pt}{4pt}{2pt}
\titlespacing*{\subsection}{0pt}{2pt}{2pt}
\titlespacing*{\subsubsection}{0pt}{2pt}{2pt}

\usepackage{booktabs}
\usepackage{subfig}
\usepackage{graphicx}
\usepackage{subcaption}
\usepackage[normalem]{ulem} 
\usepackage{array}
\usepackage{algorithm}
\usepackage[noend]{algpseudocode}
\usepackage{bold-extra}
\usepackage{tikz}
\usepackage{amsmath}
\usepackage{xspace}
\usepackage{filecontents}
\usepackage{listings}
\usepackage{etoolbox}
\usepackage{longtable}
\usepackage{lipsum} 

\usepackage{pifont}
\usepackage{multirow}
\usepackage{paralist}
\usepackage{enumitem}
\usepackage{url}

\usepackage{titlesec}

\AtBeginEnvironment{equation}{\small}
\AtBeginEnvironment{equation*}{\small}

\newcommand{\captionfonts}{\bf \small}
\usepackage[normalem]{ulem}

\makeatletter  
\long\def\@makecaption#1#2{%
	\vskip\abovecaptionskip
	\sbox\@tempboxa{{\captionfonts #1: #2}}%
	\ifdim \wd\@tempboxa >\hsize
	{\captionfonts #1: #2\par}
	\else
	\hbox to\hsize{\hfil\box\@tempboxa\hfil}%
	\fi
	\vskip\belowcaptionskip}
\makeatother   

\everypar{\looseness=-1}
\microtypecontext{spacing=nonfrench}

\makeatletter
\g@addto@macro\normalsize{%
  \setlength\abovedisplayskip{2pt}%
  \setlength\belowdisplayskip{2pt}%
  \setlength\abovedisplayshortskip{2pt}%
  \setlength\belowdisplayshortskip{2pt}%
}
\makeatother

\begin{document}
\clearpage
\onecolumn
\clearpage
\twocolumn
\setlength{\abovedisplayskip}{4pt}  
\setlength{\belowdisplayskip}{4pt}

\title{MLCC: A Congestion Control Technique to Accelerate ML Training}

\author{Anton A. Zabreyko$^{*}$ \quad Sanjoli Narang$^{*}$ \thanks{$^{*}$ Equal contribution} \quad  Sudarsanan Rajasekaran \quad  Manya Ghobadi 
\\ \\ MIT CSAIL
}

\pagestyle{plain}

\setcounter{page}{1}
\maketitle

\begin{abstract}
We present \name, a novel technique to augment today's congestion control algorithms to accelerate DNN training jobs in shared GPU clusters in a fully distributed manner. At the heart of \name lies a straightforward principle: DNN training flows should scale their sending rate to shift other flows' communication into their compute periods, achieving interleaving. We show that integrating this principle into today's congestion control protocols is simple (requiring less than 60 lines of code for a given protocol) and enables DNN jobs to interleave within a few training iterations,  thereby reducing network contention and improving job completion times. 
Our testbed demonstrates that \name accelerates the average and 99$^{th}$ percentile training iteration times by up to 1.9$\times$ and 2.7$\times$, respectively. Through extensive packet-level simulations, we observe a 1.35$\times$ improvement in training throughput on a 36-node, 288 GPU fat-tree topology.

\end{abstract}
\section{Introduction}

The relentless growth of artificial intelligence (AI) and machine learning (ML) workloads on multi-tenant shared GPU clusters is increasingly straining network resources. Prior papers demonstrate that DNN training clusters often experience network congestion, leading to longer training times and GPU underutilization~\cite{topoopt, mudigere2021highperformance, blueconnect, crux}. A recent study of Alibaba's multi-tenant production clusters reports that 36.3\% of DNN training jobs suffer from network contention, leaving expensive accelerators idle~\cite{crux}.

Prior work employed two main strategies to reduce the likelihood of network congestion in DNN clusters: ($i$) reducing the size of data to transmit over the network through quantization or compression techniques~\cite{ako, qsgd, lin2017deep, parameter_propagation}; ($ii$) reducing the exposed communication time by overlapping the communication period of the same job with its computation phase~\cite{pipedream, gpipe, terapipe, pipelining_1, pipelining_2, bytescheduler}. 
Despite these defenses, network contention remains inevitable at scale due to failures, bandwidth oversubscription, and sub-optimal routing (such as ECMP hash collisions)~\cite{crux, topoopt, syndicate, vclos}. Whenever congestion builds, good old congestion control protocols stand as the last resort.

\begin{figure*}[t]
\centering
\includegraphics[width=0.92\textwidth]{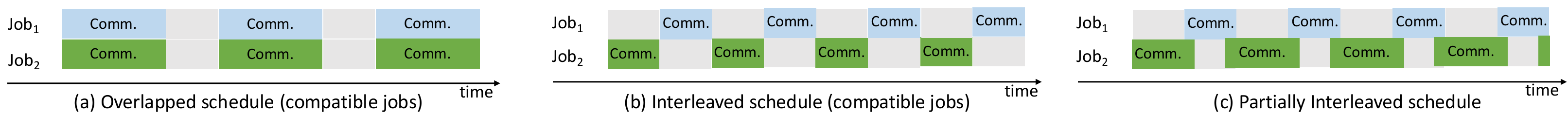}
    \vspace{-0.4cm}
    \caption{The concept of interleaving.}
    \vspace{-0.5cm}
\label{fig:concept_interleaving}
\end{figure*}

However, the vast majority of congestion control approaches are designed to enforce fair-sharing, ensuring that all jobs receive an equal portion of available bandwidth. While effective in traditional networking contexts, this strategy is ill-suited for DNN workloads because it results in alternating periods of network under-utilization and congestion, elongating the iteration time of all competing jobs~\cite{cassini_hotnets, mltcp_hotnets, cassini_nsdi24}. This is because DNN jobs exhibit distinctive periodic traffic patterns, with each iteration involving repeated phases of computation and communication. They also have strict temporal dependencies between flows, where flows in a given iteration cannot start until all flows from the preceding iteration have completed. \looseness=-1

To address this challenge, prior work proposed using centralized schedulers to control the placement and timing of flows on network links for minimizing congestion. In particular, Muri~\cite{muri} and Cassini~\cite{cassini_nsdi24} demonstrated that \textit{interleaving} the communication phases of one DNN job with the compute phases of other jobs sharing the link results in better network utilization and accelerates training for all competing jobs.
However, these approaches are not practical because they incur substantial computational overhead due to the need to implement complex algorithms, which significantly limits their scalability. Moreover, we show that their scope is limited to a small set of \textit{compatible} job scenarios where the compute and communication phases of competing jobs perfectly fit inside each other~(\S\ref{sec:motivation}, \S\ref{sec:evaluation}).

In this paper, we demonstrate that it is possible to omit centralized schedulers and achieve communication interleaving of DNN jobs by relying only on congestion control protocols. To do so, we propose \name, a novel \textit{technique} that augments today's congestion control mechanisms for accelerating DNN jobs. We show that with a straightforward change to the flows' congestion control algorithm, DNN training jobs automatically converge to an interleaved state in a distributed manner without the need for any hardware changes, priority queues, switch support, or centralized control.

To achieve the above goals, \name uses a two-pronged approach. First, flows using a \name-augmented congestion control scheme adjust their congestion window sizes (or sending rates) according to the ratio of bytes successfully sent during the current training iteration, normalized by the maximum number of bytes in any communication period of the job

~(\S\ref{sec:design}). At any point in time, a job $j$ that is closer to the end of its communication phase is allowed to send data into the network more aggressively, thereby reducing the available bandwidth for competing jobs. We find that this creates a positive cascading effect, shifting the competing jobs' communication into $j$'s compute phase. Over iterations, such shifts accumulate, eventually pushing all the jobs into an interleaved state. Rate adaptation alone can still get stuck in suboptimal communication interleavings. To close this gap, our second technique adds a simulated annealing phase to congestion control. When a job’s observed iteration time exceeds its ideal, \name injects randomized delays with probability proportional to this gap, perturbing the current schedule so jobs can escape poor local optima and re-converge to a better interleaving. 

To implement \name, we augment six state-of-the-art congestion control protocols, including Reno~\cite{reno}, Cubic~\cite{cubic}, DCTCP~\cite{dctcp}, DCQCN~\cite{dcqcn}, Swift~\cite{swift}, and Homa~\cite{homa} with just 30--60 lines of code to implement their corresponding \name variants; e.g. \nameshort-Reno, \nameshort-Cubic etc. 
We evaluate these \name variants using extensive packet-level simulations and testbed experiments. Our testbed consists of 12 servers, each equipped with one NVIDIA A100 GPU~\cite{a100} and one 50~Gbps RDMA NIC. Our experiments with real-world DNN models (VGG16~\cite{vgg16}, VGG19~\cite{vgg19}, WideResNet101~\cite{wideresnet}, BERT~\cite{bert_orig, megatronlm}, RoBERTa~\cite{roberta}, CamemBERT~\cite{camembert}, GPT-1~\cite{gpt_1}, GPT-2~\cite{gpt_2}, GPT-3~\cite{gpt_3}, DeepSpeed-MoE~\cite{gpt_moe} and Llama2~\cite{llama, colossalai}) 
show that \name flows converge to an interleaved state after $\approx$30 training iterations while improving the mean and 99$^{th}$ percentile tail iteration times of jobs by up to 1.9$\times$ and 2.7$\times$, respectively. \new{We also show that \name reduces the average number of dropped or ECN-marked packets by up to 28.87$\times$ (\S\ref{sec:evaluation}). Our extensive packet-level simulations in a fat-tree topology show \name improves the throughput of training jobs by up to 1.35$\times$ (\S\ref{sec:pkt_sim_topo}) over varying load and arbitrary DNN traffic patterns. 
We evaluate \name across a range of network topologies, including fat-tree architectures and Alibaba’s HPN~\cite{hpn}, under different load-balancing schemes like Crux~\cite{crux} and HPN~\cite{hpn}. In oversubscribed topologies, \name rivals packet spraying, outperforming it when paired with Crux and delivering a tail $1.35\times$ improvement over ECMP (\S\ref{sec:pkt_sim_lb})}.

\begin{figure*}[t]
    \centering
    \begin{minipage}{0.73\textwidth}
    \centering
    \includegraphics[width=0.98\linewidth]{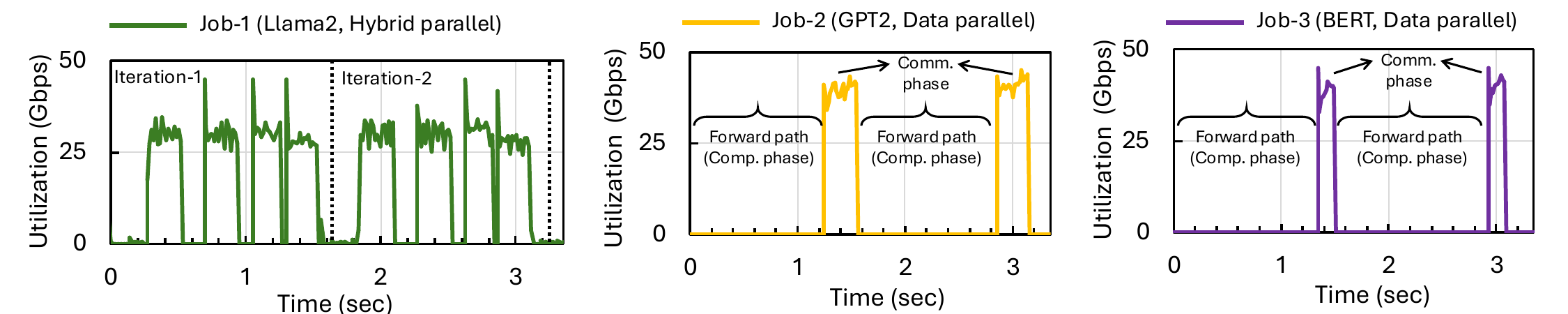}
    \vspace*{-4mm}
        \caption{The periodic traffic pattern of jobs in isolation.
        }
    \vspace{-0.1cm}
    \label{fig:jobs_pattern}
    \end{minipage}
    \begin{minipage}{0.26\textwidth}
    \centering
    \includegraphics[width=0.98\linewidth]{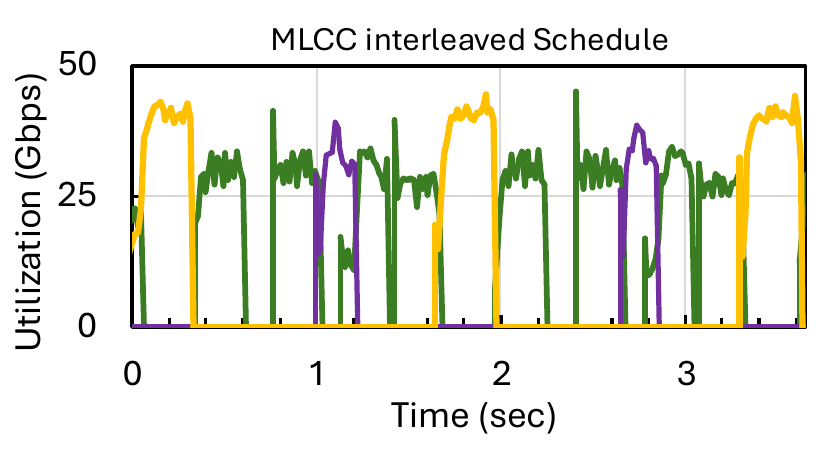}
    \vspace*{-4mm}
        \caption{\namebf's schedule.
        }
    \vspace{-0.1cm}
    \label{fig:mlcc_motiv}
    \end{minipage}
\end{figure*}

An earlier version of this paper appeared in a workshop~\cite{mltcp_hotnets}. However, it lacked several key features. First, it only considered TCP Reno, which is not commonly used in GPU clusters. Second, its proposed approach could only interleave simple on-off communication patterns, making it impractical for complex workloads. Third, its algorithm fails for sets of jobs that are not fully compatible. Finally, it did not provide any details about its fairness or convergence properties. In short, impractical, narrow in scope, and leaves many questions about its characteristics open. We will open-source our entire codebase, including Linux kernel modules, testbed scripts, and packet-level simulator. 
\section{Background and Motivation}
\label{sec:motivation}

\begin{figure*}[t]
    \centering
    \includegraphics[width=0.89\textwidth]{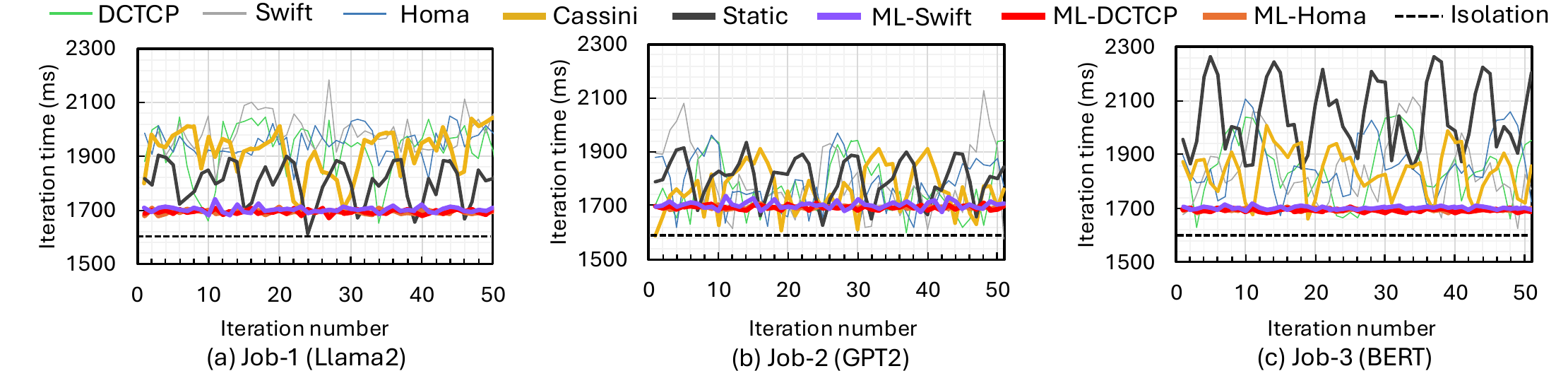}
        \vspace*{-0.4cm}
        \caption{Evolution of iteration time of competing jobs for different protocols, their \namebf variants, and scheduling approaches.
        }
    \label{fig:itertime_vs_iternum}
    \vspace{-0.5cm}
\end{figure*}

This section motivates the need for \name by demonstrating the limitations of traditional congestion control algorithms and existing centralized schemes using a combination of testbed experiments and packet-level simulations.

\para{DNN Traffic Pattern.} Prior work has shown that distributed DNN training jobs exhibit a periodic traffic pattern that repeats iteration after iteration~\cite{topoopt, cassini_nsdi24, cassini_hotnets, muri, echelon,mltcp_hotnets}. During the communication phase of each iteration, a distributed DNN job creates several flows, where the arrival of each flow depends on the completion of the previous one~\cite{topoopt}. 
This property distinguishes DNN flows from traditional datacenter flows~\cite{dc_in_the_wild}, where flow arrivals tend to be independent. In addition, DNN flow sizes range several orders of magnitude, from a few MB to tens of GB, unlike typical datacenter traffic, which is dominated by short flows with a heavy tail. \new{We use the following terminology throughout the paper: a \textit{DNN training iteration} is forward and backward pass required to fully process a batch of data; \textit{iteration time} is the duration of one such iteration; \textit{training time} is the total duration of training across thousands of iterations; and \textit{slowdown} is the ratio of a job’s iteration time to its iteration time in isolation.} 

\para{Compatibility and Interleaving.} Prior work~\cite{cassini_hotnets,cassini_nsdi24,muri,mltcp_hotnets} introduced the idea of interleaving as placing one job's communication during others’ compute to avoid colliding their communications and hence avoid network contention, as shown in Figure~\ref{fig:concept_interleaving}(a). Figure~\ref{fig:concept_interleaving}(b) illustrates two jobs in a perfectly interleaved schedule. This perfect interleaving is feasible only for \textit{compatible} jobs whose communication phases fit entirely within others’ compute every iteration. Prior approaches largely operate only when jobs are compatible, which is rare in practice. We generalize this concept by introducing \textit{partial interleaving} where certain bandwidth allocations enable jobs to fit as much of their communication as possible within the compute of the others with minimal network contention (Figure~\ref{fig:concept_interleaving}(c)). This broadens the benefit of periodicity of DNN jobs beyond the narrow class of compatible jobs. \new{Note that an interleaved schedule is \textit{stable} when it has a repetitive pattern with a finite period that does not drift over time -- possibly spanning multiple iterations of each job}.

\para{Experiment Setup.} To demonstrate the need for DNN-specific congestion control protocols, we train three DNN models on a 12-server testbed and use the traffic patterns to feed into our packet-level simulator. Job-1 trains a Llama2~\cite{llama} (2 billion parameters) hybrid parallel model from the ColossalAI framework~\cite{colossalai} using 2-way data, 2-way tensor, and 2-way pipeline parallelism, requiring a total of 8 GPUs to run. Job-2 trains a GPT-2~\cite{gpt_2} (350 million parameters) model with 2-way data parallelism, and Job-3 trains a BERT~\cite{bert} (250 million parameters) model with 2-way data parallelism. We perform five categories of experiments: ($i$) each job runs separately with complete isolation from the others; ($ii$) all three jobs compete on a single 50~Gbps link using traditional congestion control schemes such as DCTCP, Swift, and Homa; ($iii$) we use the approach proposed in ~\cite{cassini_hotnets} to enforce static unfair bandwidth allocation to jobs; ($iv$) we use Cassini~\cite{cassini_nsdi24} centralized scheduler to force the communication patterns of the jobs to interleave as much as possible; ($v$) we augment the congestion control schemes in ($ii$) with \name.

\para{Isolation Experiments.} Figure~\ref{fig:jobs_pattern} shows the traffic pattern of the three DNN jobs in isolation. The iteration time of Job-1 varies between 1.5-1.65 seconds, while that of Job-2 and Job-3 is nearly 1.55 seconds. Importantly, we note that the communication patterns of these jobs are not compatible and cannot be perfectly interleaved.

\para{Traditional Congestion Control Experiments.} 
Figure~\ref{fig:itertime_vs_iternum} shows the time series graph of iteration times of the three jobs when they compete for network bandwidth using different schemes. As shown, traditional congestion control schemes struggle to interleave the communication patterns of these jobs, resulting in delayed iteration times for all three jobs.

\para{Static Unfairness}. Prior work~\cite{cassini_hotnets} proposes a decentralized method for temporarily interleaving the communication demands of DNN jobs using static unfairness. By assigning fixed, albeit unfair, bandwidths to the jobs, it enables their communication patterns to slide past each other, helping them interleave. The ``Static'' line in Figure~\ref{fig:itertime_vs_iternum} shows that the interleaving is unstable, causing oscillations between interleaved and overlapped states and fluctuations in iteration times.  

\para{Centralized Scheduler}. Cassini~\cite{cassini_nsdi24} computes per-flow interleaved schedules via a compute-intensive ILP. Although Cassini captures the periodic nature of DNN jobs, its ILP provides good performance for fully compatible cases, where each job’s communication phase fits exactly within others’ compute phases, enabling perfect interleaving. \new{The same is not true for partially compatible jobs. Cassini's ILP is built on the \textit{a priori assumption} of a pre-computed time period (perimeter of unified circle) over which the jobs will have an interleaved schedule. Cassini obtains this time period by taking the LCM of the jobs' iteration times profiled in \textit{isolation} and computes start-times of jobs assuming those traces remain fixed~\cite{cassini_nsdi24}. This maintains interleaving when jobs remain stationary with respect to each other.  In practice, communication overlap is asymmetric, so jobs contend with different subsets of others and incur unequal iteration time changes. These differential changes cause jobs' subsequent iteration start times to deviate from Cassini's planned schedule over time. While Cassini's runtime corrections may try to fix small drift, it is still chasing a moving target with its assumed schedule, leading to further churn and limited gains. Cassini focuses only on interleaving and not on the stability of its schedule}. We observed the same in our motivation example in Figure~\ref{fig:itertime_vs_iternum}(b): Cassini fails to sustain interleaving beyond the first two iterations, causing massive fluctuations in iteration times of jobs and poor performance like the fair sharing baselines. We provide a detailed explanation of this limitation of Cassini with simpler examples in Appendix~\S\ref{sec:cassini_limitation}.

\para{Our Proposal.} In contrast, \name-augmented congestion control protocols discover and maintain a partially interleaved state for all three DNN jobs. All 3 variants of \name stabilize to a \textasciitilde1.7 seconds iteration time, within 10\% of their ideal iteration time in isolation. \name speeds up iteration times by up to 15\% compared to traditional congestion control. \name guides the jobs to explore all possible bandwidth allocations on their own by sliding them relative to each other, until they converge to an interleaved schedule (Figure~\ref{fig:mlcc_motiv}). Unlike static unfairness schemes, \name does not arbitrarily bias bandwidth sharing. It gradually nudges flows to the nearest interleaved schedule and stabilizes them there. \new{Rather than assuming a fixed unified circle as Cassini does, \name converges to a stable interleaved schedule allowing the correct period, and thus the right circle, to emerge.}

\begin{figure}[t]
\centering
\includegraphics[width=0.43\textwidth]{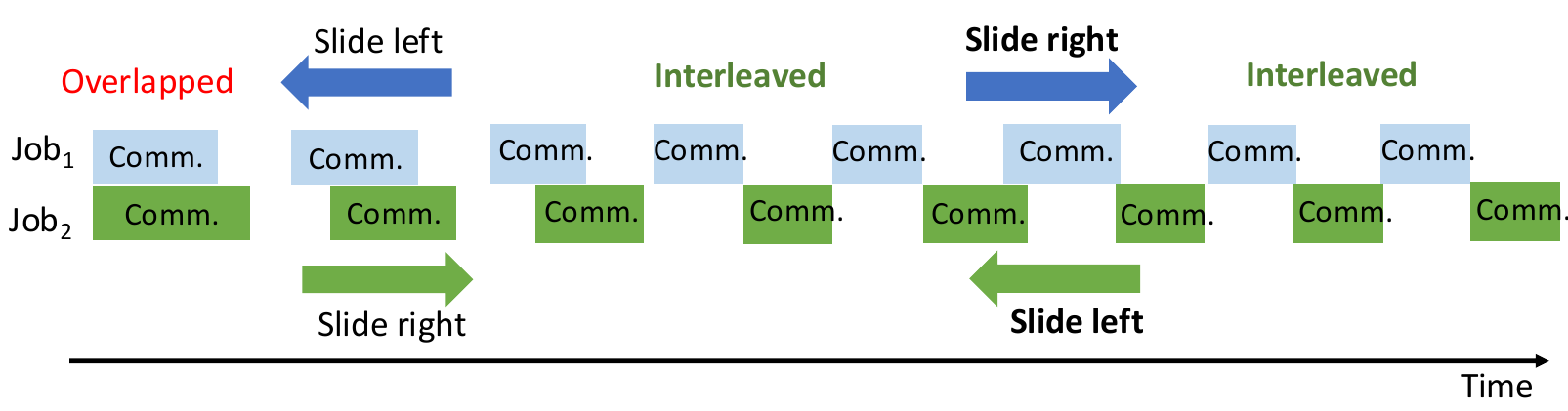}
    \vspace{-0.4cm}
    \caption{\namebf slides jobs to stable interleaved state.}
\label{fig:concept_sliding}
\end{figure}

\section{\namebf Design}
\label{sec:design}

\begin{figure*}[t]
\centering
\includegraphics[width=0.88\textwidth]{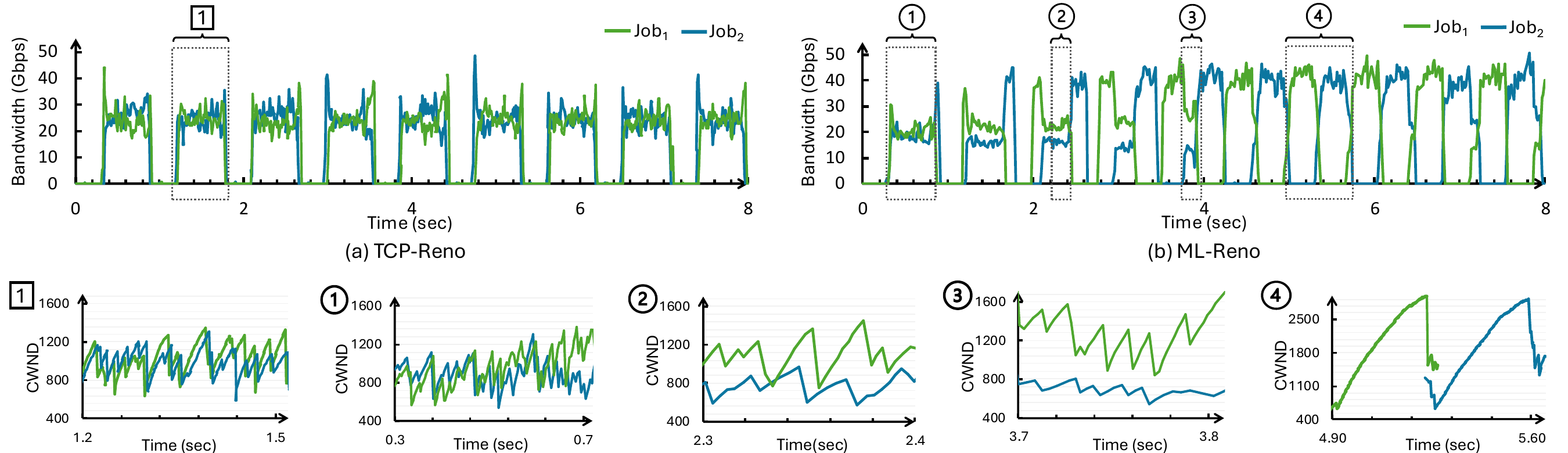}
\vspace{-0.4cm}
\caption{Comparison between Reno and \nameshortbf-Reno.} 
\vspace{-0.45cm}
\label{fig:job_favoring_cwnd_plot}
\end{figure*}

\name creates an unfair bandwidth allocation that favors jobs closer to completing their communication by adjusting the congestion window or rate of competing flows based on the percentage of bytes sent in the current iteration. This induces a controlled sliding effect, shifting the start times of the competing jobs to minimize the overlap between their communication phases. The sliding naturally disappears in fully interleaved states (no congestion). In partially interleaved states, successive shifts cancel across iterations, leading to a stable schedule unlike Cassini. Figure~\ref{fig:concept_sliding} illustrates \name's sliding effect for two jobs. Unlike static unfairness schemes, \name dynamically adjusts bandwidth allocations to immediately restore interleaving when perturbed, making it stable. To avoid convergence to suboptimal interleaved states, \name introduces a simulated annealing perturbation that promotes exploration and helps escape from local optima. Our evaluations show that most competing DNN jobs converge to an interleaved schedule (optimal or not) within \textasciitilde 30 iterations using \name's sliding effect (\S\ref{sec:eval_convergence_benchmarks}).

Next, we provide simple modifications to augment a suite of congestion control algorithms to generate their \name variants. Finally, we discuss the full algorithmic implementation of \name with more details on the annealing heuristic. We discuss more realistic settings of jobs with complex communication patterns in Section~\S\ref{sec:discussion}. 

\subsection{Adjusting the Sending Rate} 
\label{sec:aggressiveness_function}

\name achieves its iteration-by-iteration sliding by dynamically adjusting the congestion window (cwnd) or sending rate of flows based on a linearly increasing \textit{bandwidth aggressiveness function}, $\f$, where $\bytes$ is the amount of bytes successfully sent during the current communication phase, normalized by the maximum total bytes in any communication phase. The normalization is essential to switch the unfairness at fixed points to reach the interleaved state faster and to accommodate the huge diversity of DNN model sizes. \new{The slope of this function determines the aggressiveness of congestion control and hence \name's sliding effect, while the intercept is a lower bound on rate control to avoid throughput loss (Details in Appendix~\S\ref{sec:f_choice})}.

\textbf{\nameshortbf-Reno.} Consider the conventional TCP Reno~\cite{reno} algorithm. Reno uses a cumulative ack mechanism to acknowledge multiple in-flight packets with a single ack. Upon the receipt of an ack, Reno increases its congestion window as:

\begin{equation}
\setlength\belowdisplayskip{2pt}
\setlength\abovedisplayskip{2pt}
\label{eq:reno_ai}
cwnd \leftarrow cwnd + {\#num\_acks}/{cwnd}
\end{equation}
To implement \name, we update Equation~\ref{eq:reno_ai} to: 
\begin{equation}
\label{eq:mlreno_ai}
\setlength\belowdisplayskip{2pt}
\setlength\abovedisplayskip{2pt}
cwnd \leftarrow cwnd + \f \times {\#num\_acks}/{cwnd}
\end{equation}
where $\bytes = {bytes\_sent}/{max\_bytes}$ in the current training iteration and $\mathcal{F}$ is linear in $bytes\_ratio$: 
\begin{equation}
\label{eq:bw_agg}
\setlength\abovedisplayskip{2pt}
\setlength\belowdisplayskip{2pt}
\f =  Slope \times bytes\_ratio + Intercept
\end{equation}

\para{Example.} 
Figure~\ref{fig:job_favoring_cwnd_plot}(a) shows the bandwidth allocation of two competing DNN jobs using the conventional TCP Reno protocol. Our results hold for the general case but for clarity of illustration, assume both jobs are identical and start at the same time. As shown, during the communication phases of each iteration, the cwnd sawtooth profiles of both jobs are identical (area marked by \reqsq{1}). This is because Reno adjusts the cwnds of competing flows based on the same additive increase equation (Equation~\ref{eq:reno_ai}). Thus, each job receives half of the link bandwidth in every iteration and leaves the network underutilized during their compute phases.

Figure~\ref{fig:job_favoring_cwnd_plot}(b) shows the bandwidth allocation of the two jobs using our proposed \nameshort-Reno under the same scenario. \name's sliding mechanism gets triggered under congestion. Due to dynamically changing additive increase coefficients, the cwnd sawtooth profile of each job changes as their $\bytes$ increases over time. Even during the very first iteration, Job$_1$ happens to send a few packets earlier than Job$_2$ making their cwnd profiles diverge towards the end of their communication phase (area marked by \req{1}). As a result, Job$_1$ finishes its first iteration slightly earlier and starts its next iteration before Job$_2$. Again since Job$_1$ gets to send more data before Job$_2$ starts, Job$_1$'s cwnd becomes even larger than Job$_2$'s cwnd, grabbing a larger bandwidth share. This domino effect causes their cwnd profiles to diverge in subsequent iterations (areas marked by \req{2} and \req{3}). This sliding effect causes the two jobs to interleave after a few iterations (\req{4}). In the interleaved state, both jobs have large cwnds because there is almost no congestion. \nameshort-Reno maintains this interleaved state of jobs for all subsequent iterations, speeding up the iteration time of both jobs by $1.5 \times$ compared to Reno. If any externality, e.g. noise, job arrival/departure disturbs the interleaved state, \name's sliding effect and dynamic job favouritism goes into action again, moving the jobs back to the interleaved state in a few iterations. Appendix~\S\ref{sec:theory_2_jobs} provides the full theoretical analysis of \name's sliding mechanism for two competing DNN jobs.

\subsection{Augmenting existing CCAs with \namebf}
\label{sec:cca_impl}

We implement \nameshort-Reno in Linux kernel as a kernel module with $\approx$50 lines of code. Similarly, we augment other window-based congestion control (CUBIC~\cite{cubic}, DCTCP~\cite{dctcp}) and delay-based congestion control Swift~\cite{swift}. We also augment DCQCN~\cite{dcqcn}, RDMA's rate-based congestion control protocol, by modifying the NIC's hardware registers.

\textbf{\nameshortbf-DCTCP.} DCTCP's additive increase algorithm is similar to Reno. Hence, our changes to DCTCP's additive increase phase are similar to \nameshort-Reno's Equation~\ref{eq:mlreno_ai}.

\textbf{\nameshortbf-CUBIC}. CUBIC's window growth algorithm computes the target cwnd using a cubic function based on the time gap from the last multiplicative decrease event:
\begin{equation}
\setlength\abovedisplayskip{2pt}
cwnd \leftarrow CUBIC(time)
\setlength\belowdisplayskip{3pt}
\end{equation}
To implement \nameshort-CUBIC, we scale the calculated time gap by $\mathcal{F}$ before applying the cubic function: 
\begin{equation}
\setlength\abovedisplayskip{2pt}
cwnd \leftarrow CUBIC(\f \times time)
\setlength\belowdisplayskip{2pt}
\end{equation}
When different flows compete for bandwidth, a smaller value of $\mathcal{F}$ dilates the time calculation, making less favored flows with smaller $\mathcal{F}$, increase their cwnd slower than those with higher values. We scale the time gap with $\mathcal{F}$ instead of the additive increase factor to avoid distorting the characteristic cubic growth function of cwnd.

\textbf{\nameshortbf-Swift}. Swift is a delay-based congestion control whose window increase phase is similar to Reno's additive increase unless a specific target delay is exceeded. We only augment Swift's window increase equation to implement \nameshort-Swift, without changing its mechanism of target delay control.

\textbf{\nameshortbf-DCQCN}. DCQCN is a rate-based congestion control algorithm commonly used in lossless RoCE networks~\cite{dcqcn}. It  periodically increases the target rate by fixed steps $R_{AI}$ during the additive increase stage:
\begin{equation}
\small
\setlength\abovedisplayskip{2pt}
target\_rate \leftarrow target\_rate + R_{AI}
\setlength\belowdisplayskip{2pt}
\end{equation}
We update the above equation to: 
\begin{equation}
\small
\setlength\abovedisplayskip{2pt}
target\_rate \leftarrow target\_rate + \f \times R_{AI}
\setlength\belowdisplayskip{2pt}
\end{equation}
\textbf{\nameshortbf-Homa}. Homa is a receiver-driven protocol that uses switch priority queues to provide low latency by ordering incoming flows. It assumes that all congestion occurs on the receiver's link, which is rarely true for ML training workloads. \new{Consequently, it assigns priorities based only on the traffic each receiver sees, blind to competing jobs on different machines.} Given this, we change how it orders flows to emulate the \name policy by changing the assigned priority:
\begin{equation}
\small
\setlength\abovedisplayskip{2pt}
priority\_level \leftarrow round(max\_priority * \f)
\setlength\belowdisplayskip{2pt}
\end{equation}
where round means round to the nearest integer.

We experiment with different bandwidth aggressiveness functions, $\mathcal{F}$ and observed that any function satisfying the following three requirements achieves \name's interleaving goals: ($i$) the range is large enough to absorb the noise (e.g., slight variations in round-trip time (RTT) or iteration times) in the network;  ($ii$) the derivative of the function is non-negative;  ($iii$) all flows employ the same bandwidth aggressiveness function (details in Appendix~\S\ref{sec:f_choice}).

\begin{algorithm}[t]
\scriptsize
\begin{algorithmic}[1]
\IOComment{\textbf{Global Parameter} \textcolor{purple}{$COMP\_TIME$}}: Min gap between communication \peaks
\IOComment{\textbf{Global Parameter} \textcolor{purple}{$IDEAL\_TIME$}}: Ideal iteration time
\IOComment{\textbf{Global Parameter} \textcolor{purple}{$ANL\_RATE$}}: Decay rate of annealing heuristic
\IOComment{\textbf{Global Parameter} \textcolor{purple}{$ANL\_FREQ$}}: Frequency of random-restart annealing
\IOComment{\textbf{Global Parameter} \textcolor{purple}{$MTU$}}: Maximum packet size used by the system
\IOComment{\textbf{Constant} $\gamma = 0.9$}: Noise absorbing factor
\IOComment{Initialize state parameters:}

    \State $\textcolor{violet}{cwnd} = 10$ \textcolor{gray}{\footnotesize $\triangleright$ Congestion window}
    \State $\textcolor{violet}{max\_bytes} = 100$ \textcolor{gray}{\footnotesize $\triangleright$ Total segment data in an iteration}
    \State $\textcolor{violet}{max\_gap} = 100$ \textcolor{gray}{\footnotesize $\triangleright$ Max compute time between communication \peaks}
    \State $\textcolor{violet}{mean\_iter} = 0$ \textcolor{gray}{\footnotesize $\triangleright$ Current iteration time measured by algorithm}
    \State  $\textcolor{violet}{\bytes} = 0$ \textcolor{gray}{\footnotesize $\triangleright$ Fraction of bytes sent}
    \State $\textcolor{violet}{bytes\_sent} = 0$ \textcolor{gray}{\footnotesize $\triangleright$ Number of successfully sent bytes}
    \State $\textcolor{violet}{prev\_ack\_tstamp} = 0$ \textcolor{gray}{\footnotesize $\triangleright$ Timestamp of the previous ack}
    \State $\textcolor{violet}{prev\_iter\_tstamp} = 0$ \textcolor{gray}{\footnotesize $\triangleright$ Timestamp of the previous iteration}
    \State $\textcolor{violet}{anneal\_counter} = 0$ \textcolor{gray}{\footnotesize $\triangleright$ Iteration count from the last annealing}

\Procedure{UPDATE\_PARAMS}{$num\_acks$} 
\IOComment{\textbf{Input Parameter} $num\_acks$}
    \State $curr\_tstamp = \Call{get\_real\_time}$ 
    \label{line:current_time}
    \State \textcolor{gray}{\scriptsize $\triangleright$ Time Gap between current and the previous ACK}
    \State $curr\_gap = curr\_tstamp - \textcolor{violet}{prev\_ack\_tstamp}$ \label{line:current_gap}
    \State $\textcolor{violet}{prev\_ack\_tstamp} = curr\_tstamp $   
    \State \textcolor{gray}{\scriptsize $\triangleright$ Increment the $bytes\_sent$ and $bytes\_ratio$}
    \State \textcolor{violet}
    {$bytes\_sent$} += $num\_acks \times \textcolor{purple}{MTU}$ \label{line:bytes_sent}
    \State $\textcolor{violet}{\bytes} = \Call{min}{1, \,\, {\textcolor{violet}{bytes\_sent}}/{\textcolor{violet}{max\_bytes}}}  $ 
    \label{line:update_ratio}
    \State \textcolor{gray}{\scriptsize $\triangleright$ Check for the start of new communication \peak}
    \If {$curr\_gap$ > $\textcolor{purple}{COMP\_TIME}$ } \label{line:if_condition}
    \State \textcolor{gray}{\scriptsize $\triangleright$ New \peak, update bytes parameters}
    \State $\textcolor{violet}{\bytes} = 0$; $\textcolor{violet}{bytes\_sent}$ = 0; \label{line:reset_start}
    \State $\textcolor{violet}{max\_bytes}$ = \Call{max}{$\textcolor{violet}{max\_bytes}$, $\textcolor{violet}{bytes\_sent}$} \label{line:maxbytes}
    \EndIf
    \State $\textcolor{violet}{max\_gap} = \Call{max}{\textcolor{violet}{max\_gap}, curr\_gap}$ \label{line:maxgap}
    \State \textcolor{gray}{\scriptsize $\triangleright$ Check for the iteration boundary}
    \If {$curr\_gap$ > $\gamma \times \textcolor{violet}{max\_gap}$}
    \State \textcolor{gray}{\scriptsize $\triangleright$ Update the mean iteration time}
    \State $\textcolor{violet}{mean\_iter} \ +={(curr\_tstamp - \textcolor{violet}{prev\_iter\_tstamp})}/{\textcolor{purple}{ANL\_FREQ}}$ \label{line:itertime}
    \State $\textcolor{violet}{prev\_iter\_tstamp} = curr\_tstamp $
    \State \textcolor{gray}{\scriptsize $\triangleright$ Random-Restart applied after every ANL\_FREQ iterations}
    \State $\textcolor{violet}
    {anneal\_counter}$ ++;
    \If {$\textcolor{violet}{anneal\_counter}$ > \textcolor{purple}{$ANL\_FREQ$}} 
    \State \textcolor{gray}{\scriptsize $\triangleright$ Compute jump probability to apply random delay}
    \State $anl\_prob = \left(1 - \frac{\textcolor{purple}{IDEAL\_TIME}}{mean\_iter} \right)\times e^{(-\textcolor{violet}{ANL\_RATE} \times curr\_tstamp)}$ \label{line:jump}
    \State $\textcolor{violet}{mean\_iter} = 0$
    \State $\textcolor{violet}{anneal\_counter} = 0;$ 
    \While {$anl\_prob$}
    \State $\Call{random\_wait}{0, \textcolor{purple}{IDEAL\_TIME}}$ \label{line:random_wait}
    \EndWhile
    \EndIf
    \EndIf
    \State \Return 
\EndProcedure
\vspace{0.1cm}
\Procedure{CONGESTION\_AVOIDANCE}{$num\_acks$} 

\IOComment{\textbf{Input Parameter} $num\_acks$}
    \State \textcolor{gray}{\scriptsize $\triangleright$ Finally update the congestion window}
    \State $cwnd = cwnd + \mathcal{F}(bytes\_ratio) \times
    (\frac{num\_acks}{cwnd})$ \label{line:cwnd_update}
    \State \Return 
\EndProcedure

\end{algorithmic}

\caption{\sc{\nameshort-Reno Update\_Params Algorithm}
\label{alg:mltcp_iteraion_boundaries}}
\end{algorithm}

\subsection{\namebf Algorithm}
\label{sec:updating_parameters}
Algorithm~\ref{alg:mltcp_iteraion_boundaries} describes the congestion control and state variables of \nameshort-Reno. Its key steps apply to all TCP-based congestion control algorithms. We use Linux's pluggable congestion module~\cite{pluggable_cc, tcp_infra_split} to hook $UPDATE\_PARAMS$ procedure, shown in Algorithm~\ref{alg:mltcp_iteraion_boundaries}, into TCP stack. \new{TCP stack invokes CONGESTION\_AVOIDANCE upon every ACK and UPDATE\_PARAMS every 50 ACKs to minimize repeated computations. The computations in UPDATE\_PARAMS are also optimized further to reduce kernel overhead.}

\para{Updating \textbf{\textit{bytes\_sent}} and $\textbf{\textit{max\_bytes}}$.} 
We update $bytes\_sent$, the number of bytes successfully sent in the ongoing communication phase with the number of packets acked so far (line~\ref{line:bytes_sent}). We measure the start of a new communication \peak by checking whether the time gaps in the consecutive ack arrivals exceed a threshold value (lines~\ref{line:current_gap} and \ref{line:if_condition}). We use \new{$COMP\_TIME = 50$ ms} as a fixed threshold for detecting the start of the communication phase; this default works well across our workloads and does not require tuning. We then reset $bytes\_sent$ and $\bytes$ upon detecting the start of a new communication \peak (line~\ref{line:reset_start}). The total number of bytes to be sent in each communication \peak in an iteration and the intermediate compute times are nearly constant for each job and depend on the size of the DNN model, the parallelization strategy, and the communication collective. In our implementation, the algorithm automatically learns these values by measuring the total amount of data sent in every communication \peak followed by taking a running maximum to compute $max\_bytes$ (line~\ref{line:maxbytes}). 

\para{Measuring iteration time using $\textbf{\textit{max\_gap}}$.} 
We set the $IDEAL\_TIME$ parameter by profiling the job's iteration time in isolation. The algorithm regularly measures the maximum compute gap between two consecutive communication \peaks and stores it as a state variable, $max\_gap$ (line~\ref{line:maxgap}). We compute the job's current iteration time as the difference between the time instants at which we observe the largest compute gap ($max\_gap$). We then compute a running average of the measured iteration time over $ANL\_FREQ$ iterations (line~\ref{line:itertime}). 

\para{Random-Restart annealing}. 
For DNN jobs having multiple communication \peaks in the same iteration instead of a simple on-off pattern~\cite{cassini_nsdi24}, we deploy a simulated annealing heuristic (lines~26 to 38) in addition to \name's sliding mechanism to avoid partially interleaved states. The heuristic probabilistically adds random delay to the job (line~\ref{line:random_wait}) with some anneal probability to restart \name's sliding effect every $ANL\_FREQ$ iterations. The anneal probability (line~\ref{line:jump}) is a function of the average iteration of the job in the last $ANL\_FREQ$ iterations compared to the ideal value ($IDEAL\_TIME$). The anneal probability decreases exponentially with time at $ANL\_RATE$ (line~\ref{line:jump}) so that the heuristic terminates at an interleaved state after running enough number of times. Both $ANL\_FREQ$ and $ANL\_RATE$ are configurable parameters depending on the characteristics of the DNN jobs. 

\begin{figure}[t]
    \includegraphics[width=0.44\textwidth]{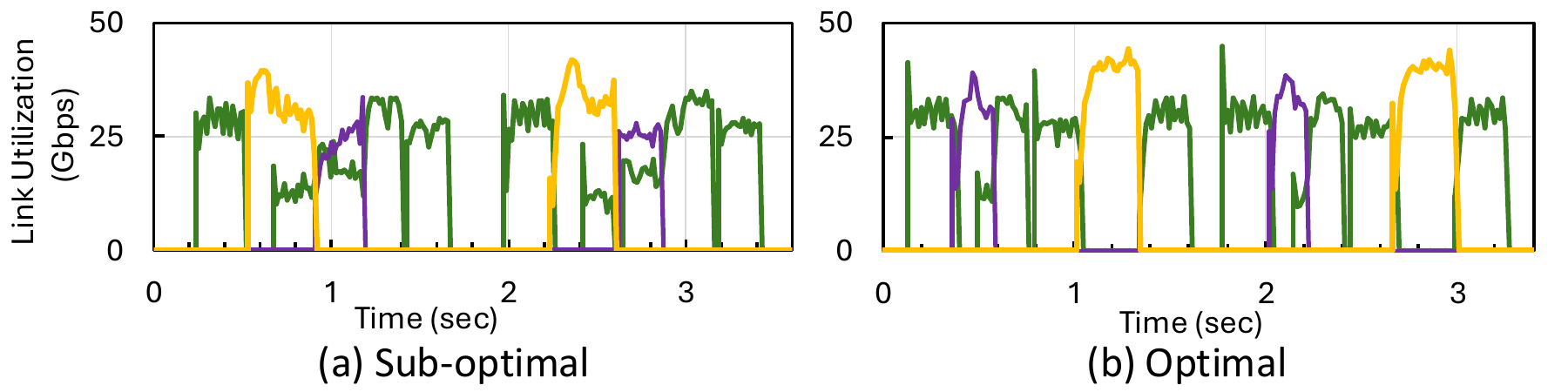}
    \vspace{-0.3cm}
        \caption{\nameshortbf-Reno's final interleaved state (a) \textbf{without} and (b) \textbf{with} Random-Restart Annealing.
        }
    \label{fig:annealing_perform}
\end{figure}

We show the effectiveness of this heuristic by running the three jobs shown in Figure~\ref{fig:jobs_pattern} (Section~\S\ref{sec:motivation}) using \nameshort-Reno. Figure~\ref{fig:annealing_perform} compares their interleaved schedules after ~150 training iterations with and without Random-Restart annealing, showing that annealing helps \nameshort-Reno achieve better interleaving. \new{We provide further evaluation of annealing in Appendix~\S\ref{sec:random_restart}, showing that it improves \name’s performance by over 1.2×, and also prove that it enables \name to converge to the best interleaved state with probability one.} 

\para{Hyperparameter Selection.} \label{sec:hyperparameter_setting} \name uses four parameters: the slope and intercept of $\fblack$, and two annealing parameters, ANL\_FREQ and ANL\_RATE. The slope and intercept depend on the augmented CCA, since different CCAs expose different control knobs through which $\fblack$ sets aggressiveness. A larger slope drives faster interleaving but can increase oscillation around the stable point. We observe that a range of Slope and Intercept values provides interleaving benefits in our evaluations (Figure~\ref{fig:heatmap} in Appendix~\ref{sec:eval_aggressiveness_function}). We chose values large enough to absorb noise in our testbed that disrupts the stability of interleaved schedule. In practice, similar CCAs such as TCP Reno and DCTCP work well with the same slope and intercept. We select these values once per CCA and use them unchanged across experiments; the chosen values are reported in Section~\S\ref{sec:eval_setup}.

The annealing parameters are not particularly sensitive. We observe that \name reaches the nearest interleaved state in less than 40 iterations even under high contention, so we fix $ANL\_FREQ=40$ throughout all experiments, independent of DNN model size. $ANL\_RATE$ determines how long annealing remains active; in practice, five annealing rounds are sufficient, corresponding to about 200 training iterations. Since our jobs run for thousands of iterations over hours, we set $ANL\_RATE$ to be a few minutes of training. Any new job, therefore, anneals only briefly to explore interleaved states with already existing jobs before stabilizing. These settings work well across all workloads we evaluate, and do not require rigorous tuning.

\section{Practical Considerations}
\label{sec:discussion}

In this section, we describe \name's behaviour for jobs with complex communication patterns, different model sizes or iteration times and multiple bottleneck scenarios.

\para{Partial Compatibility}.

We define \textit{compatibility score} in the same manner as Cassini~\cite{cassini_nsdi24} using its unified circle geometric abstraction as $(1 - average(Excess(demand)))$. It ranges from 1 when a fully interleaved schedule exists to 0 when all jobs require full link bandwidth at all times. 
Note that jobs with different iteration times may still be compatible with each other as long as their communication patterns can be fully interleaved with no network contention.    
In practical settings, the communication patterns of DNN jobs are not fully compatible with each other due to parallelization strategy, hyper-parameter settings, and model sizes.  

\begin{figure}[t]
\centering
\includegraphics[width=0.45\textwidth]{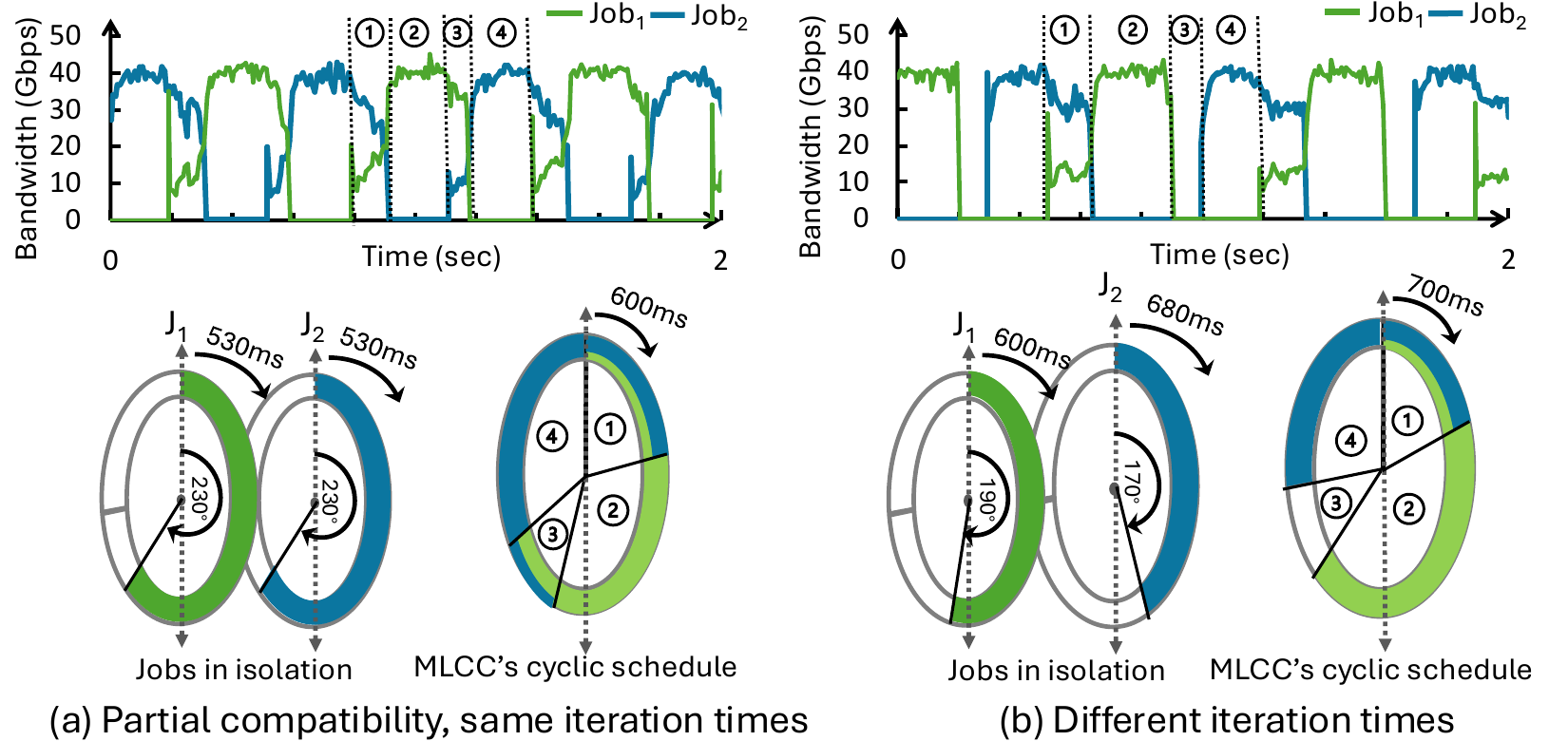}
\vspace{-0.4cm}
\caption{Illustration of \name's partially interleaved states}
\label{fig:partial_comp}
\vspace{-0.2cm}
\end{figure}

Computing a partially interleaved schedule for such jobs is intractable because overlap changes job iteration times, causing the schedule to depend on quantities that are themselves determined by the schedule. Nevertheless, \name's dynamic favouritism converges to a partially interleaved state with only brief congestion episodes. Figure~\ref{fig:partial_comp}(a) shows two GPT-2 data parallel jobs with similar iteration times (\textasciitilde530ms) and 0.75 compatibility converging to a partially interleaved schedule that maintains full network utilization. The period of \name's schedule (\textasciitilde600ms) slightly exceeds each job’s isolation time (same as the perimeter of Cassini's unified circle) due to the unavoidable overlapped communication phases.

\begin{figure}[t]
\centering
\includegraphics[width=0.42\textwidth]{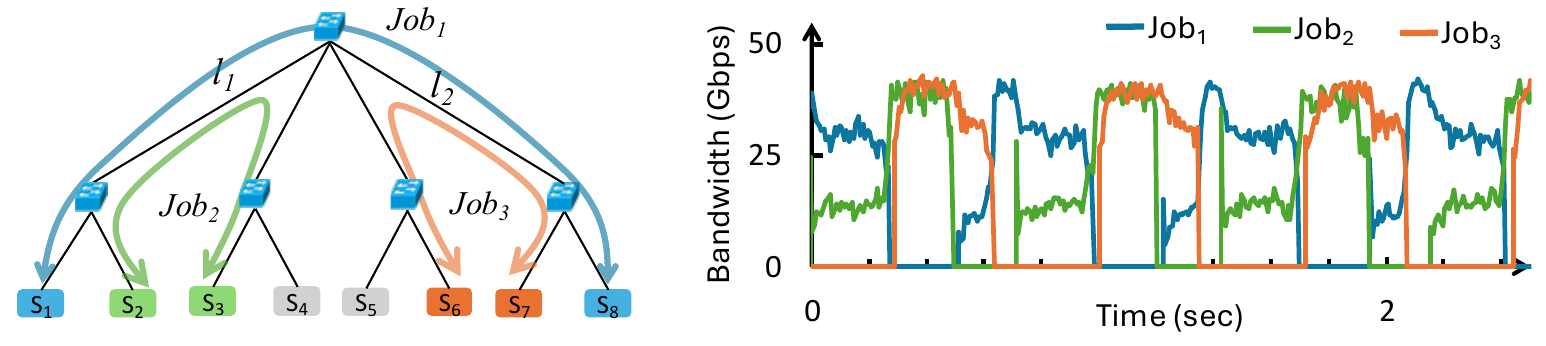}
\vspace{-0.4cm}
\caption{Interleaved state for jobs with multiple bottlenecks}
\label{fig:multi_btlnk}
\end{figure}

\begin{figure}[t]
\centering
\includegraphics[width=0.43\textwidth]{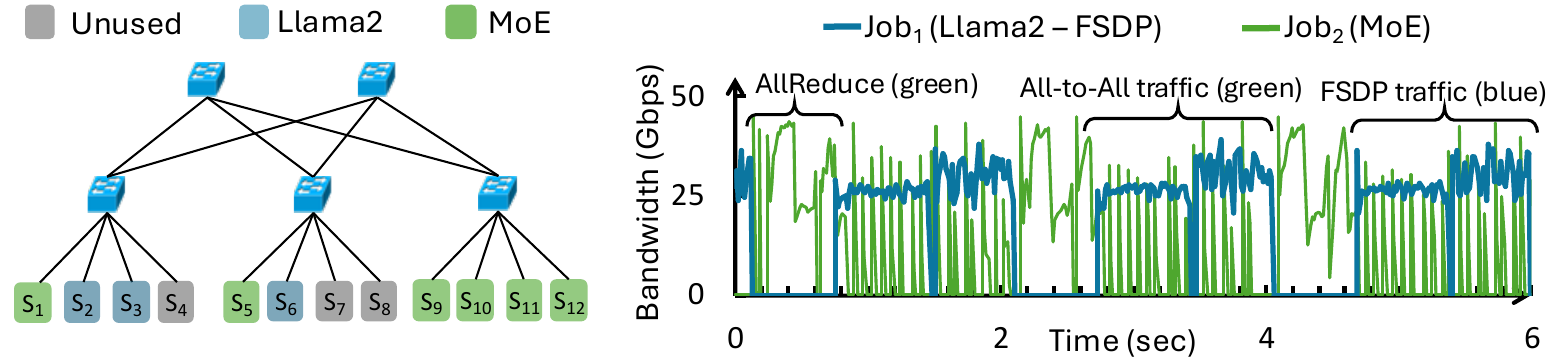}
\vspace{-0.4cm}
\caption{Interleaving MoE with FSDP Llama2}
\label{fig:moe_illus}
\end{figure}

\para{Delta in Iteration Times}. Small differences in job iteration time serve as an additional constant drift in addition to \name's sliding mechanism. In such a case, a fully interleaved state is unstable. Instead, the jobs converge to a partially interleaved schedule in which \name’s induced shift balances this drift (Appendix~\ref{sec:theory_2_jobs}). Figure~\ref{fig:partial_comp}(b) shows two GPT-2 jobs with different batch sizes and hence different iteration times (\textasciitilde600ms and 680ms respectively) converging to a partially interleaved schedule with period 700ms. As the iteration-time difference increases, \name no longer maintains per-iteration interleaving and instead yields multi-iteration interleavings similar to Cassini. \new{For highly incompatible jobs, no stable interleaved schedule may exist at any time scale. In such cases, there is no meaningful notion of convergence. \name exhibits \textit{ergodic} behavior: the jobs continue sliding past each other, and \name's long-term behavior approaches that of fair sharing, thereby preserving baseline performance at the least.} 

\para{Persistent Network Traffic}. \new{Pure} tensor parallel jobs (with no other parallelism degrees) are incompatible as they actively use the network throughout the training iteration, thereby limiting the opportunity of temporal interleaving. \new{\name's behavior for these jobs is close to fair sharing with no added gains.} Hybrid parallel jobs with at least one more non-tensor parallelism degree (e.g., Llama2  in Figure~\ref{fig:jobs_pattern}(a)), on the other hand, do have phases of high and low network demands~\cite{cassini_nsdi24} and hence benefit from interleaving.

\para{Multiple Bottlenecks}.

\name explores all possible bandwidth sharing states on the links by sliding the jobs to yield an interleaved schedule stable across all bottlenecks. Figure~\ref{fig:multi_btlnk} illustrates an interleaved schedule for three GPT-2 jobs (iteration time 530ms, 600ms and 680ms) that contend on different links. The resulting interleaved schedule maintains full network utilization throughout. 

\para{Expert Parallelism}. Expert parallelism (EP) introduces all-to-all communication with potentially variable volume. In principle, such variability could destabilize \name's interleaving. In practice, however, training stacks such as Megatron-DeepSpeed~\cite{megatron_git} suppress it through padding, fixed expert capacity, load balancing, and constrained token routing. A second challenge is that all-to-all traffic may contend with itself. In this regime, \name interleaves EP traffic with other jobs' traffic when they compete, but reverts to fair sharing for intra-job flows, which carry the same aggressiveness ($\fblack$); thereby preserving the best of both worlds.

Figure~\ref{fig:moe_illus} shows a partially interleaved schedule when a DeepSpeed-MoE job (250 million parameters, Data + Expert parallel)~\cite{gpt_moe} with an iteration time of 1.9 seconds competes with Llama2 (2 billion parameters, Fully Sharded Data Parallel)~\cite{llama} with an iteration time of 1.6 seconds under \nameshort-Reno on our 12-server 50G link testbed. Despite frequent all-to-all traffic in MoE, \nameshort-Reno interleaves the communication-heavy All-Reduce phases of both jobs and overlaps Llama2’s communication with MoE’s light all-to-all traffic, minimizing the slowdown of both.

\para{Packet Spraying}. Packet spraying is a well-known technique that distributes packets evenly across available outgoing links at each hop, thereby reducing congestion in the network. Since \name generates a time-domain interleaving of jobs, it works well with ECMP. On the other hand, packet spraying causes many more jobs to compete, limiting the effectiveness of a time-based approach like \name. Although effective with full-bisection, the merits of packet spraying reduce compared to \name in oversubscribed topologies. \new{We evaluate these scenarios in our large-scale simulations in Section~\S\ref{sec:pkt_sim_lb}}.

\section{Testbed Experiments}
\label{sec:evaluation}
In this section, we evaluate \name's performance, convergence, and robustness using testbed experiments.

\subsection{Evaluation Methodology}
\label{sec:eval_setup}
\textbf{Setup.} We augment four congestion control schemes (Reno, CUBIC, DCTCP, and DCQCN) with \name using kernel modules, driver scripts, and NCCL libraries. Our prototype includes 12 ASUS ESC4000A-E10 servers connected via a 64-port Tofino switch. Each server has one A100 Nvidia GPU~\cite{a100} 40~GB of HBM2 memory and one 50~Gbps Mellanox ConnectX5 NIC. We experiment with both TCP and RoCEv2. For RoCE experiments, we enable DCB~\cite{dcb} and PFC to support a lossless fabric for RDMA.  The servers run Ubuntu 20.04~LTS. Our experiments use PyTorch versions 1.13 and 1.8, CUDA version 11.7, and NCCL version  2.14.3.

\begin{figure}[t]
    \centering
    \includegraphics[width=0.9\columnwidth]{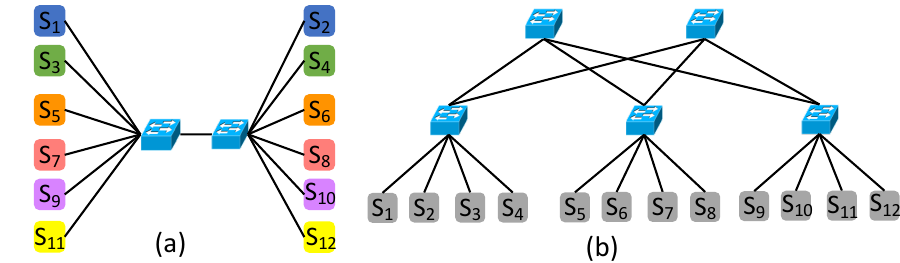}
    \vspace{-0.3cm}
    \caption{Our testbed topologies.}
    \label{fig:topology}
\end{figure}

\textbf{Topologies.} We use two topologies in our experiments. For benchmarking, we use a single-bottleneck dumbbell topology, shown in Figure~\ref{fig:topology}(a), where servers are grouped into six pairs: $\{S_1, S_2\}$, $\{S_3, S_4\}$, $\{S_5, S_6\}$, $\{S_7, S_8\}$, $\{S_9, S_{10}\}$, $\{S_{11}, S_{12}\}$. Each pair of servers trains a different DNN job and communicates gradient updates via the bottleneck link. We change the number of active server pairs throughout the experiments. For performance measurements, we use a hierarchical oversubscribed topology, shown in Figure~\ref{fig:topology}(b), to capture the impact of multiple bottlenecks in the network. Each link has a capacity of 50 Gbps. 

\textbf{DNN workloads.} We use nine real-world DNN models: VGG16~\cite{vgg16}, WideResNet101~\cite{wideresnet}, RoBERTa~\cite{roberta},  CamemBERT~\cite{camembert}, GPT-1~\cite{gpt_1}, GPT-2~\cite{gpt_2}, GPT-3~\cite{gpt_3}, BERT~\cite{bert_orig, megatronlm}, Llama~\cite{llama, colossalai}. Table~\ref{tab:model_parameters} summarizes the models and batch sizes used in our experiments.

The default TCP-based protocols cannot saturate 50~Gbps bandwidth due to kernel overheads~\cite{tcp_overhead, stack_overhead}. To overcome this, we create multiple sockets per job on each server to parallelize the transmissions using Google's NCCL fastsocket library~\cite{nccl_fastsocket}. In our experiments, we aggregate statistics across all sockets that belong to the same job. Following prior work~\cite{incast} we set the RTO minimum to 1~$ms$. For CUBIC, we scale the bic\_scale parameter by 10$^{10}$ to enable it to work on smaller RTT scales. We implement the following schemes:

\begin{itemize}[align=left, leftmargin=0pt, labelindent=0pt, listparindent=\parindent, labelwidth=0pt, itemindent=!]
\itemsep0em 
\item \textbf{\namebf variants}. 
We create pluggable kernel modules to implement \name's variants of Reno, CUBIC and DCTCP following Algorithm~\ref{alg:mltcp_iteraion_boundaries}. Since the DCQCN algorithm is implemented inside Mellanox NICs, we do not directly change the NIC implementation. The NICs API exposes a parameter called $rp\_ai\_rate$ corresponding to DCQCN's $R_{AI}$ rate. We use a C++ program that continuously monitors the amount of data sent by reading the NIC's hardware counter registers and updates $rp\_ai\_rate$ based on \nameshort-DCQCN's algorithm. Finally, we adjust the function $\mathcal{F}$'s aggressiveness by setting its parameters as follows: \nameshort-Reno and \nameshort-DCTCP use $Slope=1.75$ and $Intercept=0.25$; \nameshort-CUBIC uses $Slope=1.0$ and $Intercept=0.5$; \nameshort-DCQCN uses $Intercept=0.267$ and $Slope=1.067$.

\item \textbf{Ideal.} For benchmarking the performance of an \name variant (e.g. ML-Reno), we compare it against the best case performance when a job is run in isolation using the base congestion control scheme (e.g. Reno).

\item \textbf{Cassini.} We use the Cassini~\cite{cassini_nsdi24} scheduler as another baseline. Cassini employs a centralized scheduler to determine a series of time-shift values, adjusting job start times to achieve interleaving. It takes each job's optimal iteration time and uses an end-host agent to enforce the time shifts on each server.   

\item \textbf{Static.} We use the approach proposed in ~\cite{cassini_hotnets} as another baseline by manually setting the congestion window parameters so that the bandwidth share between the competing jobs is statically configured to be unfair. 

\end{itemize}

\begin{table}
\scriptsize
\centering
\renewcommand{\arraystretch}{0.95} 
\linespread{1.05}\selectfont\centering
\begin{tabular}{|p{2.2cm}|p{1cm}|p{2.4cm}|p{1.0cm}|} 
\hline
DNN       & B.Sz./GPU  & Parallelization strategy & Type \\\hline
VGG16~\cite{vgg16} & 1400  & Data Parallel & Vision  \\
WideResNet101~\cite{wideresnet} & 800 & Data Parallel & Vision \\
RoBERTa~\cite{roberta}  & 28 & Data Parallel &Language \\
CamemBERT~\cite{camembert}  &  28  & Data Parallel &  Language \\
LLAMA2 ~\cite{llama}  &  4-16  & Hybrid 3D Parallelism & Language \\
BERT ~\cite{bert}  &  10-64  & Data Parallel & Language \\
GPT-1~\cite{gpt_1}  &  31  & Data Parallel & Language \\
GPT-2~\cite{gpt_2} &  5-44  & Hybrid Parallel & Language \\  
DeepSpeed-MoE ~\cite{gpt_moe}  &  3  & Data/Tensor/Expert  & MoE \\ 
\hline
\end{tabular}
\vspace{-1mm}
\caption{DNN models used in our experiments.}
\label{tab:model_parameters}
\end{table}

\begin{figure*}[t]
    \centering
    \vspace*{-0.2cm}
    \includegraphics[width=0.8\textwidth]{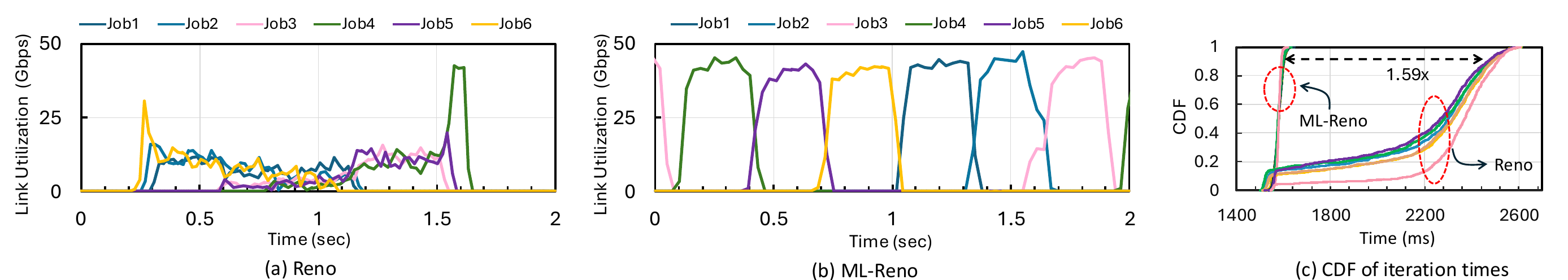}
    \vspace{-0.42cm}
    \caption{Bandwidth allocation when six jobs share the bottleneck link.}
    \label{fig:bw_6_jobs}
    \vspace{-0.15cm}
\end{figure*}

\begin{figure*}[t]
\centering
\begin{minipage}{0.29\textwidth}
    \centering
    \scriptsize
    \renewcommand{\arraystretch}{0.9} 
    \linespread{1.02}\selectfont\centering
    \captionsetup{type=table} 
    \begin{tabular}{|p{2cm}|p{0.6cm}|p{0.6cm}|p{0.6cm}|}
    \hline 
    Competing jobs (batch size) &  Num. servers & Parallel-ism & Comp-atibility   \\ \hline
     WideResNet101~(800)  &  5  & data & 0.88 \\
            VGG16~~(1400)  &  5  & data  & \\
    \hline
     CamemBERT~(28)  &  5   & data & 0.9  \\
            RoBERTa~~(28)  &  5  & data  & \\
    \hline
     GPT-1~~(31)  &  4 & data & 1   \\
           GPT-1~~(31)  &  4  & data & \\
           GPT-2~~(15) &  4  & data &  \\
    \hline
    GPT-3~~(3)  &  4 & hybrid & 1 \\ 
           GPT-3~~(3)  &  4 & hybrid & \\
    \hline
    \end{tabular}
    \caption{Snapshots of competing jobs.}
    \label{table:snapshot}
\end{minipage}
\hspace{0.4cm}
\begin{minipage}{0.3\textwidth}
    \centering
    \includegraphics[width=.88\textwidth]{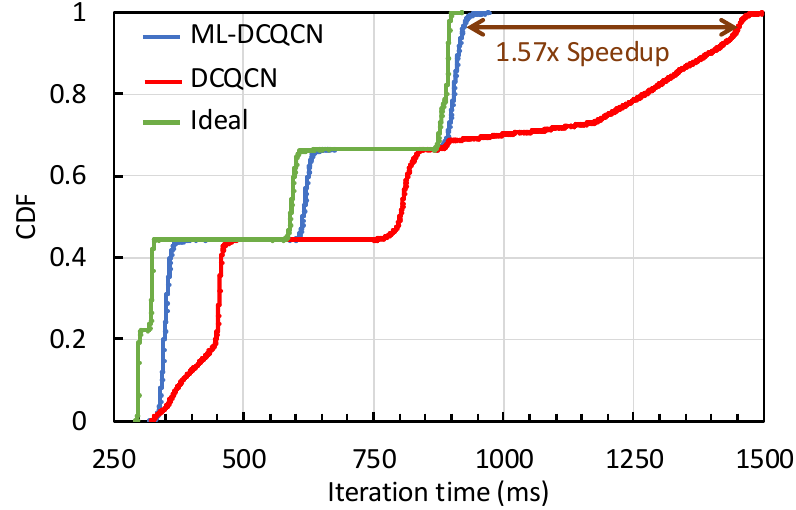}
    \vspace{-0.4cm}
    \caption{\centering Performance of \nameshort-DCQCN for jobs in Table~\ref{table:snapshot}.}
    \label{fig:mlqcn_multi_job}
\end{minipage}
\hspace{0.2cm}
\begin{minipage}{0.3\textwidth}
    \centering
    \vspace*{-5mm}
    \includegraphics[width=0.95\textwidth]{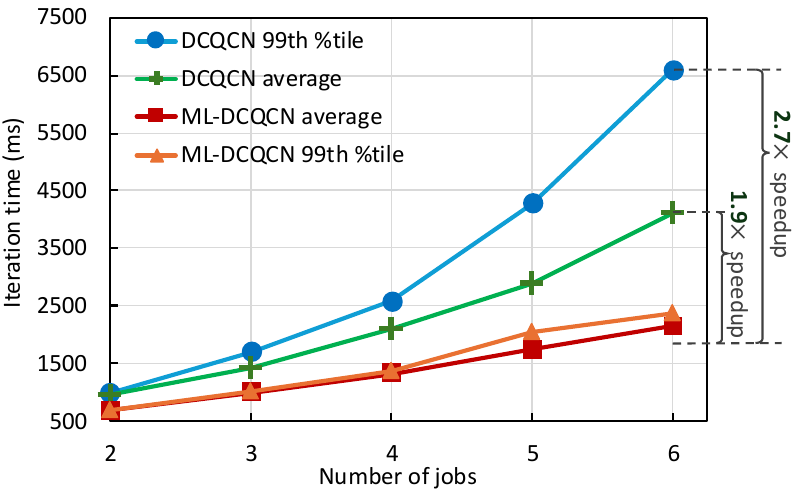}
    \vspace{-0.5cm}
    \caption{\centering Comparison of iteration times as the number of jobs increases.}
    \label{fig:iter_time_vs_num_jobs_dcqcn} 
    \vspace{-0.4cm}
\end{minipage}
\vspace{-0.5cm}
\end{figure*}

\subsection{Overall Performance}

\label{sec:overall_performance}

\begin{figure*}[t]
        \centering
        \includegraphics[width=0.88\textwidth]{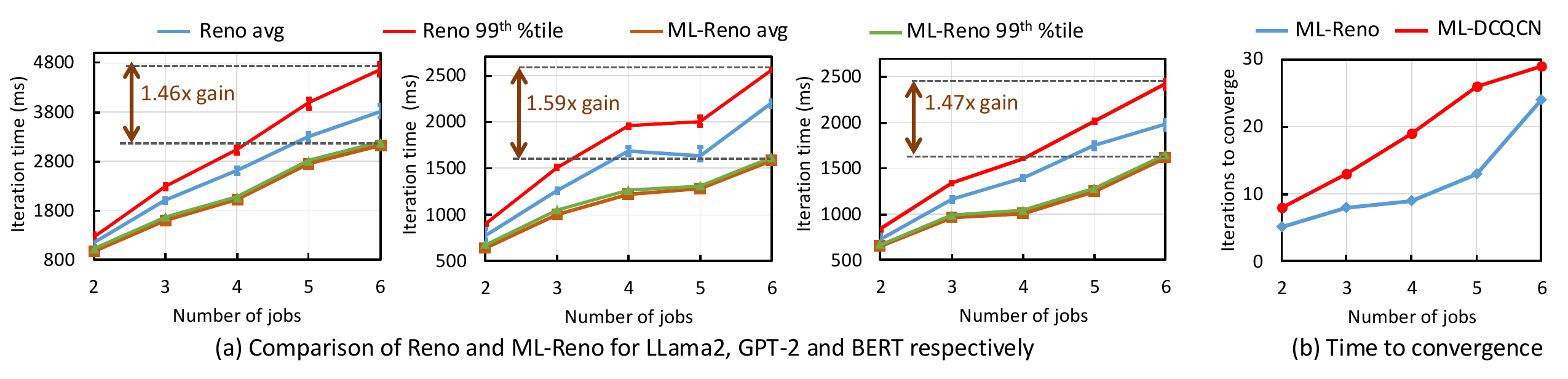}
        \vspace{-0.55cm}    
        \caption{Comparison of iteration times achieved by Reno and \nameshort-Reno for different DNN models.}
        \vspace{-0.3cm}    
        \label{fig:iter_time_vs_num_jobs}         
\end{figure*}

We begin our evaluations by comparing the training iteration times when six GPT-2 training jobs share our dumbbell topology (Figure~\ref{fig:topology}). Each job uses data-parallelism on the Book Corpus dataset~\cite{corpus}. We train each job for 1000 iterations ($\approx$30 minutes).  Figure~\ref{fig:bw_6_jobs}(a) illustrates the bottleneck link's utilization of each job under the default TCP Reno protocol. As shown, each job takes only one-sixth of the available bandwidth. In contrast, Figure~\ref{fig:bw_6_jobs}(b) shows \nameshort-Reno achieves an interleaved schedule, enabling each job to fully utilize the bottleneck link one at a time. Figure~\ref{fig:bw_6_jobs}(c) plots the CDF of iteration times between the two congestion control variants, demonstrating that \nameshort-Reno accelerates the average and 99$^{th}$ percentile iteration time by 1.4$\times$ and 1.59$\times$, respectively.

We then follow prior work's snapshot trace~\cite{cassini_nsdi24} and generate different combinations of several DNN training jobs (such as GPT-2, GPT-3, VGG16, RoBERTa, and WideResNet) sharing our hierarchical testbed topology, i.e. the Clos topology in Figure~\ref{fig:topology}(b). 
Table~\ref{table:snapshot} lists the DNN model names and details of each snapshot in this experiment. Figure~\ref{fig:mlqcn_multi_job} plots the CDF of the iteration times of all our snapshot configurations. We train each job for 500 iterations for each snapshot and collect the training iteration times with \nameshort-DCQCN and DCQCN as the congestion control algorithm.  
Compared to DCQCN, \nameshort-DCQCN improves the average and 99$^{th}$ percentile training iteration times by 1.33$\times$ and 1.57$\times$, respectively. \nameshort-DCQCN's average and tail iteration times are within $5\%$ and $4\%$ of that of ideal interleaved schedule. 

Figures~\ref{fig:iter_time_vs_num_jobs_dcqcn} and \ref{fig:iter_time_vs_num_jobs} compare the average and 99$^{th}$ percentile training iteration times of DCQCN and Reno with \nameshort-DCQCN and \nameshort-Reno as we increase the number of \new{compatible GPT-2/BERT/Llama2} jobs competing on the bottleneck link. As the number of jobs increases, \nameshort-DCQCN and \nameshort-Reno's performance benefits also increase. When six jobs compete for bandwidth, \nameshort-DCQCN accelerates average and 99$^{th}$ percentile tail iteration time by up to 1.9$\times$ and 2.7$\times$, respectively. Similarly, \nameshort-Reno accelerates average (and tail) iteration times by up to 1.23$\times$ (1.46$\times$), 1.4$\times$ (1.59$\times$), and 1.23$\times$ (1.47$\times$), for Llama, GPT-2, and BERT models, respectively. 

\subsection{Convergence Benchmarks}
\label{sec:eval_convergence_benchmarks}

\para{Time to convergence.} We now evaluate the convergence of \name-enabled congestion control algorithms when multiple compatible GPT-2 training jobs share our dumbbell topology (Figure~\ref{fig:topology}(a)). We run this experiment with different congestion control algorithms and count the number of iterations until the iteration time gets within 10\% of that of the Ideal scheme. 
As shown in Figure~\ref{fig:iter_time_vs_num_jobs}(b), the number of iterations that a \name-enabled congestion control takes to converge increases with the number of competing jobs. When there are six jobs competing, both \nameshort-Reno and \nameshort-DCQCN converge within the first $\approx$30 iterations. This is reasonable, given that DNN training jobs last for thousands of iterations. Besides, in a typical DNN training setting with ECMP load balancing, it is unlikely that a large number of DNN jobs would contend for bandwidth on a single link.

\noindent \para{Performance Benchmarks}. To dive deeper into the performance of different congestion control protocols, we repeat the previous experiment with two GPT-2 jobs and measure the training iteration time and packet loss/ECN marks of various congestion control schemes.

\begin{figure*}[t]
    \centering
    \includegraphics[width=0.89\textwidth]{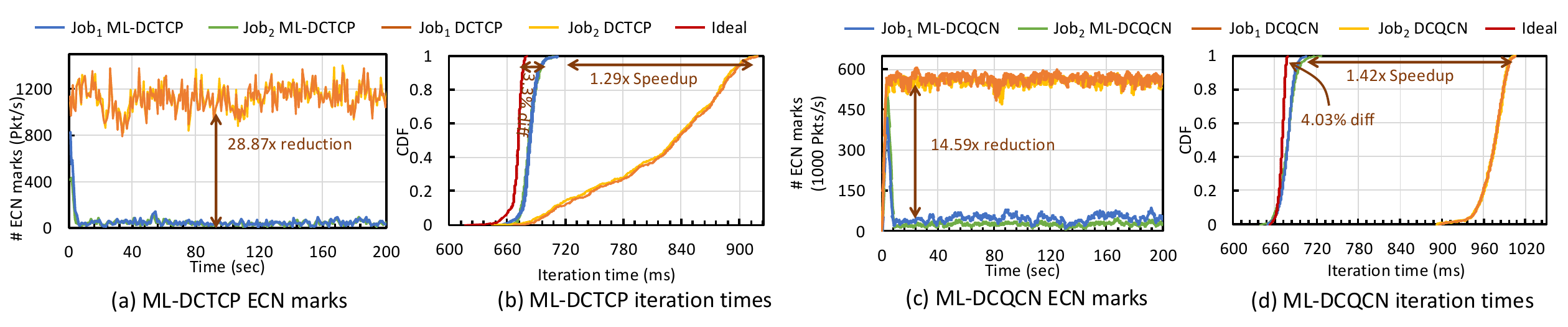}
    \vspace{-0.55cm}
    \caption{Benchmarking performance of \nameshortbf-DCTCP and \nameshortbf-DCQCN for two competing jobs.}
    \vspace{-0.3cm}
    \label{fig:gpt2_all_together}
\end{figure*}

\para{\nameshortbf-DCTCP.} \nameshort-DCTCP obtains a $1.29\times$ iteration speedup compared to normal DCTCP, within 3.3\% of the ideal scheme. 
Compared to DCTCP, \nameshort-DCTCP reduces the number of ECN marks by 28.87$\times$, accelerating the average and 99$^{th}$ percentile training iteration times by 1.2$\times$ and 1.29$\times$, respectively. 

\para{\nameshortbf-DCQCN.} The previous experiments show the performance of the traditional TCP/IP stack augmented with \name. With RoCE, we achieve even better results, as shown in Figures~\ref{fig:gpt2_all_together}(c) and (d): \nameshort-DCQCN jobs experience 14.59$\times$ less ECN marks, and improve the average and 99$^{th}$ percentile iteration times by 1.34$\times$ and 1.47$\times$, respectively. 

\new{Appendix~\S\ref{sec:app_benchmarks} provides similar benchmarking results for \nameshort-Reno and \nameshort-CUBIC.}

\begin{figure}[t]
    \centering
    \includegraphics[width=0.43\textwidth]{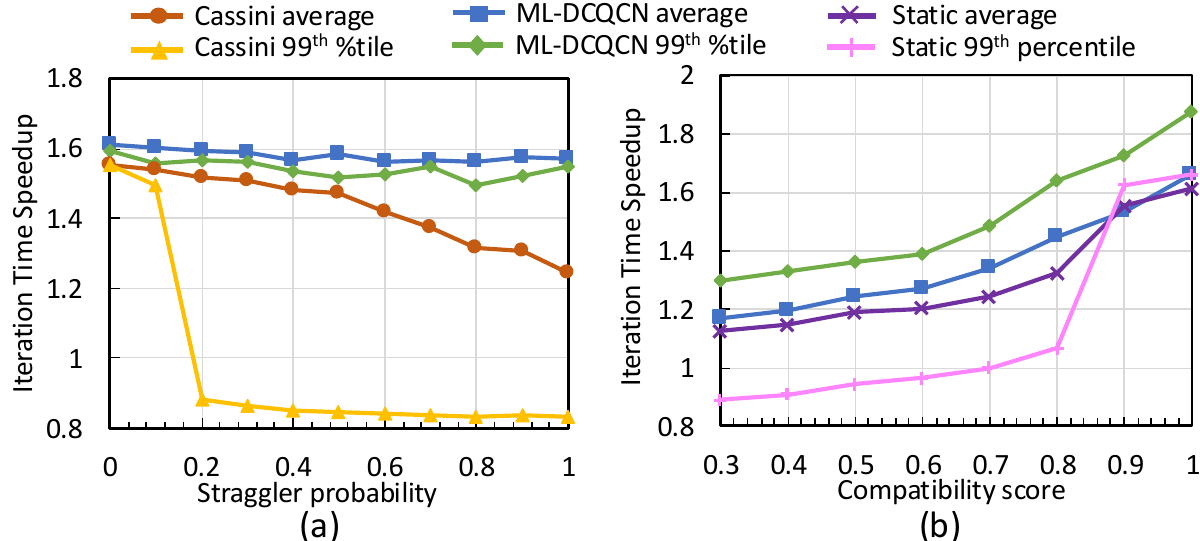}
    \vspace{-0.45cm}    
    \caption{Robustness to stragglers and partial compatibility.}
    \label{fig:straggler_compatibility_together} 
\end{figure}

\subsection{Robustness Experiments}
\label{sec:eval_robustness_to_stragglers}

\para{Stragglers}. To demonstrate \name's robustness to noise in practical settings, we train two GPT-2 jobs sharing a bottleneck link and synthetically insert sleep commands in each server's training code based on a uniform straggling probability. If a server chooses to lag at a particular iteration, it sleeps for a period uniformly randomly chosen between 5\% and 10\% of the ideal iteration time of the job in isolation. Frequent straggling events break the flow schedules for a few iterations, forcing the jobs to compete for bandwidth. 
Figure~\ref{fig:straggler_compatibility_together}(a) shows the training iteration speedups of \nameshort-DCQCN and Cassini compared to default DCQCN as the straggler probability increases. Cassini's tail iteration time speedup drops sharply as the straggler probability goes beyond 10\%. This is because Cassini's end-host agent forces the straggling server to skip an iteration, hoping that the noise in the system will be resolved in the next iteration and the jobs will return to an interleaved state. In contrast, \nameshort-DCQCN's bandwidth aggressiveness function dynamically reacts to the straggling server to re-converge to the interleaved state quickly. As a result, \nameshort-DCQCN's speedup remains more or less constant even if straggler probability increases.  

\para{Partial Compatibility}. To study the robustness of \name when partially compatible jobs compete for bandwidth, we train three GPT-2 jobs with different batch sizes (hence, different iteration times) in accordance with their compatibility score. 
Figure~\ref{fig:straggler_compatibility_together}(b) compares the average and 99$^{th}$ percentile training iteration time speedups of \nameshort-DCQCN and Static approaches over default DCQCN. 
Increasing the compatibility score naturally increases the speedup of both approaches as their communication patterns interleave better. Still, \nameshort-DCQCN outperforms 99$^{th}$ percentile iteration times of Static, at all compatibility scores. \nameshort-DCQCN uses its aggressiveness function to dynamically adjust the sending rate, making it tolerant to small noise in the system. In contrast, the Static unfairness approach is unaware of the network's conditions. When the compatibility score is below 0.7, the Static approach performs extremely poorly, where the 99$^{th}$ percentile training iteration times are lower than the default DCQCN approach.

\section{Packet Level Simulations}
\begin{figure*}[t]
    \centering
    \begin{minipage}{0.48\textwidth}
        \centering
        \includegraphics[width=\linewidth]{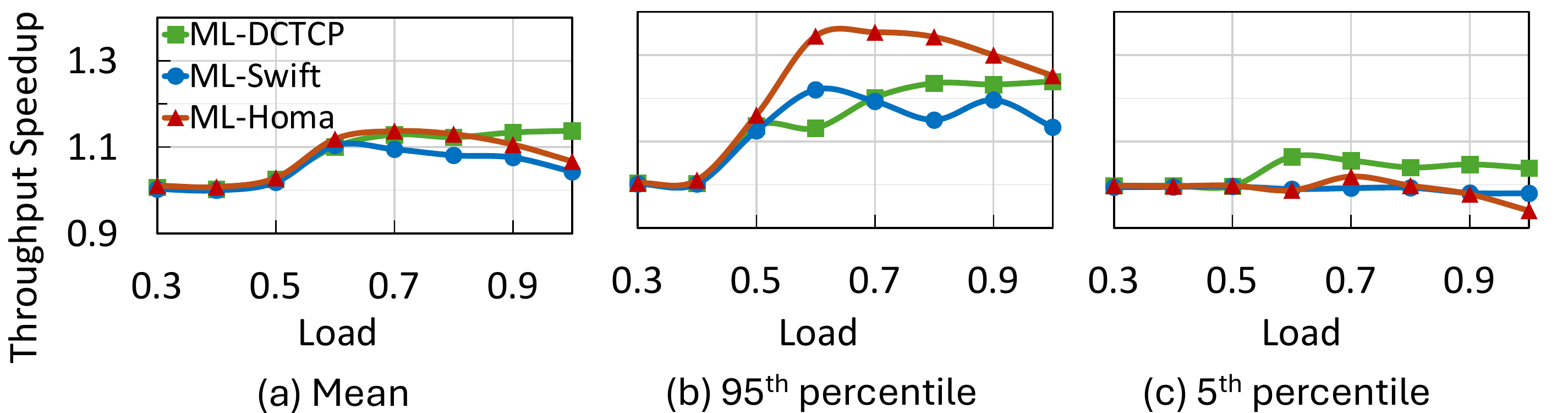}
        \vspace{-0.8cm}
        \caption{\name's performance under varying load} 
        \vspace{-0.4cm}
        \label{fig:mltcp_sim_topology}        
    \end{minipage}
    \begin{minipage}{0.49\textwidth}
        \centering
        \includegraphics[width=\linewidth]{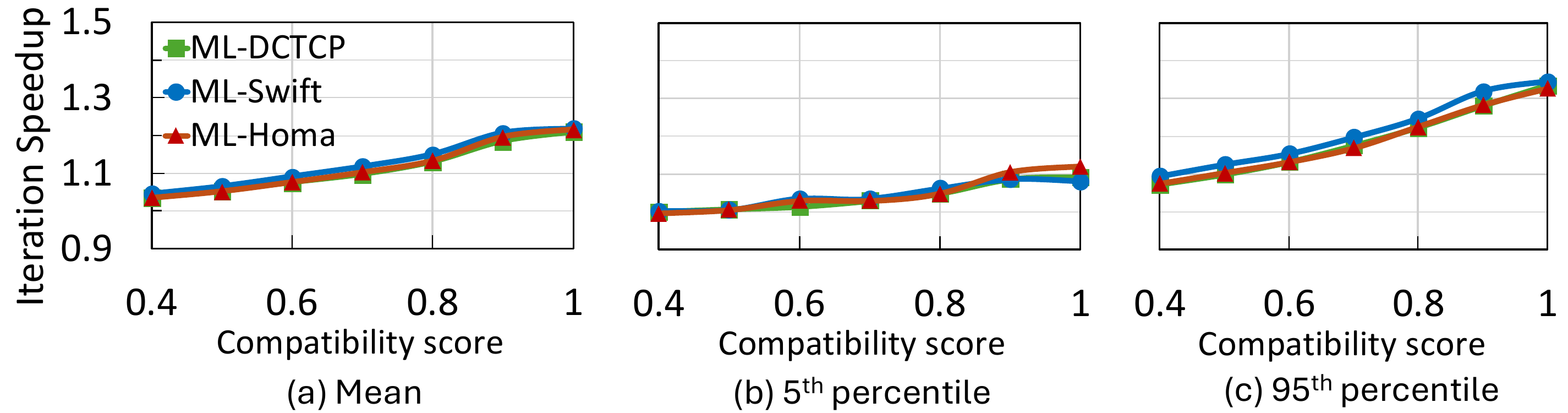}
        \vspace{-0.8cm}
        \caption{\name's performance under varying job-compatibility}
        \vspace{-0.35cm}
        \label{fig:mltcp_sim_perf}
    \end{minipage}
    \vspace{-2mm}
\end{figure*}
In this section, we evaluate \name's performance and fairness properties using packet-level simulation. We built on top of the YAPS simulator~\cite{phost}, adding \textasciitilde 8k lines of code. We also evaluate additional congestion control protocols unavailable in our testbed, namely Swift~\cite{swift} and Homa~\cite{homa}. 

\subsection{Large-Scale Evaluation}
\label{sec:pkt_sim_topo}
We study \name's performance on a 36-node (288 GPUs) Fat-Tree with full-bisection bandwidth. We sample hundreds of job placements at varying loads from a scheduler running on randomized job traces/snapshots generated by a Poisson process. The workload includes BERT, Llama2, and GPT models, whose iteration times and compatibility scores depend on job placement, thereby replicating a real-world setup. 
Figure~\ref{fig:mltcp_sim_topology} reports the speedup in \new{the \textit{training throughput}--measuring the number of jobs completed per unit time}--for the mean, 95$^{th}$ and 5$^{th}$ percentile job scenarios across hundreds of snapshots, capturing average, best, and worst-case performance. Note that the speedup of a \name variant is measured relative to the non-\nameshort version of the congestion control.

As load rises, contention between concurrent jobs increases, reducing the throughput of conventional congestion control, which cannot interleave traffic. A higher load causes more congested links per job, making interleaving itself more challenging. Hence, the performance speedup does not change monotonically with load. \nameshort-Homa achieves up to 1.35$\times$ performance improvement at the 95th percentile while degrading throughput by at most 6\% at the 5th percentile. Both \nameshort-DCTCP and \nameshort-Swift achieve a performance speedup of 1.2$\times$ at the 95th percentile.

\nameshort-Homa uses packet priorities to interleave; this makes \nameshort-Homa an extremely aggressive form of \name that accelerates interleaving. However, this aggressiveness can backfire by pausing many flows in the network to clear out one path. For \nameshort-Swift and \nameshort-Homa, gains taper beyond 70\% load. \nameshort-Swift maintains a delay target by trading off throughput, so under heavy contention, its ability to interleave degrades. \nameshort-DCTCP, which supports partial bandwidth sharing unlike \nameshort-Homa and increases its window aggressively unlike \nameshort-Swift, delivers consistent gains at all loads.

\subsection{Compatibility and Starvation Analysis}
\label{sec:pkt_sim_speedup}
\para{Arbitrary Communication Patterns}. We evaluate \name’s single-bottleneck performance on randomly sampled jobs with arbitrary periodic traffic as a function of their compatibility scores using hundreds of simulations. Figure~\ref{fig:mltcp_sim_perf} reports mean, 95$^{th}$ and 5$^{th}$ percentile iteration speedups relative to each protocol’s non-\nameshort baseline, capturing average, best-case, and worst-case behavior.
Each simulation samples a new set of competing jobs to capture workload variability. At full compatibility, \name achieves up to 1.35$\times$ iteration speedup across all variants. As compatibility decreases, performance degrades smoothly towards the baseline congestion control.

\begin{figure}
    \centering
    \includegraphics[width=0.45\textwidth]{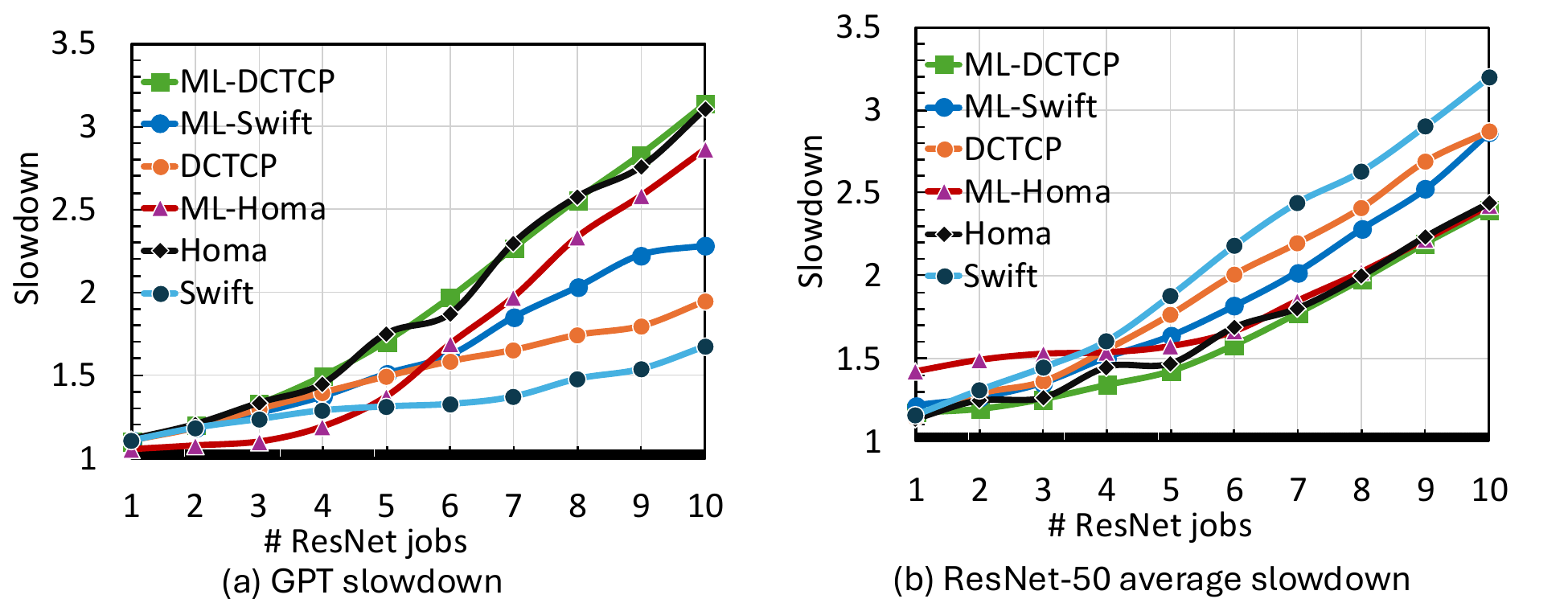}
    \vspace{-0.4cm}
    \caption{Analysis of \name variants and baseline congestion control when elephant competes with mice.}   
    \label{fig:mltcp_sim_starvation}
\end{figure}

\para{Starvation among ML jobs.} To observe the unfairness and starvation behavior of \name, we run a single GPT-2 (elephant, 1.6 GB per iteration) job competing with an increasing number of ResNet-50(mice, 98 MB per iteration) jobs. Figure~\ref{fig:mltcp_sim_starvation} reports the average iteration slowdowns of GPT-2 and ResNet under \name compared to their isolation iteration times. Because \name prioritizes jobs by $\bytes$, it naturally favors smaller jobs, but only until the large job has transmitted \textasciitilde 30\% of its data (Appendix~\ref{sec:theory_fairness}) after which it gets more bandwidth share. 
\nameshort-Swift hurts GPT-2 by 6\% while boosting the ResNet-50 jobs by 6\% when number of competing jobs is less than 4. Beyond this point, \nameshort-Swift and \nameshort-DCTCP speed up ResNet-50 by up to 1.2$\times$ and 1.1$\times$ respectively, while slowing GPT-2 by at most 1.6$\times$. \nameshort-Homa reduces GPT-2’s slowdown by up to 8\% without hurting ResNet-50 by interleaving via priority queues—alternating communication between jobs. Unlike SRPT, \nameshort-Homa’s \bytes-based ordering lets large jobs gain bandwidth over time, preventing starvation. Accordingly, although \name is less fair than baseline CC, even with 10 competing mice jobs, GPT-2 continues to make progress. Note that in realistic training clusters, it is unlikely for so many jobs to compete on a single link. \looseness=-1

\begin{figure*}[t]
    \centering
    \begin{minipage}{0.6\textwidth}
        \centering
        \includegraphics[width=\linewidth]{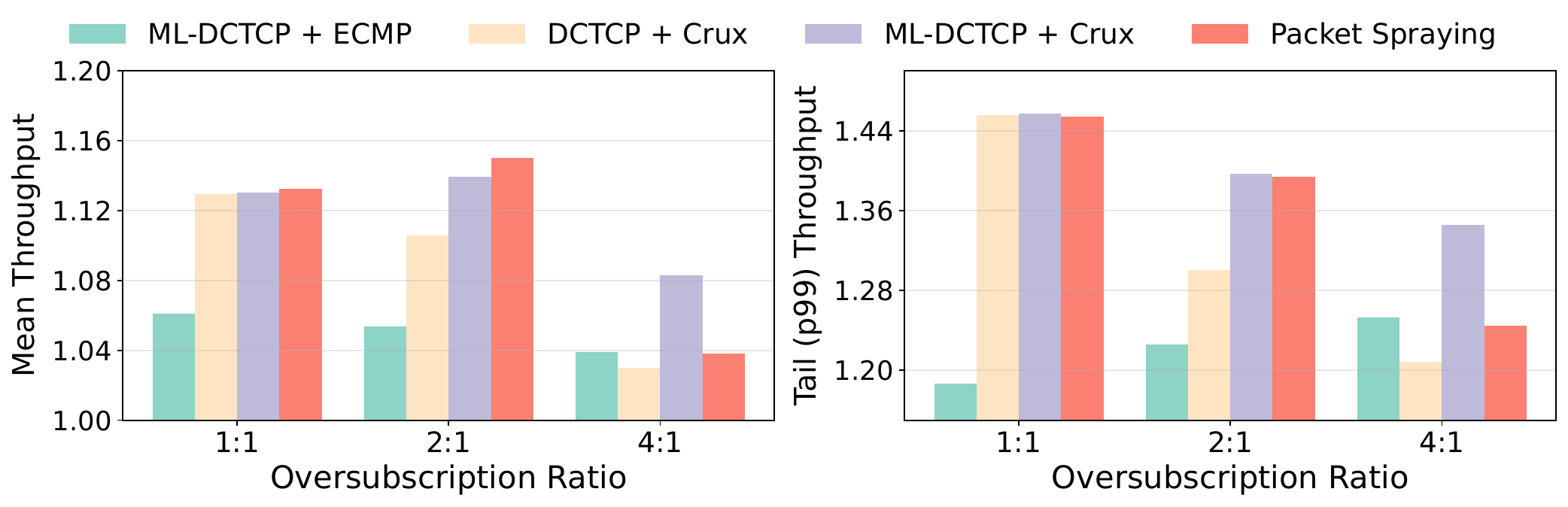}
        \vspace{-0.9cm}
        \caption{\name's performance on fat-tree with and without various load balancing schemes (results normalized by DCTCP+ECMP baseline).}
        \vspace{-0.45cm}
        \label{fig:load_balance}
    \end{minipage}
    \hspace{5mm}
    \begin{minipage}{0.33\textwidth}
        
        \centering
        \includegraphics[width=0.83\linewidth]{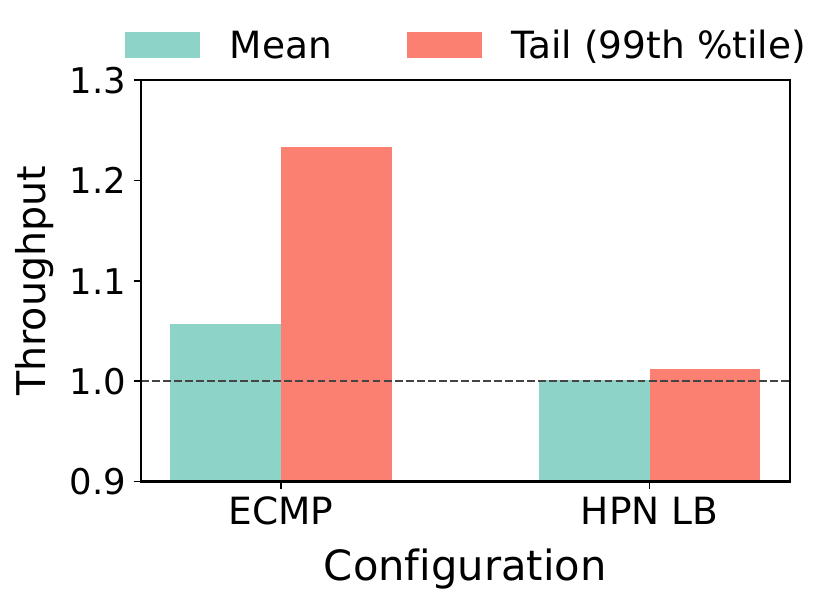}
        \vspace{-0.5cm}
        \caption{\namebf's performance on HPN topology  (results normalized by DCTCP).}
        \vspace{-0.45cm}
        \label{fig:hpn1}
    \end{minipage}
\end{figure*}

\subsection{Load Balancing \& Topology Ablation}
\label{sec:pkt_sim_lb}
In this section, we pair \name with existing schedulers and load balancing schemes on different topologies to study the benefits of interleaving. We use normalized \textit{training throughput} defined in Section~\S\ref{sec:pkt_sim_topo} as our comparison metric.  For both experiments, we sample job placement configurations from varying loads, with an average cluster occupancy of 80\%. We use the same language models as in \S\ref{sec:pkt_sim_topo}.

\para{\namebf with Load Balancing on Fat-tree.} \new{We evaluate the performance of ECMP, Crux (with DCTCP), and compare them against their combinations with \name. We also compare them against per-packet spraying without \name, since the two are incompatible. Figure~\ref{fig:load_balance} shows mean and tail throughput speedups (normalized by the DCTCP+ECMP baseline) across many runs and job placements as a function of network oversubscription in a 72-host (576-GPU) Clos topology. Overall, \name consistently improves the paired load-balancing schemes over the fair-sharing ECMP baseline (DCTCP+ECMP) across all contention levels. We observe similar trends in the average and tail performance plots. 

At high contention (4:1 oversubscription), Crux and packet spraying improve tail throughput over DCTCP+ECMP by $1.21\times$ and $1.25\times$ respectively. \name further improves both as load balancing has a limited effect on the ToR--aggregation bottleneck: once path diversity is exhausted, interleaving becomes necessary to further reduce congestion. \name+ECMP already outperforms Crux alone at this congestion level, achieving a $1.25\times$ tail improvement over the DCTCP+ECMP baseline, while \name+Crux outperforms packet spraying and emerges as the best overall design. A similar trend holds at 2:1 oversubscription, though the gains from \name are reduced because path load balancing is more effective in this regime.

At full bisection, \name improves ECMP's tail throughput by $1.18\times$. However, Crux and packet spraying alone already deliver about $1.13\times$ mean speedup and $1.45\times$ tail speedup through effective path isolation and low congestion, leaving little additional room for improvement by \name.
}

\para{HPN dual-TOR Topology. } \new{Beyond traditional FatTree/Clos topologies, we also evaluate \name on Alibaba's new HPN topology~\cite{hpn} designed for LLM training. We study both ECMP and HPN's proposed load balancing scheme on this topology with and without \name. Figure~\ref{fig:hpn1} plots the mean and tail throughput speedups of \name relative to DCTCP on the HPN topology~\cite{hpn}. While \name continues to provide $1.23\times$ gains over ECMP, its gains over HPN's load balancer are limited because interleaving is fundamentally less compatible with HPN's unpredictable multi-path spreading of flows. In contrast, Crux selects single, stable paths, making it more amenable to \name's interleaving.}

\subsection{Coexistence with other Traffic}
\label{sec:pkt_sim_cca}

\begin{figure}[t]
    \centering
        \includegraphics[width=0.9\linewidth]{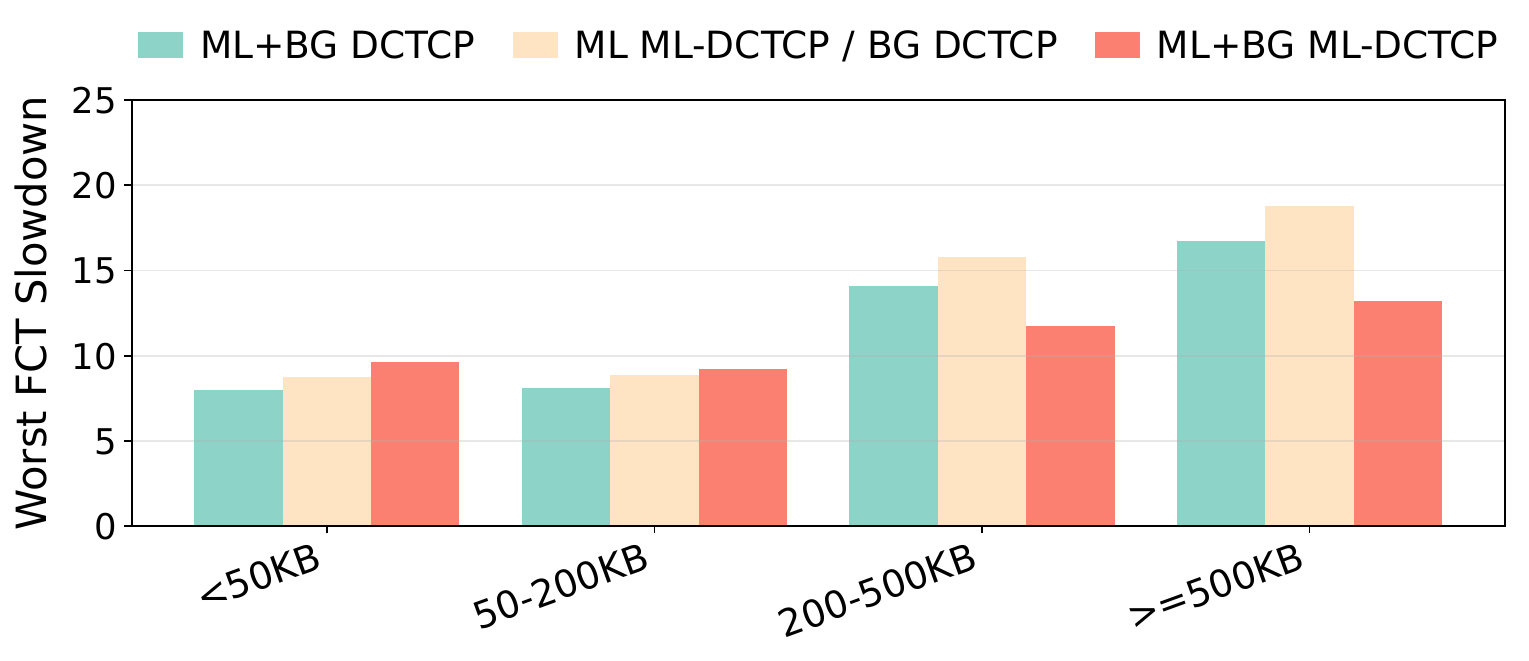}
        \vspace{-0.5cm}
        \caption{Impact of ML Jobs on Background Flow Slowdown.}
        \label{fig:mixed_traffic}
\end{figure}

\new{To analyze the impact of ML jobs using \name on legacy datacenter traffic, we run non-periodic flows in the background along with ML jobs on a 72-host(568 GPU) Clos topology with full bisection bandwidth. We use the same job configuration placements as in \S\ref{sec:pkt_sim_lb}. The flows are generated in a closed loop manner and their sizes sampled uniformly between 10 KB and 1 MB along with their source-destination pairs. 
Figure~\ref{fig:mixed_traffic} plots the tail slowdowns experienced by the background traffic under different flow size buckets for the following three cases: (1) Both ML and background traffic using DCTCP; (2) ML traffic using \nameshort-DCTCP, while background traffic using DCTCP; (3) Both ML and background traffic using \nameshort-DCTCP. Since \nameshort-DCTCP's aggressiveness varies with $\bytes$, the bandwidth available to competing flows depends on when they arrive relative to the lifecycle of a \nameshort flow. Flows that arrive near the completion of an ongoing \nameshort transfer can experience reduced bandwidth, as the \nameshort flow may momentarily dominate the link. As a result, when \name is deployed only for ML training jobs, the tail completion time of coexisting classic flows increases by up to $1.12\times$.

In contrast, when both ML and classic traffic use \name, flows larger than $200$ KB outperform DCTCP, with tail latency reductions of up to $26$\%. Smaller flows (< $200$ KB), however, experience up to $1.21\times$ performance degradation because the increased aggressiveness of \name-enabled flows sometimes leads to higher queuing delays, to which short flows are particularly sensitive. When ML jobs run \name, overall behavior remains consistent with \S\ref{sec:pkt_sim_lb} with speedup as high as 1.2$\times$, as the larger training flows are primarily throughput-bound.}

\label{sec:pkt_sim}
\section{Related Work}
\label{sec:related_work}
\para{Congestion control.} There is a vast literature on congestion control. Many rely on feedback signals indicating congestion in the network and reduce their sending rate~\cite{10.1145/52325.52356, cubic, vegas, bic, westwood, swift, 10.1145/2829988.2787510, DX, illinois, yeah, dctcp, dcqcn}. Others are deadline-aware~\cite{d3, d2tcp}, router-assisted~\cite{pfabric, xcp, rcp}, and receiver-based~\cite{ndp, phost}. We augment DCTCP, DCQCN, and Swift because they are commonly used today. We believe that our \name is applicable to any feedback-based protocol without significant change. 

\para{Flow scheduling.} There have been many workload-specific flow scheduling techniques focusing on efficient utilization of network resources, achieving high bisection bandwidths~\cite{hedera, coflow1, coflow2, coflow3, fp} and minimizing flow completion times~\cite{premtive_schdle, pfabric, FCT_schdl}. Many proposals require centralized schedulers with extensive knowledge about active flows~\cite{coflow1, coflow2, coflow3, hedera, fp}. Others implement heuristics like SRPT~\cite{SRPT_1966, SRPT_optimality_1968}, SJF, LAS, and EDF, requiring switch suppor~\cite{pfabric, FCT_schdl, premtive_schdle, homa} to schedule flows. In contrast, our approach creates an interleaved flow schedule with no scheduler or switch support.

\para{CoFlows}. 
CoFlow schedulers~\cite{coflows, varys, aalo, coda, sincronia, coflows_dependencies, scheduling_coflows_sigmetric} prioritize the completion of an entire related set of flows rather than optimizing individual flow completion times. 
However, most of the flows generated by a DNN application are interdependent, meaning that one flow cannot begin until the previous flow has finished, making them different from coflows.

\para{Periodic resource scheduling.} Real-time and embedded systems have periodic task scheduling problems, but these have explicit deadlines and independent inter-arrival times~\cite{real1, real2}. Cyclic scheduling is a well-studied problem in mathematics~\cite{cyclic1, cyclic2}. The most general form of this problem~\cite{cyclic2} expresses our scenario but is shown to be NP-hard. 
Our approach tackles this by iteratively improving its current state to minimize congestion.

\para{Accelerating DNN Training.} Prior work demonstrated that generic flow schedulers are not optimal for DNN training jobs~\cite{echelon, mlnet}. DNN-specific flow schedulers have been developed to meet this demand~\cite{taccl, byteps, tictac, bytescheduler, syndicate}. Alternatively, intra-job pipelining overlaps the compute and communication phases of the \textit{same training job}~\cite{pipedream, gpipe, terapipe, pipelining_1, pipelining_2, bytescheduler, terapipe}. These approaches only optimize a single job's performance, while we target multi-tenant cluster performance. Job placement schedulers try to minimize multi-job contention. Many only consider the network so far as to try and schedule workers for a job close together~\cite{gandiva, themis, pollux, tiresias, shockwave, optimus, sia}. Our work complements these schedulers.

\para{Resource Interleaving and Optimization.} Muri~\cite{muri} introduced the idea of multi-resource interleaving for DNN training, but required all jobs to share the same GPUs. Cassini~\cite{cassini_nsdi24} exploited the opportunity to overlap the computation and communication of different jobs using an ILP. Both are centralized schedulers that would struggle to scale in a real system. In contrast, our implementation of \name performs a robust distributed live optimization. Crux~\cite{crux} improves GPU utilization in DNN training clusters by routing flows to balance load and avoid contention, whereas \name performs no routing and is complementary to Crux.

\section{Conclusion}
This paper introduces a technique to augment a family of congestion control algorithms to interleave DNN training jobs. The key idea is to dynamically adjust the flow congestion window (or sending rate) of the congestion control algorithms to induce a sliding effect between the jobs. Through extensive simulations and testbed experiments, we evaluate \name's performance, convergence, and fairness and demonstrate that \name accelerates the average and 99$^{th}$ percentile training iteration times by up to 1.9$\times$ and 2.7$\times$, respectively.

\label{bodypage}

\bibliographystyle{abbrv} 
\begin{small}
\bibliography{reference}
\end{small}

\appendix
\clearpage
\nobalance
\section*{Appendix}

\section{Cassini for Partially Compatible Jobs}
\label{sec:cassini_limitation}
\begin{figure}[h]
\centering
\includegraphics[width=0.47\textwidth]{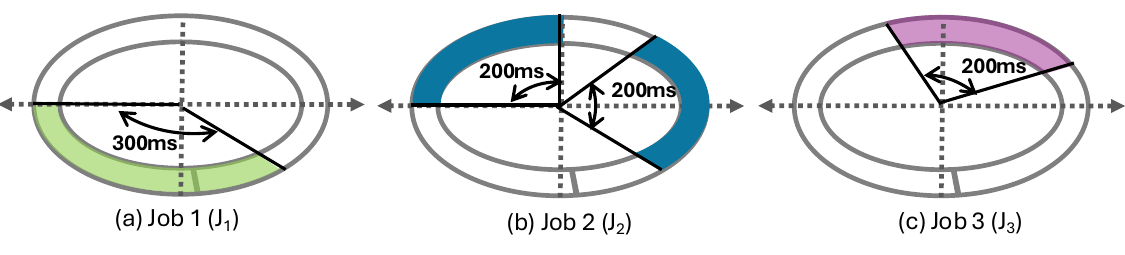}
\vspace{-0.1cm}
\caption{Jobs in isolation (rotated according to Cassini),s}
\label{fig:cassini_iso}
\end{figure}

We illustrate the limitation of Cassini's static unified-circle abstraction with a simple example of three partially compatible jobs, each with an 800\,ms iteration time. Their communication phases together span 900\,ms, making at least 100\,ms of overlap unavoidable.  Figure~\ref{fig:cassini_iso} shows a single iteration of each job in isolation on the unified circle of 800 ms perimeter, rotated according to the Cassini's schedule to minimize overlap. 

\begin{figure}[h]
\centering
\includegraphics[width=0.47\textwidth]{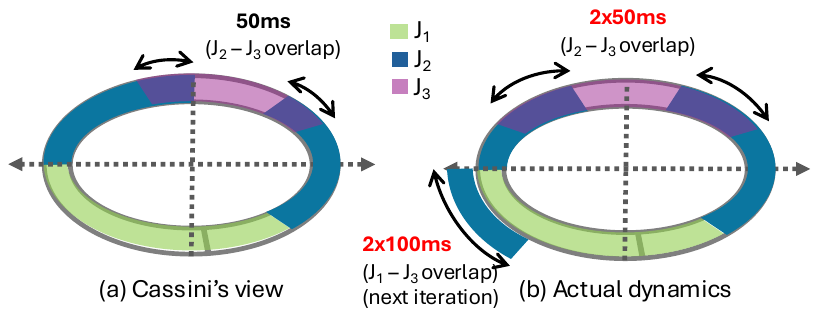}
\vspace{-0.1cm}
\caption{Jobs' schedule under contention}
\label{fig:cassini_partial_comp}
\end{figure}

Figure~\ref{fig:cassini_partial_comp}(a) shows Cassini's hypethetical intended schedule of the jobs together on a single unified circle of perimeter 800 ms, where each job's Figure~\ref{fig:cassini_partial_comp}(b) shows the realized execution under fair sharing, where overlap elongates the communication phases. Because the 100\,ms overlap is unavoidable, the very first iteration incurs unequal delays of 0\,ms, 200\,ms, and 100\,ms across the three jobs, with job \(J_2\)'s communication also spilling 100\,ms into the second iteration of \(J_1\). These unequal delays induce differential drift in the next-iteration start times, so the expected interleaving is not a stable fixed point and degrades over time as jobs slide past one another. This pathology does not appear with two jobs, or under perfectly symmetric overlap, since identical iteration elongation cancels out in relative time shifts.

\section{Analysis of \namebf's Convergence for two simple DNN jobs}
\label{sec:theory_2_jobs}
In this section, we model the bandwidth-time dynamics of two periodic jobs competing one a bottleneck link using \name's bandwidth aggressiveness function as a controlled dynamics system, $\mathcal{D}$. We mathematically study the optimality, convergence and flow-level and iteration-level fairness properties of our approach for the two jobs case. For the simplicity of analysis, we assume that the AIMD congestion control algorithm stabilizes jobs to a bandwidth allocation following the ratio of their additive increase values almost instantaneously. In practice, it takes a few RTTs before the flows using AIMD settle to the desired bandwidth allocation. We neglect this microsecond latency of a few RTTs as it does not alter the functional model of \name. We make this assumption only for simplicity of analysis.

\subsection{Analysis Setting}
The use of bandwidth aggressiveness function $\mathcal{F}(bytes\_ratio)$ in \name creates an instantaneous unfair bandwidth allocation to the jobs causing them to "shift" relative to each other.  The \textit{Shift} function governs the evolution of the jobs' start times toward a partially or fully interleaved state. The goal of interleaving is to start the communication \peaks of the competing jobs at the desired time points, so that the communication \peak of one job is always within in the compute \peak of the other. 

Consider a set of two compatible DNN jobs, $J_1$ and $J_2$ competing on a bottleneck link of capacity $B$ and each having an \textit{ideal iteration time} $I$. This ideal iteration time is achieved when each job is executed in isolation, as shown in  Figure~\ref{fig:DS_illus}(a). The compute \peak lasts $T_1^{comp}$ and $T_2^{comp}$ seconds for the two jobs respectively. Their communication \peak is of durations $T_1$ and $T_2$ seconds when each uses the full link capacity $B$ to send data. We denote the difference in the start times of the $i^{th}$ iterations of the two jobs by $\Delta _i$. Figure~\ref{fig:DS_illus}(b) shows the $i^{th}$ and $(i+1)^{th}$ iterations of the two jobs competing using \name. In this scenario, the first job receives more than half of the bandwidth allowing it to complete its current iteration earlier while the second job incurs more delay due to lower bandwidth. This phenomenon causes the difference in the start times of the next iteration of the two jobs to increase to $\Delta_{i+1}$ where $\Delta_{i+1} = \Delta_i + Shift(\Delta_i)$. We refer to the increase in the jobs' start time difference in the next iteration compared to the previous one as the $Shift$ introduced by \name's bandwidth aggressiveness. This $Shift$ induces a sliding effect in the communication pattern of the two jobs which, over multiple iterations, aids in interleaving their communication demands. 

\subsection{A Discrete Time Dynamic System Model}
\label{sec:ds_appendix}
\begin{figure}[t]
    \centering
    \includegraphics[width=0.96\linewidth]{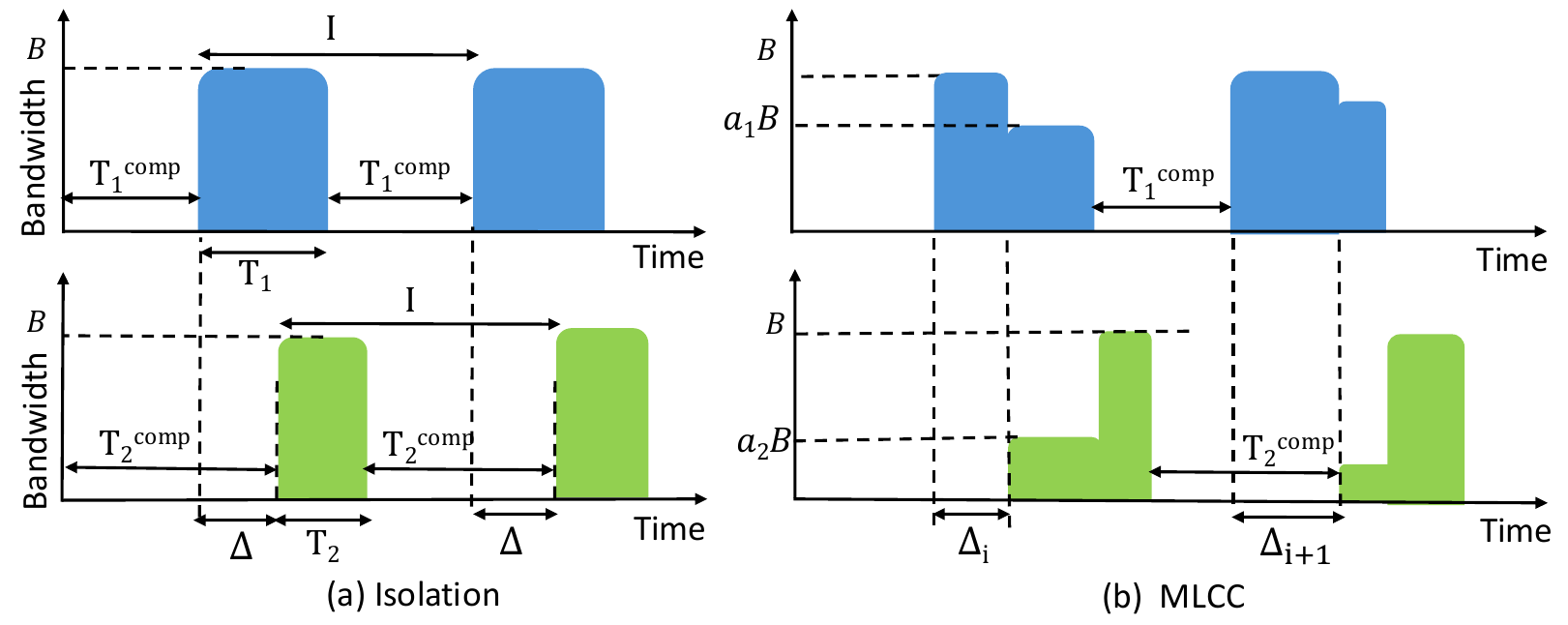}
    \caption{Illustration of two competing jobs each with single communication \peak per iteration}
    \label{fig:DS_illus}
\end{figure}

We use the recipe of \textit{Shift} and the relative start-time difference $\Delta_i$ at the $i^{th}$ iteration to formally define a dynamic system, \textbf{$\mathcal{D}(\Delta, \Phi)$} for the case of two DNN jobs. 
To analyze the dynamics of the jobs' relative start times, we observe how the value of $\Delta_i$ changes every iteration considering it as the \textit{state} of the system $\mathcal{D}$. The "trajectory" of the sequence of $\{{\Delta_i}\}_{i=1}^{n}$ over $n$ iterations describes the way the system converges to an interleaved state given the initial state, i.e., jobs' relative start-time difference in their first iteration, $\Delta_1$. For simplicity, we assume both the jobs to have the same iteration time $I$ (Figure~\ref{fig:DS_illus}). However, the analysis is applicable to the case of different iteration times, where the granularity of a "discrete step" in the dynamics changes from an iteration to the least common multiple of the jobs' iteration times. Thus, we describe the system by its infinite state space $\Delta_i \in \Delta$ and an evolution function $\Phi(x_i)$
\begin{equation}
    \label{eq:evolution}
    \Delta_{i+1} \ \ = \ \ \Phi(\Delta_i) \ \ \ \ \ \ \ where \ \ \Phi(\Delta_i) \ = \ \Delta_i + Shift (\Delta_i)
\end{equation}
 We compute the \textit{Shift} function using the bandwidth-sharing properties of AIMD congestion control algorithm. If the competing flows have different values of additive increase coefficients at any point in time, the bandwidth sharing settles to the ratio of their additive increase values. This principle allows us to find the instantaneous bandwidth received by each competing job in that iteration, and compute the relative start time difference for the next iteration. 
 
 To find the possible interleaved states to which the jobs converge, we find the equilibrium points of the system $\mathcal{D}(\Delta, \Phi)$. The equilibrium points of the system $\mathcal{D}$ are nothing but the fixed points of its evolution function. We set the next state in the evolution equation~\ref{eq:evolution} to be the same as the current state. Let $\mathcal{E}$ be the set of equilibrium states of $\mathcal{D}(\Delta, \Phi)$, then $\mathcal{E}$ is also the set of fixed points of the function $\Phi$. 
\begin{equation}
    If \ \ \Delta_{i+1} = \Delta_i  \ \ \ or \ \ \Phi(\Delta_i) = \Delta_i \ \ \ or \ \ Shift(\Delta_i) = 0 \implies \Delta_i  \in  \mathcal{E} 
\end{equation}
Mathematically, the evolution function $\Phi$ may not have any fixed points if the jobs are incompatible. If the set $\mathcal{E}$ is empty, the system shows \textit{limit cycle} behavior, meaning that the jobs will only visit a finite number of states $\{{\Delta_k}\}_{k \in \mathcal{S}}$ periodically at steady state without staying in one state for long. Here $\mathcal{S}$ is the set of \textit{periodic points} of the evolution function $\Phi$ of period $|\mathcal{S}|$ where $|\mathcal{S}|$ denotes the cardinality of set $\mathcal{S}$. The system shows a limit cycle behavior of least period $|\mathcal{S}|$ if
\begin{equation}
    \Phi^{|\mathcal{S}|}(\Delta_o) = \Delta_o  \ \ \ \forall \ \ \Delta_o \in \mathcal{S}
\end{equation}
$\Phi^{|\mathcal{S}|}$ denotes the $|\mathcal{S}|^{th}$ iterate of the function $\Phi$. Note that the fixed points of the function $\Phi$ are actually its periodic points with period 1.

\subsection{Evolution of Two Jobs}
\label{sec:ds_2jobs_appendix}
With the system model in mind, let's get back to the two jobs $J_1$ and $J_2$ in Figure~\ref{fig:DS_illus} with communication durations $T_1$ and $T_2$ and compute durations $T_1^{comp}$ and $T_2^{comp}$ respectively. For simplicity, let's assume them to have the same iteration time, $I$. Let the size of the jobs in terms of data sent in their communication phase be $D_1$ and $D_2$ respectively such that $D_1 = B\times T_1$ and $D_2 = B\times T_2$. Without losing generality, assume $T_1 > T_2$ (i.e. $D_1 > D_2$). Henceforth, we will call $J_1$ as the "big" job and $J_2$ as the "small" job. The goal is to find the bandwidth allocation of the two jobs as a function of time, given that they use \name's bandwidth aggressiveness function ($\fblack$) for modifying their additive increase. The bandwidth allocation will decide the iteration time of each job and their relative shift.\\~\\
Denote by $a$ and $b$, the $Slope$ and $Intercept$ of function $\fblack$. If the data sent by a job, $J_i, i=1,2$, till a time $t$ in the current iteration is denoted by $d_i(t), i=1,2$ and the total amount of data to be sent every iteration is denoted by $D_i, i=1,2$, then
\begin{equation}
    \mathcal{F} (t) = b + a \times \frac{d_i(t)}{D_i}
\end{equation}
\begin{equation}
    cwnd \xleftarrow{} cwnd + \cfrac{\mathcal{F}(t)}{cwnd}
\end{equation}
\subsection{Computation of shift function}
\begin{figure}[!h]
    \centering
    \includegraphics[width=0.99\linewidth]{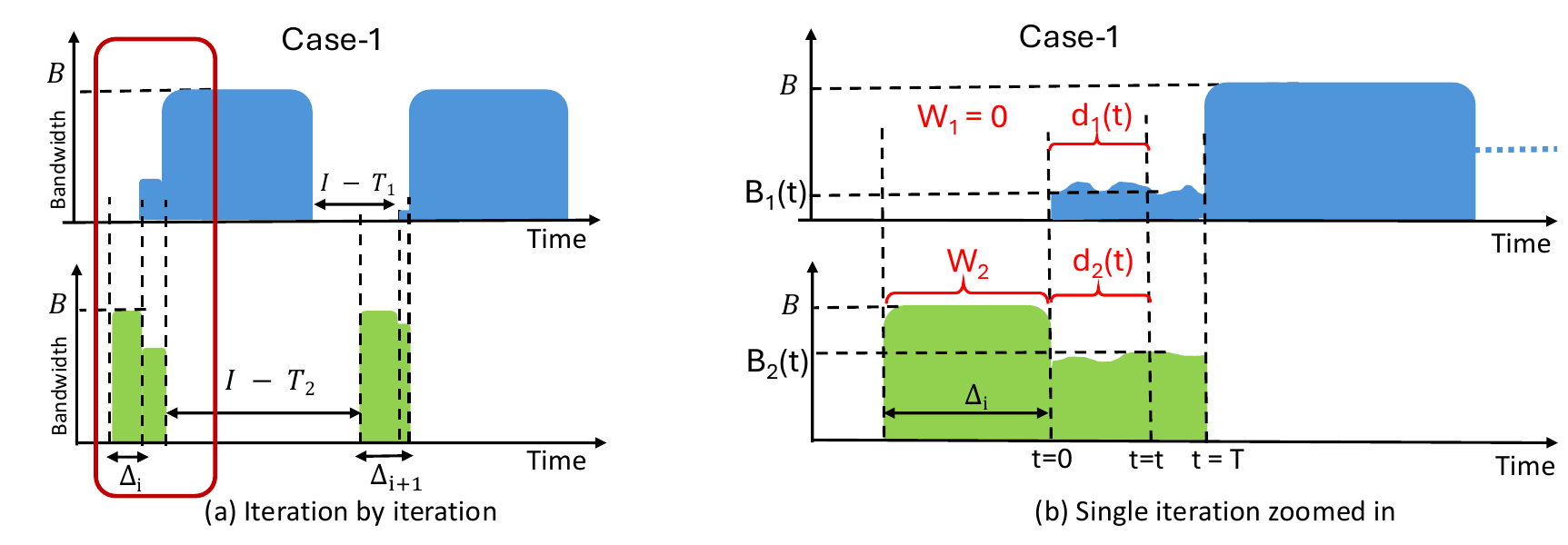}
    \caption{Case-1 of communication overlap}
    \label{fig:case1}
\end{figure}

The bandwidth aggressiveness function, $\fblack$ solely governs the bandwidth allocation to the two jobs. To compute shift, we need to find the bandwidths of the two jobs as a function of time in their current (let's say $i^{th}$) iteration. We consider the start of network contention between the jobs in the current iteration as the origin of the time axis (Figure~\ref{fig:case1}(b)). The contention period ends after $T$ seconds. Consider a time point $t$ in the period of contention. Let $W_1$ and $W_2$ denote the data sent by the two jobs in the current $i^{th}$ iteration before contention began. Let \textbf{$d_1(t)$} and \textbf{$d_2(t)$} be the data sent by the two jobs between time 0 to $t$. Let $B_1(t)$ and $B_2(t)$ denote the instantaneous bandwidths of the two jobs at that time instant $t$. 
$$ B_1(t) = \cfrac{d(d_1(t))}{dt} \ \ \ \ , \ \ \ \  B_2(t) = \cfrac{d(d_2(t))}{dt}$$
Also, during the period of contention, the sum of the bandwidths allocated to the two jobs at any point in time must equal the link capacity $B$ : 
\begin{equation}
    \label{eq:bt}
    d_1(t) + d_2(t) = B \times t
\end{equation}
Note that $W_1$ and $W_2$ are not a function of time. The ratio of bandwidths at time $t$ is the same as the ratio of the additive increase factors of the two jobs, 
\begin{equation}
\begin{split}
    & \cfrac{B_1(t)}{B_2(t)} = \cfrac{d(d_1(t))/dt}{d(d_2(t))/dt} = \cfrac{\mathcal{F}(\cfrac{W_1 + d_1(t)}{D_1})}{\mathcal{F}(\cfrac{W_2 + d_2(t)}{D_2})} \\
    & = \cfrac{b + a \times W_1/D_1 + a\times \frac{d_1(t)}{D1}}{b + a \times W_2/D_2 + a \times \frac{d_2(t)}{D_2}}
\end{split}
\end{equation}

Integrating both sides :
\begin{equation}
\begin{split}
     & \int_{0}^{t} \cfrac{d(d_1(t))}{dt} \left( \cfrac{1}{b + a\times W_1/D_1 + a \times \cfrac{d_1(t)}{D_1}} \right) 
     \\
     & = \int_{0}^{t} \cfrac{d(d_2(t))}{dt} \left( \cfrac{1}{b + a\times W_2/D_2 + a \times \cfrac{d_2(t)}{D_2}} \right)
\end{split}
\end{equation}\\

\begin{equation}
\footnotesize
    \implies D_1 \times \log_e \left( 1 + \cfrac{a\times d_1(t)}{b\times D_1 + W_1}\right) = D_2 \times \log_e \left( 1 + \cfrac{a\times d_2(t)}{b\times D_2 + W_2}\right)
\end{equation}\\

\begin{equation}
    \label{eq:b1b2}
    \boxed{\textcolor{blue}{\cfrac{\log_e \left( 1 + \cfrac{a\times d_1(t)}{b\times D_1 + a \times W_1}\right)}{\log_e \left( 1 + \cfrac{a\times d_2(t)}{b\times D_2 + a \times W_2}\right)} = \cfrac{D_2}{D_1} }}    
\end{equation}
The above equation will be our recipe for finding the shift function and the set $\mathcal{E}$ using simple boundary conditions!
\begin{figure}[t]
    \centering
    \includegraphics[width=0.5\textwidth]{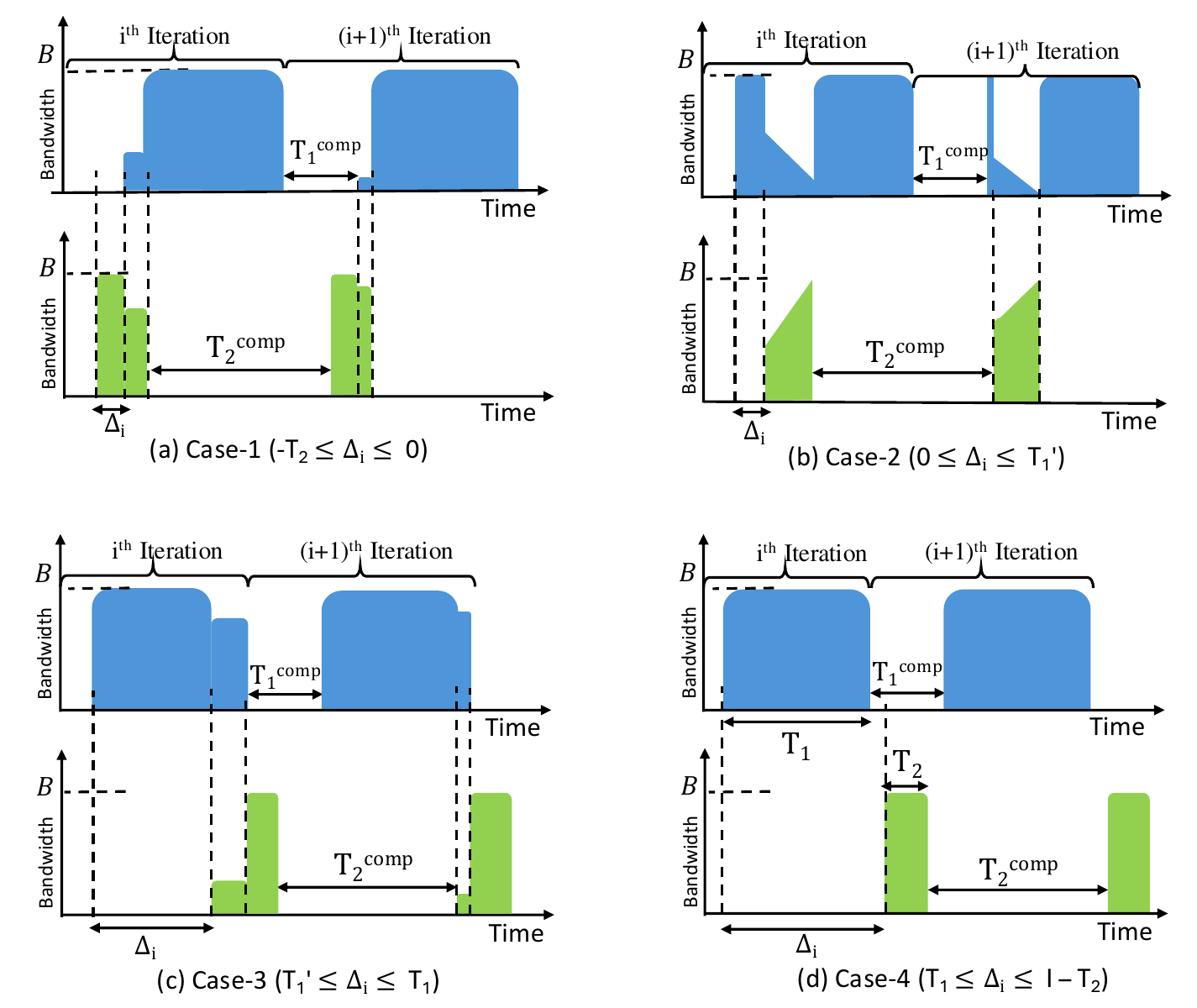}
    \caption{Different scenarios of communication overlap between two periodic jobs}
    \label{fig:cases}
\end{figure}

$\Delta_i$ is the start time difference between the two jobs at the $i^{th}$ iteration. We define $\Delta_i$ to be $Start\_time(J_2) - Start\_time(J_1)$. Due to the nature of $\fblack$ to favor jobs that sent a larger fraction of their total data, there are four possible scenarios of bandwidth allocation, depending on this value of $\Delta_i$. These are :
\begin{itemize}
    \item Case-1 : $-T_2 < \Delta_i < 0$ ; $J_2$ starts before $J_1$
    \item Case-2 : $0 < \Delta_i < T_1^\prime$ ; $J_2$ starts and finishes inside $J_1$ 
    \item Case-3 : $T_1^\prime < \Delta_i < T_1$ ; $J_2$ starts inside $J_1$ but finishes after $J_1$
    \item Case-4 : $T_1 < \Delta_i < I - T_2$ ; $J_2$ starts after $J_1$ finishes
\end{itemize} 
Figure~\ref{fig:cases} illustrates these four cases. These include all possible points of relative alignment of the two jobs' start times in their current iteration. $T_1^\prime$ is a quantity that we calculate later in this section.  
 
We define the Shift function $Shift(\Delta_i)$ as the additional relative time difference in the next iteration as a function of the start time difference $\Delta_i$ between the jobs in the current iteration.
$$Shift(\Delta_i) = \Delta_{i+1} - \Delta_i$$

\para{Case-1 ($-T_2 < \Delta_i < 0$)}.

In this case, the $i^{th}$ iteration of $J_2$ starts earlier than $J_1$. So, $J_2$ has already sent $W_2 ( = B \times \Delta_i)$ amount of data before the contention starts, as shown in Figure~\ref{fig:case1}. However, $J_1$ has not sent any data before the contention period, hence $W_1 = 0$. At time $t = T$, the contention period ends as $J_2$ finishes its communication phase.
Putting the values of $W_1$ and $W_2$ in Equation~\ref{eq:b1b2}, at time $t=t$, we get 
\begin{equation}
    \label{eq:case1}
     \cfrac{\log_e \left( 1 + \cfrac{a\times d_1(t)}{b\times D_1}\right)}{\log_e \left( 1 + \cfrac{a\times d_2(t)}{b\times D_2 + a \times B \times |\Delta_i|}\right)} = \cfrac{D_2}{D_1}   
\end{equation}
\\
We solve the Equations~\ref{eq:bt} and \ref{eq:case1} for $t=T$ when the contention period ends. In that case, $J_2$ has sent all its data, hence $W_2 + d_2(T) = D_2$. Since $d_2(T)$ is known, $d_1(T)$ is given by Equation~\ref{eq:case1} : 
\begin{equation}
    \label{eq:d1t}
\begin{split}
    & d_1(T) = \cfrac{b}{a} \times D_1 \times \left[ \left( 1 + \cfrac{a\times (D_2 - B \times |\Delta_i|)}{b\times D_2 + a\times B \times |\Delta_i|} \right)^\frac{D_2}{D_1} - 1\right] \\
    & = \cfrac{b}{a} \times D_1 \times \left[ \left( 1 + \cfrac{a\times (1 -  \cfrac{|\Delta_i|}{T_2})}{b\times + a\times \cfrac{|\Delta_i|}{T_2}} \right)^\frac{D_2}{D_1} - 1 \right]
\end{split}
\end{equation}
Note that we can solve for $T$ as well using Equation~\ref{eq:bt} and \ref{eq:d1t} but we don't actually need the value of $T$ for finding the $Shift(\Delta_i)$. We compute $Shift(\Delta_i)$ as the change in the relative start time of $J_2$ compared to $J_1$
\begin{equation}
\begin{split}
    & Shift(\Delta_i) = (T - \cfrac{d_2(T)}{B}) - (T - \cfrac{d_1(T)}{B}) = - \cfrac{(d_2(T) - d_1(T))}{B}
\end{split}
\end{equation}
\begin{equation}
\begin{split}
    & \implies Shift(\Delta_i)  =  \ \ -\cfrac{D_2 - B\times |\Delta_i|}{B} \\
    &+ \cfrac{b}{a} \times \cfrac{D_1}{B} \times \left[ \left( 1 + \cfrac{a\times (1 -  \cfrac{|\Delta_i|}{T_2})}{b + a\times \cfrac{|\Delta_i|}{T_2}} \right)^\frac{D_2}{D_1} - 1 \right]
\end{split}
\end{equation}
\begin{equation}
\small
    \label{eq:case1_shift}
    \boxed{\textcolor{blue}{Shift(\Delta_i) = |\Delta_i| - T_2 +  T_1 \times \cfrac{b}{a} \times \left[ \left( 1 + \cfrac{a\times (1 -  \cfrac{|\Delta_i|}{T_2})}{b\times + a\times \cfrac{|\Delta_i|}{T_2}} \right)^\frac{D_2}{D_1}- 1 \right]}}
\end{equation}
The Equation~\ref{eq:case1_shift} provides the $Shift(\Delta_i)$ function for $-T_2 < \Delta_i < 0$. There is no state $\Delta_i$ in this case where the $Shift (\Delta_i)$ becomes zero, implying that there are no fixed points of $\Phi$ in this case.

\para{Case-2 ($0 \leq x \leq T_1'$)}.

\begin{figure}[!h]
    \centering
    \includegraphics[width=0.99\linewidth]{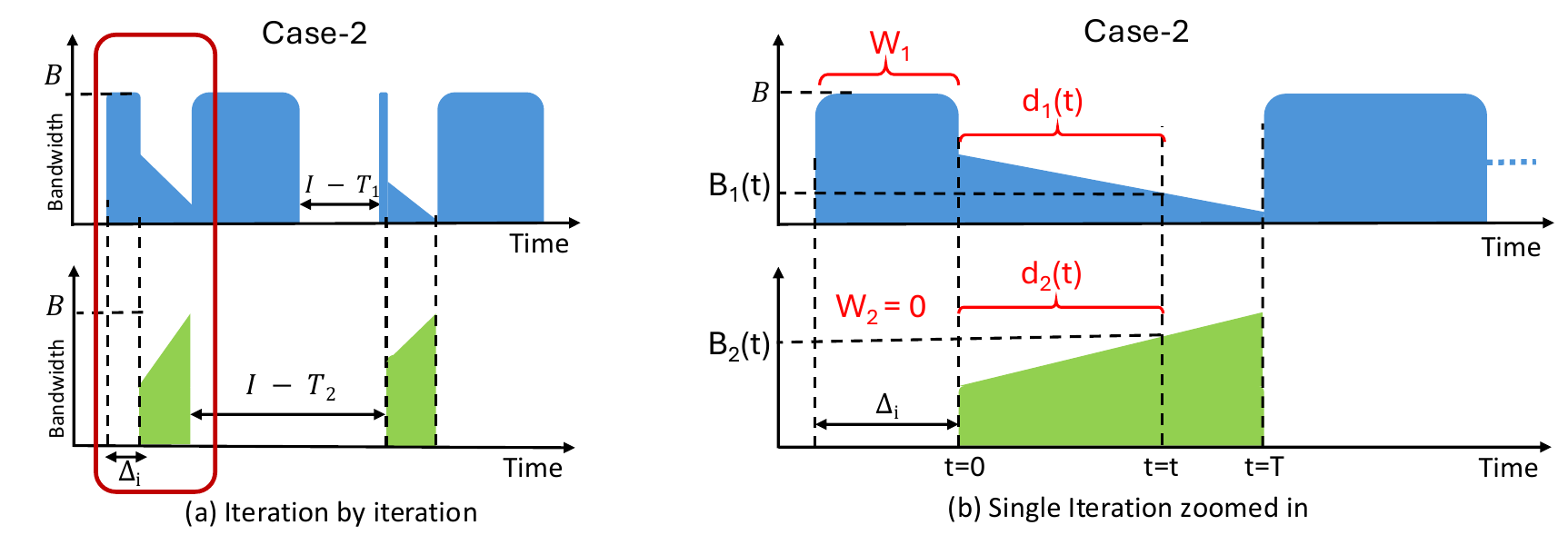}
    \caption{Case-2 of communication overlap}
    \label{fig:case2}
\end{figure}

In this case, the communication \peak of $J_2$ starts and ends within the communication \peak of $J_1$ (Figure~\ref{fig:case2}). $J_2$ spends its entire communication \peak contending with $J_1$, hence $d_1(T) = D_2$ irrespective of the start time difference $\Delta_i$. However, this condition holds until $J_2$ manages to finish before $J_1$ finishes, which provides the value of the mysterious $T_1^\prime$. Note that $T_1^\prime < T_1$. In this case, $W_1 = B \times \Delta_i$ and $W_2 = 0$. Using Equations~\ref{eq:b1b2} and \ref{eq:bt}, with the new boundary conditions, 
\begin{equation}
    \label{eq:case2}
    \boxed{\textcolor{blue}{Shift(\Delta_i) = T_1 \times \left( \cfrac{b}{a} + \cfrac{\Delta_i}{T_1} \right)\times \left[ \left( 1 + \cfrac{a}{b} \right)^\frac{D2}{D1} - 1\right] - T_2}}
\end{equation}
\textbf{Finding the value of $T_1^\prime$}\\
To find the value of $T_1'$, we need to consider the limiting scenario when both the jobs finish their communication phase at the same time, i.e. $d_2(T) = D_2$ but $d_1(T) + W_1 = D_1$ as well. Using these values in Equation~\ref{eq:b1b2} and solving for $\Delta_i = T_1^\prime$, 
\begin{equation}
    \cfrac{\log_e \left( 1 + \cfrac{a\times (D_1 - B\times \Delta_i)}{b\times D_1 + a \times B\times \Delta_i}\right)}{\log_e \left( 1 + \cfrac{a\times D_2}{b\times D_2}\right)} = \cfrac{D_2}{D_1}
\end{equation}

\begin{equation}
  \label{eq:T1p}
  \implies \boxed{\Delta_i = T_1' =  T_1  \times \left(\cfrac{a + b}{a \times \left(1 + \frac{a}{b} \right)^\frac{D_2}{D_1}} - \cfrac{b}{a} \right)}
\end{equation}
Note that $T_1' \xrightarrow{\frac{D_2}{D_1} \xrightarrow{} 0} T_1$\\
Since, there is maximum overlap in the communication \peaks of the two jobs in this case, both the jobs see a maximum slowdown in their communication \peaks for some states $\Delta_1$ and $\Delta_2$ in this case.\\~\\
\textbf{Finding the maximum slowdown of the big job : } \\
Denote by $T_1^{ML}$ the time taken by $J_1$'s communication \peak to complete : 
\begin{equation}
\begin{split}
 & T_1^{ML} = \cfrac{W_1}{B} + T + \cfrac{D_1 - W_1 - d_1(T)}{B}  =   \Delta_i + T + T_1 - \Delta_i - \cfrac{d_1(T)}{B} \\
 & = T_1 + T - \cfrac{d_1(T)}{B}    
\end{split}
\end{equation}
Using Equation~\ref{eq:bt} at time $t=T$ and putting $d_2(T) = D$, 
\begin{equation}
    d_1(T) + D_2 = B \times T \ \ \ \implies T_2 = T - \cfrac{d_1(T)}{B}
\end{equation}
Substituting, 
\begin{equation}
    T_1^{ML} = T_1 + T_2
\end{equation}
$T_1^{ML}$ is independent of $\Delta_i$.
This means that the big job $J_1$'s communication phase is always slowed down by a factor of $(1+\frac{T_2}{T_1})$ irrespective of the value of $\Delta_i$\\~\\
\textbf{Finding the maximum slowdown of the small job : } \\
For $J_2$, the slowdown depends on the value of $\Delta_i$ unlike the big job $J_1$. The job $J_2$ spends its entire communication in contention, hence, its communication time $T_2^{ML} = T$. Using Equation~\ref{eq:b1b2} to find $d_1(T)$ and substituting it in Equation~\ref{eq:bt} to find $T$,
\begin{equation}
    B\times T = D_2 + D_1\times \left( \cfrac{b}{a} + \cfrac{\Delta_i}{T_1} \right)\times \left[ \left( 1 + \cfrac{a}{b} \right)^\frac{D2}{D1} - 1\right] 
\end{equation}
\begin{equation}
    \label{eq:J2sl}
    \implies \boxed{T_2^{ML} = T = T_2 + T_1\times \left( \cfrac{b}{a} + \cfrac{\Delta_i}{T_1} \right)\times \left[ \left( 1 + \cfrac{a}{b} \right)^\frac{D2}{D1} - 1\right]}
\end{equation}
Under the limit $D_1 >> D_2$, Equation~\ref{eq:J2sl} simplifies to :
\begin{equation}
    \lim_{\frac{D_2}{D_1}\to0} \cfrac{T_2^{ML}}{T_2} = 1 + \lim_{\frac{D_2}{D_1}\to0} \cfrac{T_1}{T_2} \times \left( \cfrac{b}{a} + \cfrac{\Delta_i}{T_1} \right)\times \left[ \left( 1 + \cfrac{a}{b} \right)^\frac{D2}{D1} - 1\right]
\end{equation}
Since we know that $\frac{T_2}{T_1} = \frac{D_2}{D_1}$, the above limit is of the form of $\lim_{y\to0} \frac{(1+c)^y - 1}{y} = \log_e (1 + c)$ where $y = \frac{D_2}{D_1}$. Hence the slowdown of $J_2$ under this limit is given by:
\begin{equation}
    \label{eq:J2approx}
    \boxed{\lim_{\frac{D_2}{D_1}\to0} \cfrac{T_2^{ML}}{T_2} = 1 + \left( \cfrac{b}{a} + \cfrac{\Delta_i}{T_1} \right) \times \log_e \left(1 + \cfrac{a}{b} \right)}
\end{equation}
The slowdown of $J_2$'s communication \peak increases as $\Delta_i$ increases, and is maximum when $\Delta_i = T_1^\prime$. Substituting $\Delta_i$ in Equation~\ref{eq:J2sl} by $T_1^\prime$ from Equation~\ref{eq:T1p},
\begin{equation}
    \label{eq:J2slmax}
    \begin{split}
    &T_2^{ML})_{max} = T_2 + T_1\times \left( \cfrac{b}{a} + \cfrac{T_1^\prime}{T_1} \right)\times \left[ \left( 1 + \cfrac{a}{b} \right)^\frac{D2}{D1} - 1\right]\\
    & = T_2 + T_1 \times \left( 1 + \cfrac{b}{a}\right) \times \left(1 - \cfrac{1}{\left( 1 + \frac{a}{b}\right)^{\frac{D_2}{D_1}}} \right)
    \end{split}
\end{equation}

\textbf{Fixed point of $\phi$}\\
In this case, there is a value $\Delta_i = \Delta_o$ at which the $Shift (\Delta_o)$ (Equation~\ref{eq:case2}) between the jobs becomes zero, i.e $\Delta_o$ is a fixed point of the evolution function $\Phi$. This means that if the jobs' start time difference is $\Delta_o$, then they would maintain that same $\Delta_o$ and stay in that configuration for all subsequent iterations. To find the value of that $\Delta_o$, we use Equation~\ref{eq:case2}
\begin{equation}
Shift_{21}(\Delta_o) = T_1 \times \left( \cfrac{b}{a} + \cfrac{x_o}{T_1} \right)\times \left[ \left( 1 + \cfrac{a}{b} \right)^\frac{D2}{D1} - 1\right] - T_2 = 0    
\end{equation}
\begin{equation}
\implies  \boxed{\Delta_o = \cfrac{T_2}{\left[ \left( 1 + \cfrac{a}{b} \right)^\frac{D2}{D1} - 1\right]} \ \ - \ \ T_1 \times \cfrac{b}{a} }   
\end{equation}
\textbf{Checking the stability of the fixed point}\\
We compute the derivative of the evolution function $\Phi(\Delta_i) = Shift(\Delta_i) + \Delta_i$ at with respect to $\Delta_i$, the point $\Delta_i = \Delta_o$, using Equation~\ref{eq:case2}
\begin{equation}
    \cfrac{d(\Phi)}{d\Delta_i} = \cfrac{d(Shift)}{d\Delta_i} + 1 =  \left[ \left( 1 + \cfrac{a}{b} \right)^\frac{D2}{D1} - 1\right] + 1 \ \ \ \ > 1
\end{equation}
The magnitude of the derivative of $\Phi$ at all points including $\Delta_o$ is greater than 1, implying that the point $\Delta_o$ is an \textbf{unstable} state. Even though this $\Delta_o$ is not an interleaved state, the jobs never converge to this point in a real system.\\
In the limit $D_2 << D_1$,
\begin{equation}
\lim_{\frac{D_2}{D_1}\to0} \Delta_o = T_1 \times \left(\cfrac{1}{\log_e\left( 1 + \cfrac{a}{b}\right)}  - \cfrac{b}{a}\right) 
\end{equation}
which is independent of $T_2$. For \textbf{a = 1.75} and \textbf{b = 0.25} used in our evaluations,
\begin{equation}
    lim_{\frac{D_2}{D_1}\to0} \Delta_o = 0.33 \times T_1    
\end{equation}
This means if the small job $J_2$ starts before the big job $J_1$ has completed $33\%$ of its communication, the small job receives higher average bandwidth as compared to the bigger job. But, if it starts later than this, the situation reverses. No matter what the values of the slope and intercept parameters $a$ and $b$ are, $\Delta_o$ never increases to $0.5 \times T_1$. However, this flow-level fairness disparity is not a concern for average iteration time fairness of \name as we show in our flow-level simulations~\S\ref{sec:fl_sim}
\\~\\

\para{Case-3 ($T_1' < \Delta_i < T_1$)}

\begin{figure}[!h]
    \centering
    \includegraphics[width=0.99\linewidth]{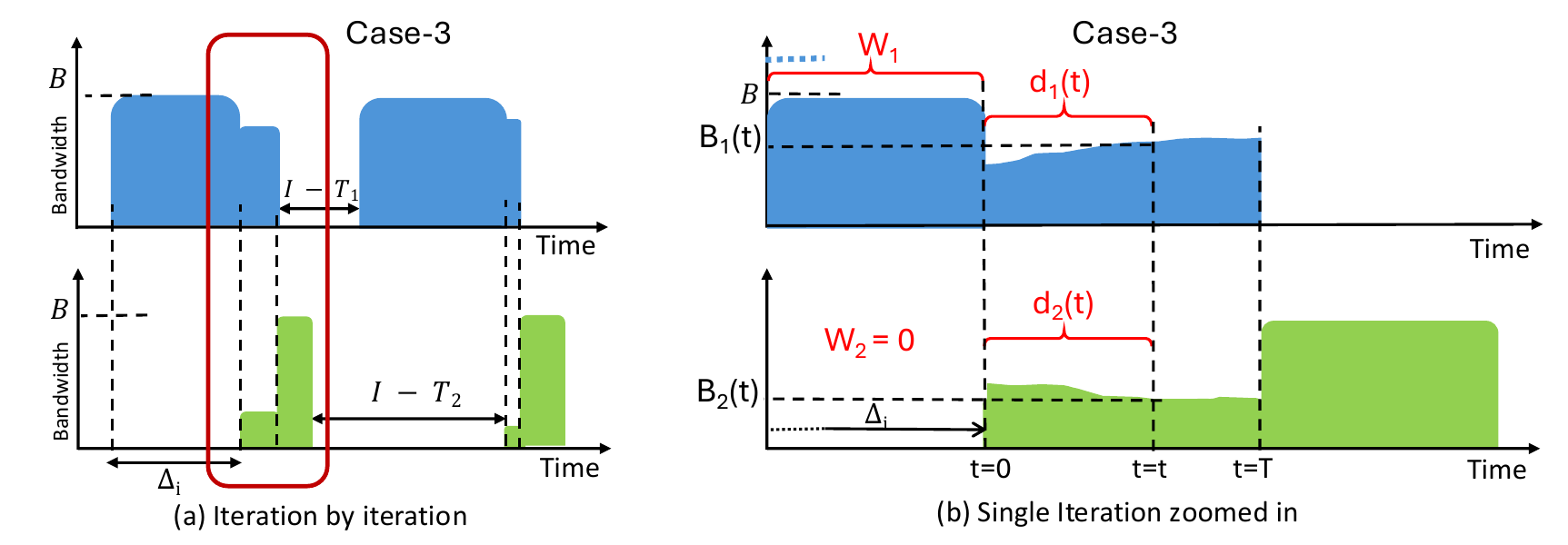}
    \caption{Case-3 of communication overlap}
    \label{fig:case3}
\end{figure}

In this case, the current iteration of $J_2$ starts before and finishes after $J_1$ ends its communication \peak. So, $J_1$ has already sent $W_1 ( = B \times \Delta_i)$ amount of data before the contention starts, as shown in Figure~\ref{fig:case3}. However, $J_2$ has not sent any data before the contention period, hence $W_2 = 0$. At time $t = T$, the contention period ends as $J_1$ has finished its communication phase.\\
Putting the values of $W_1$ and $W_2$ in Equation~\ref{eq:b1b2}, at time $t=t$, we get 
\begin{equation}
    \label{eq:case3}
     \cfrac{\log_e \left( 1 + \cfrac{a\times d_1(t)}{b\times D_1 + a \times B \times \Delta_i}\right)}{\log_e \left( 1 + \cfrac{a\times d_2(t)}{b \times D_2}\right)} = \cfrac{D_2}{D_1}   
\end{equation}
Putting boundary conditions at $t=T$, when $d_1(T) = (D_1 - B \times \Delta_i$), and using Equation~\ref{eq:bt} to find $d_2(T)$, we compute the $Shift(\Delta_i)$
\begin{equation}
    \boxed{\textcolor{blue}{Shift(\Delta_i) = T_1 - \Delta_i -  T_2 \times \cfrac{b}{a} \times \left[ \left( \cfrac{a + b}{b + a\times \cfrac{\Delta_i}{T_1}} \right)^\frac{D_1}{D_2} - 1 \right]}}
\end{equation}
\\
\textbf{Fixed point of $\Phi$}\\
Note that the shift is positive in this case, which means that the bigger job is favored compared to the smaller job. The shift becomes zero at \textcolor{blue}{$\Delta_i = T_1$}, which is an interleaved state where the jobs' communication \peaks don't interfere at all. The jobs would not move any further after this point, unless noise or external factors cause a relative shift between them. Hence, $\Delta_i = T_1$ is also a fixed point of the function $\Phi$.\\
\textbf{Checking the stability of the fixed point}\\
We compute the derivative of the evolution function $\Phi(\Delta_i) = Shift(\Delta_i) + \Delta_i$ at with respect to $\Delta_i$ at the point $\Delta_i = T_1$, using Equation~\ref{eq:case3}
\begin{equation}
\begin{split}
    & \cfrac{d(\Phi)}{d\Delta_i} = \cfrac{d(Shift)}{d\Delta_i} + 1 \\
    & = -T_2\times \cfrac{b}{a}  \times \cfrac{D1}{D2} \times \cfrac{(a + b)}{(b + a\times \cfrac{\Delta_i}{T_1})^2} \times \cfrac{a}{T_1} \times \left( \cfrac{a + b}{b + a\times \cfrac{\Delta_i}{T_1}} \right)^{(\frac{D_1}{D_2}-1)}
\end{split}
\end{equation}
\begin{equation}
    \implies \left(\cfrac{d(\Phi)}{d\Delta_i} \right)_{\Delta_i = T_1}  = -\cfrac{b}{a+b}
\end{equation}
The magnitude of the derivative of evolution function is less than 1 at $\Delta_i = T_1$, implying that it is a \textbf{stable} equilibrium state. This proves that the state corresponding to fully interleaved state is stable under \name. 

\subsubsection{\large{Case-4 ($T_1 < \Delta_i < I - T_2$)}}
In this case, there is no overlap between the jobs' communication phases, hence shift is zero. This case may or may not exist depending on the iteration times of the jobs and their communication patterns. $\fblack$ does not have any effect, as the jobs are already interleaved and there is no network contention (Figure~\ref{fig:cases}(d)).
All the states in this case are fixed points of the evolution function $\Phi$. But they are all \textbf{saddle} points; they are neither stable nor unstable. 

After this case, if $\Delta_i$ increases further, the prior cases will repeat in the same order, i.e. Case-1, then Case-2, and so on in a cyclic fashion.

\subsection{Stability of interleaved state}
\label{sec:appendix_stability}
\begin{figure}[t]
    \centering
    \includegraphics[width=0.9\linewidth]{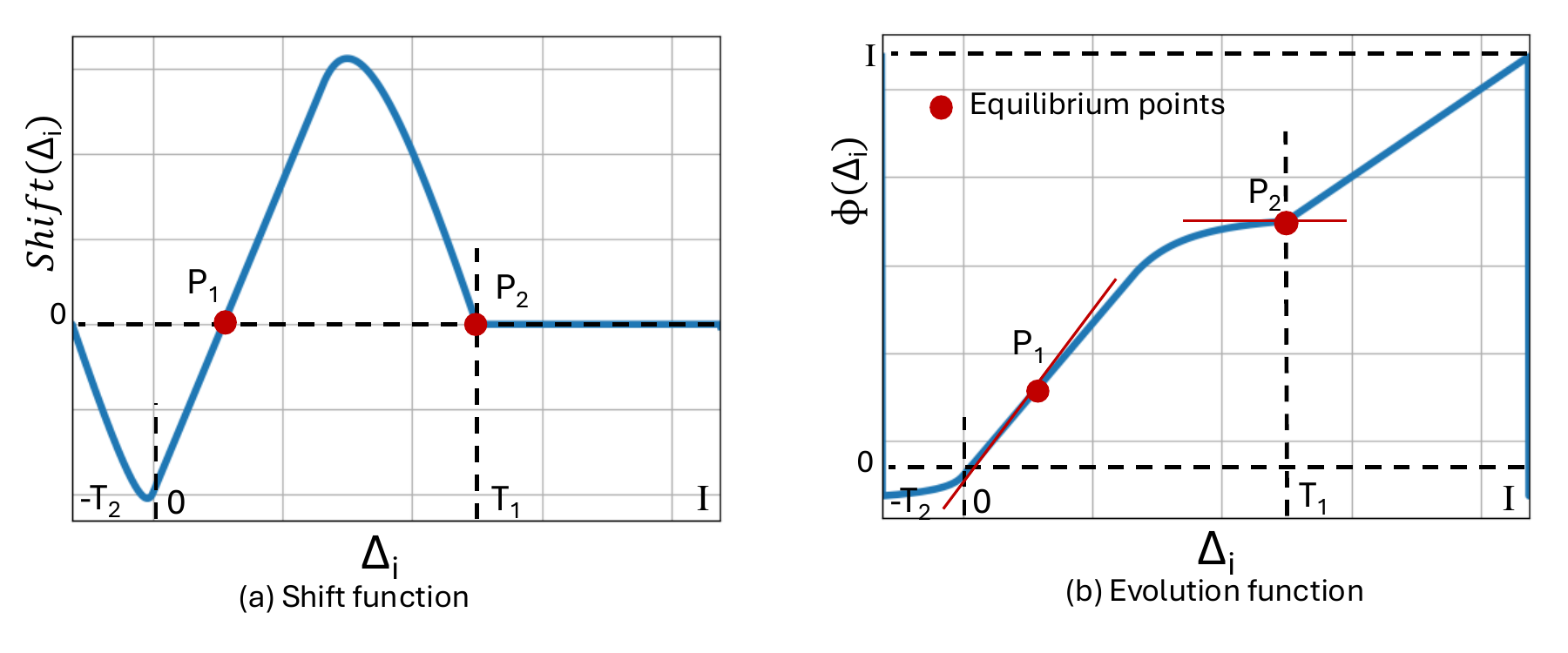}
    \caption{System evolution function for two compatible jobs}
    \label{fig:shiftevolfn}
\end{figure}

Figure~\ref{fig:shiftevolfn} shows the $Shift$ and $\Phi$ functions for the two jobs, combining all the four cases. Note that the scenarios repeat in a circular manner, giving rise to a periodic shift function, with period $I$. Note that the shift becomes zero at $\Delta_i =  0.33 \times T_1$ from case-2 (the point $P_1$), $\Delta_i = T_1$ from case-3 (the point $P_2$) and all the states belonging to case-4. But, as we mentioned earlier, only the fully interleaved state corresponding to the point $P_2$ (where the magnitude of the derivative of the evolution function is less than 1) is stable.

\subsection{Fairness}
\label{sec:theory_fairness}
\begin{figure*}[t]
    \centering
    \includegraphics[width=0.98\textwidth]{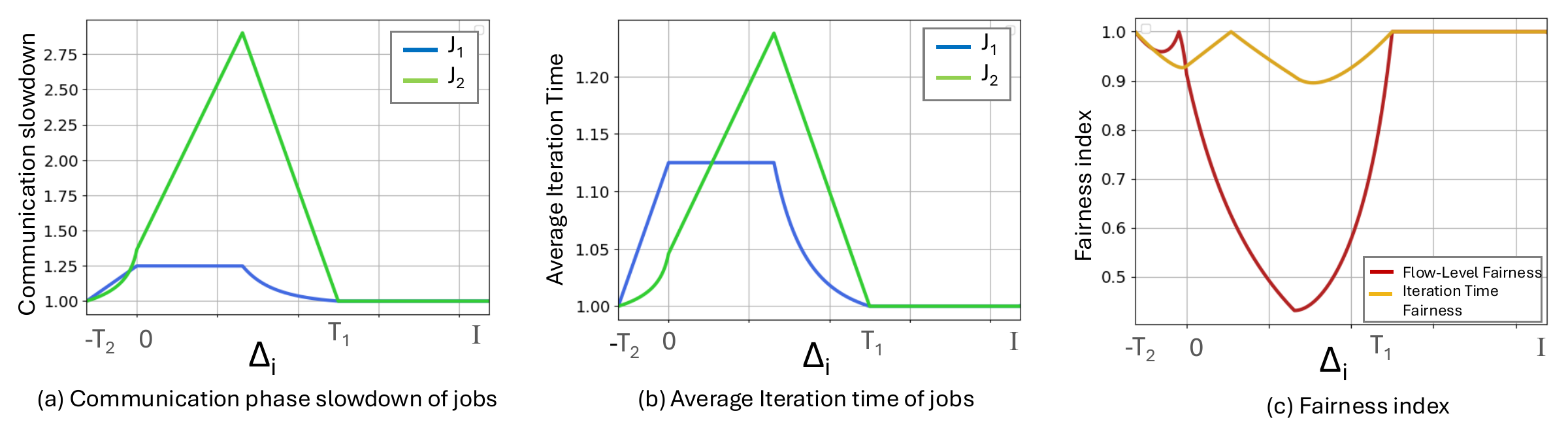}
    \caption{Illustration of flow-level fairness and iteration time fairness for two simple DNN jobs}
    \label{fig:sldnplot_2_jobs}
\end{figure*}

The slowdown of the communication \peak of the jobs $\beta_i (x) = \frac{T_i^{ML}}{T_i}$ as a function of the current relative start time difference $\Delta_i$ is also shown in Figure~\ref{fig:sldnplot_2_jobs}(a). The maximum slowdown of the bigger job $J_1$ is bounded by $(1 + \frac{T_2}{T_1})$ which is less than 2.0 since $T_2 < T_1$ by definition. The maximum slowdown of the communication \peak of the smaller job is significantly higher and given by the expression in Equation~\ref{eq:J2approx} which is nearly equal to 3.0 for the \name's default parameter values of $a=1.75$ and $b=0.25$. 

Figure~\ref{fig:sldnplot_2_jobs}(b) provides the average normalized iteration time of the jobs as a function of $\Delta_i$. Although there is a significant unfairness in the regular terms of flow-level fairness, it is still irrelevant to the goal of \name. This is because we use unfair bandwidth allocation to create the "\textit{Shift}" that eventually slides the jobs into an interleaved state improving the iteration time of all the jobs. Hence, the conventional definition of instantaneous flow rate fairness is not aligned with the goal of interleaving. Recently, there have been calls to consider new notions of fairness in the network. Zaplatel et al.~\cite{unfair1} argue that flow completion time rather than bandwidth fairness should be the metric for fairness. We used this metric to redefine the notion of fairness for periodic DNN jobs. Arslan et al.~\cite{unfair2} follow a similar line of reasoning to conclude that unfairness can also lead to higher energy efficiency. Recent work~\cite{cassini_hotnets} demonstrated that introducing unfairness in congestion control has a surprising payoff that accelerates the training time of DNN jobs in shared GPU clusters. An unfair bandwidth sharing between competing DNN training jobs creates a desirable side-effect of "\textit{Shift}" that nudges the communication phase of one job to interleave with the other, resulting in better overall network utilization and accelerating the training time of both jobs.

We therefore, define a new notion of \textit{Average Iteration Time Fairness} as the ratio of the normalized average iteration time of the faster and slower jobs following the prior work on slowdown metric for fairness~\cite{unfair1}. Since DNN jobs have different iteration times, we normalize all average iteration times to the best possible iteration time for that DNN job when it is run in isolation \cite{pfabric}. Note that the fair sharing congestion control algorithms have an Average Iteration Time Fairness index of 1. Figure~\ref{fig:sldnplot_2_jobs}(c) plots the flow-level fairness of the jobs as the ratio of the slowdowns of the communication phases of the faster and the slower job. Figure~\ref{fig:sldnplot_2_jobs}(c) also plots the Iteration Time Fairness metric for the two jobs using \name as the ratio of the normalized iteration time of the faster job and the slower job as a function of $\Delta_i$. The value of this metric is within 10\% of the baseline fair congestion control for all values of start-time difference $\Delta_i$. This means that the iteration time of the jobs remain within 10\% of each other in our setting of two jobs, despite the large difference in their maximum communication \peak slowdowns or flow-level unfairness. In our flow-level simulations (\S\ref{sec:fl_sim}), we observe a maximum disparity of 20\% between the average iteration times of the jobs even with multiple communication \peaks per iteration. (\S\ref{sec:fl_sim}). 

\section{More on Bandwidth Aggressiveness Function}
\label{sec:f_choice}
\label{sec:eval_aggressiveness_function}
In this section, we discuss the impact of using different bandwidth aggressiveness functions on the convergence and performance of \name. Many aggressiveness functions achieve \name's interleaving goals as long as they satisfy the following three requirements: ($i$) the range is large enough to absorb the noise (e.g., slight variations in round-trip time (RTT) or iteration times) in the network;  ($ii$) the derivative of the function is non-negative;  ($iii$) all flows employ the same bandwidth aggressiveness function. 

We illustrate the performance of different bandwidth aggressiveness functions using a testbed experiment. We train three GPT-2~\cite{gpt_2} models using \nameshort-Reno on a dumbbell network topology. We repeat this training experiment for six different $\fblack$ functions as follows:
\begin{itemize}[align=left, leftmargin=0pt, labelindent=0pt, listparindent=\parindent, labelwidth=0pt, itemindent=!]
\itemsep0em 
\item $\mathcal{F}_1 = 1.75 (\bytes) + 0.25$
\item $\mathcal{F}_2 = 1.75 (\bytes)^2 + 0.25$
\item $\mathcal{F}_3 = {1}/({-3.5 (\bytes) + 4})$
\item $\mathcal{F}_4 = -1.75 (\bytes)^2 + 3.5 (\bytes) + 0.25$
\item $\mathcal{F}_5 = -1.75 (\bytes)+ 2$
\item $\mathcal{F}_6 = -1.75 (\bytes)^2+ 2$
\end{itemize}

\begin{figure}[t]
    \centering
    \includegraphics[width=0.4\textwidth]{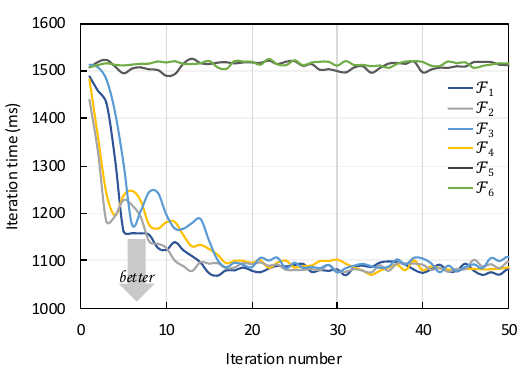}
    \caption{Performance of different $\fblack$ functions.}    
    \label{fig:f_functions} 
\end{figure}

All these functions have the same range (0.25 -- 2) but $\mathcal{F}_1,..., \mathcal{F}_4$ are increasing and $\mathcal{F}_5$ and $\mathcal{F}_6$ are decreasing. Figure~\ref{fig:f_functions} shows the average training iteration time of different iterations as the jobs start their training process. As shown, the iteration time of \nameshort-Reno with $\mathcal{F}_1,..., \mathcal{F}_4$ starts to decrease, as the communication demands interleave after $\approx$20 iterations. Even though different increasing functions take slightly different numbers of iterations to interleave the jobs, they eventually achieve the interleaved state. On the other hand, the iteration times of \nameshort-Reno with $\mathcal{F}_5$ and $\mathcal{F}_6$ do not improve. The functions that don't comply to the requirements of \name's favoritism policy are not able to achieve interleaving and may even have worse performance than baseline congestion control. For simplicity and ease of implementation, we use the linear function $\mathcal{F}_1$ in $bytes\_ratio$ as the bandwidth aggressiveness function of \name throughout this paper.  \looseness=-1

\begin{figure}[t]
    \centering
    \includegraphics[width=0.45\textwidth]{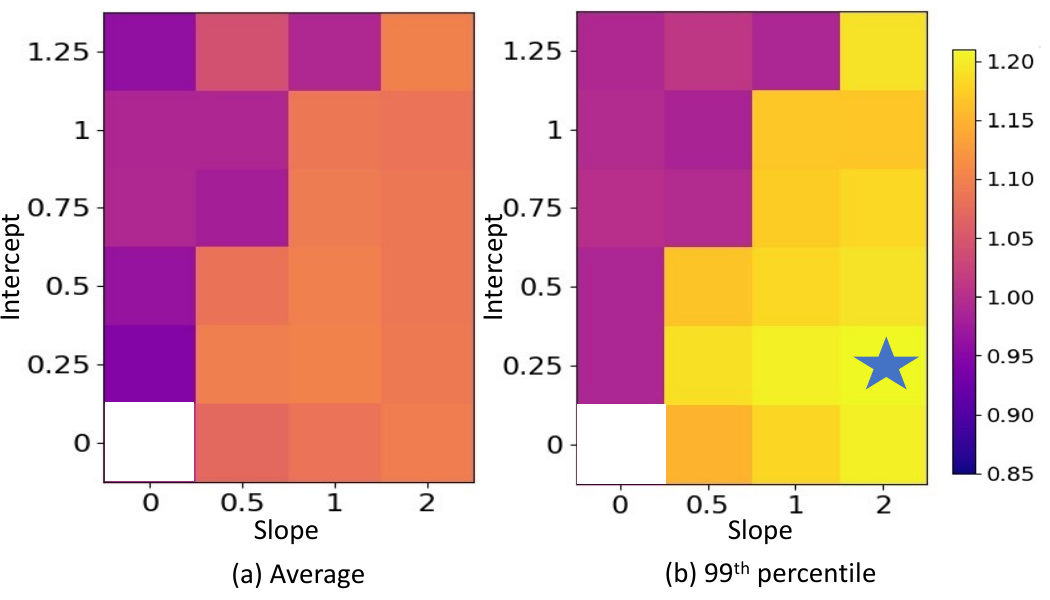}
    \caption{Heatmap of \nameshortbf-Reno's speedup with different slopes and intercepts.}
     \label{fig:heatmap}
\end{figure}

\textbf{Choosing Slope and Intercept.} Next, to study the impact of slope and intercept parameters on $\mathcal{F}$, we sweep a series of possible parameters and measure the average and 99$^{th}$ percentile training iteration time speedups of \nameshort-Reno over default Reno, shown in Figure~\ref{fig:heatmap}. The color in the heatmap represents the speedup value. The heatmaps show that \nameshort-Reno is able to achieve speedups for a wide range of slope and intercept parameters. The star marks the set of parameters with the largest speedup in this experiment.

\section{Random-Restart Annealing}
\label{sec:random_restart}
We saw previously that there may be multiple stable interleaved states to which the jobs using \name may converge depending on the initial state. As the communication patterns of the jobs become complex, the possibility of \name converging to a partially interleaved state also increases. We propose a simulated annealing heuristic to escape from partially interleaved states. We theoretically study our proposed heuristic, \textit{Random-Restarts} in this section.

Simulated annealing is a popular metaheuristic used for escaping local minima in optimization problems with large search spaces. Although the method of \textit{Random-Restarts} in \name is motivated from simulated annealing, there are subtle differences. First, we have causal system, once we make a decision to add random delay, we cannot freely revert back to the older equilibrium state if the new state is worse. Second, we jump to any random state, unlike simulated annealing which always moves the systems to a neighborhood point. To analyze it theoretically, consider an arbitrary set of \textbf{M} periodic jobs competing on a bottleneck link using \name congestion control. Let $\mathcal{E}$ be the set of stable partially or fully interleaved states for this set of jobs.  Let us assume that there are $N$ such stable equilibrium states, given by $\{\Delta_k\}_{k=1}^{N}$. 
$$ \Delta_k  \in  \mathcal{E} \ \ \ \forall \ 1 \leq k \leq N$$
If the jobs are compatible, then there is at least one state (let's say $\Delta_k$) corresponding to the fully interleaved state.

\para{Domain of Convergence.} Each of these states owns a local region centered around them called the \textit{Domain of Convergence ($\mathcal{C})$}. If the jobs initial start-time difference lies in the \textit{Domain of Convergence}, $\mathcal{C}_i$ of the state $\Delta_i$, then \name converges to the state $\Delta_i$. This convergence behavior that is visible in the vicinity of a stable state is also termed as \textit{local convergence}. 

An important property of simulated annealing is exploring the state space in a stochastic manner where each step in exploration is taken with some probability measure. In the case of periodic jobs, the probability of taking another "step" in the exploration by adding a random delay in the job's start time, depends on the quality of the current interleaved state and the sizes of the Domain of Convergence of the other interleaved states. Hence, the probability of exploring the state space is a product of two quantities: a \textit{land probability} and a \textit{jump probability}. 

\para{Jump Probability.} By design of \name, the probability of leaving the current interleaved state, depends on how much larger the current iteration time of the job is, compared to the ideal. Let us denote the jump probability of state $\Delta_k$ by $q_k$
\begin{equation}
    q_k \propto max\left( 1 - \cfrac{Ideal \ Iteration \ time \ of \ i^{th} \ Job}{Iteration \ time \ of \ i^{th} \ Job}\right)_{i=1}^M 
\end{equation}
The higher the iteration time of a job compared to the ideal, the larger is the jump probability. For simplifying the analysis, we assume that we add random delay to all jobs once we decide to take the jump. Taking a maximum across the jobs ensures that the worst-doing job in the current state dictates the jump probability. 

\para{Land Probability.} As we add a random delay to the jobs, the probability of the system to land in the domain $\mathcal{C}_k$ of state $\Delta_k$ is directly proportional to the size of the region $\mathcal{C}_k$. Let's denote the probability to land in a state $\Delta_k$ by $l_k$ and the size of the domain $\mathcal{C}_k$ by $c_k$, then 
\begin{equation}
    l_k \propto c_k \ \ \implies \ \ l_k = R \times  c_k  \ \ \ and \ \ \ \sum_{k=1}^N l_k = 1
\end{equation}
where $R$ is the constant of proportionality.

\para{Transition Probability.} We denote the probability of leaving the current interleaved state, $\Delta_i$ to enter a new interleaved state, $\Delta_j$ by $p_{ij}$. This "explore" probability depends on the jump probability of $\Delta_i$, $q_i$ and land probability of $\Delta_j$, $l_j$
\begin{equation}
\label{eq:explore}
    p_{ij} = l_j \times q_i
\end{equation}
Taking a maximum across the jobs ensures that the worst-doing job in the current state dictates the jump probability.

\para{Annealing Schedule.} To ensure that the jobs remain stable in an optimal interleaved state after full exploration of state space, we let them explore the state space freely at the start and then gradually slow down the exploration pace to achieve stability. This calls for a decay of the annealing process to terminate it at the optimal interleaved state. Hence, we decay the exploration probabilities exponentially with time in Equation~\ref{eq:explore}.
\begin{equation}
\label{eq:explore_exp}
    p_{ij}^{(t)} = l_j \times q_i \times e^{-\alpha t} \ \ \ \ \ \ with \ \ i \neq j
\end{equation}
where $\alpha$ is the decay rate of the annealing schedule.

\begin{figure}[t]
    \centering
    \includegraphics[width=0.95\linewidth]{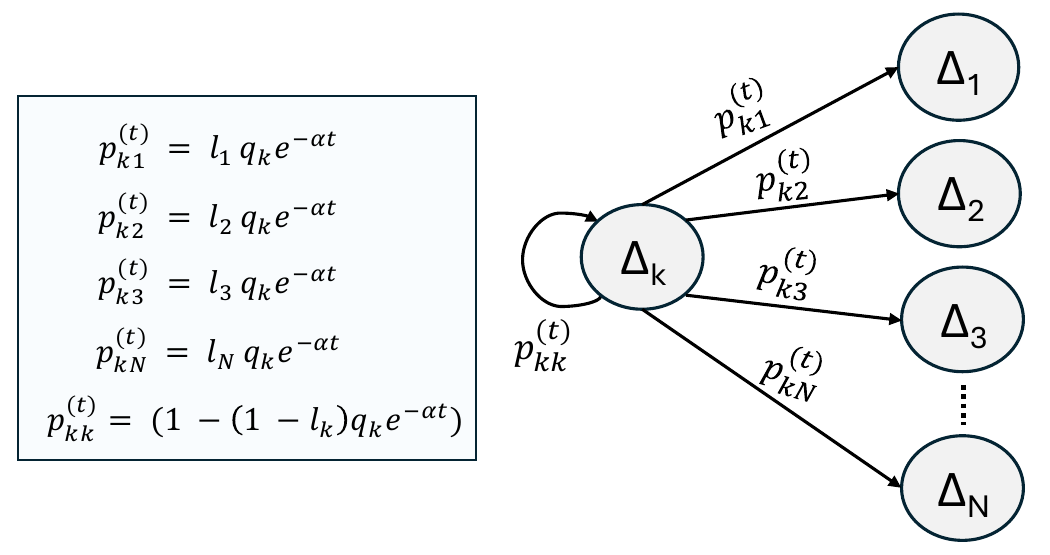}
    \caption{State transition diagram for Random-Restart annealing}
    \label{fig:restart_mc}
\end{figure}

\para{Markov Chain Model.} The probability of going into a new state $\Delta_j$ from the previous state $\Delta_i$ given by $p_{ij}$ is independent of the history of states that jobs saw in the past and the trajectory of states that the system explored. Hence, the exploration of \name using annealing is essentially a Markov Chain with state space $\mathcal{E}$. We use the explore probabilities in Equation~\ref{eq:explore_exp} to define the transition probabilities of a time-dependent Markov Chain with state space consisting of all stable interleaved states, $\mathcal{E}$. Figure~\ref{fig:restart_mc} illustrates the annealing process as a \textit{Markov Chain} and shows its state transition diagram for one of the states, $\Delta_k$. The transition probability matrix $\mathcal{P}(t)$ is given by $\mathcal{P}(t) = [p_{ij}^{(t)}]$.  

\para{Stationary distribution.} We need to find the stationary or limiting distribution of the Markov Chain to find the final interleaved state that \name achieves when annealing terminates. The stationary distribution provides the likelihood of each interleaved state $\Delta_k$ as the final state to which the jobs converge eventually. Let us denote the stationary distribution of the Markov Chain as $\Pi = [\pi_1 \ \pi_1 \ ... \ \pi_N]$ where $\pi_k$ denotes the likelihood of the state $\Delta_k$ as the final interleaved state. The stationary distribution of a Markov Chain has the following property: 
\begin{equation}
    \label{eq:st_dist_eq}
    \Pi = \Pi \times \mathcal{P}(t)
\end{equation}
Solving Equation~\ref{eq:st_dist_eq} gives the likelihood of interleaved states as the final state : 
\begin{equation}
    \label{eq:st_dist_anneal}
    \Pi = \Pi \times \mathcal{P}(t) \ \ \implies \ \ \pi_k = \cfrac{l_k}{q_k} \times \cfrac{1}{\sum_{m=1}^N \frac{l_m}{q_m}}
\end{equation}
The stationary distribution exists and is also independent of time. The likelihood of a state $\Delta_k$ depends inversely on its jump probability $q_k$ and directly on its land probability $l_k$. Also, the jump probability $q_k$ is close to zero for a fully interleaved state. Taking the limit $q_k \rightarrow 0$ in Equation~\ref{eq:st_dist_anneal},
\begin{equation}
    \label{eq:st_lim}
    \lim_{q_k \rightarrow 0} \pi_k  =  \lim_{q_k \rightarrow 0} l_k \times \cfrac{1}{\sum_{m=1}^N \frac{l_m \times q_k}{q_m}} = l_k \times \cfrac{1}{l_k} = 1
\end{equation}
Thus, the likelihood of a fully interleaved state to be the final converged state is 1. This also means that the likelihood of all other partially interleaved states is nill if a fully interleaved state exists in the state space $\mathcal{E}$

\para{Conclusion.} 
Equation~\ref{eq:st_lim} proves that a fully interleaved state is reachable with probability 1, using Random-Restart annealing, even if many stable partially interleaved states exist in the state space. However, a real environment has noise and all the jobs don't transition from the current state together, further adding to the randomness of the outcome. Thus, although this method improves \name's performance speedup further for DNN jobs, it does not guarantee optimal interleaved schedule in all scenarios.

\begin{figure}
    \centering
    \includegraphics[width=0.35\textwidth]{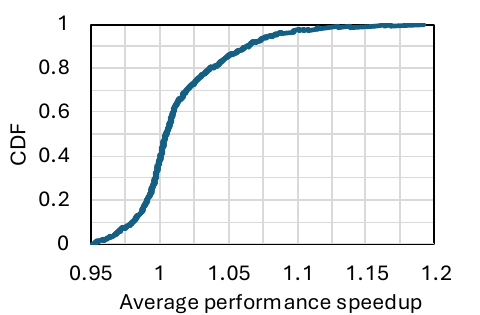}
    \caption{Distribution of performance improvement of \namebf with Random-Restarts enabled relative to when disabled.}
    \label{fig:annealing_eval}
\end{figure}

\para{Simulation Evaluation.} We evaluate the impact of Random-Restarts on \name using hundreds of flow-level simulations, each consisting of two randomly generated jobs competing on a single bottleneck link. We run each scenario with \name both with and without annealing. Figure~\ref{fig:annealing_eval} plots the CDF of \name’s performance improvement across these runs, showing that Random-Restarts Annealing improves performance by up to 1.2× through better interleaving. 

\section{Throughput-Loss Formulation for \nameshortbf-Reno}
\label{sec:tput_formula}

In this section, we provide a relation between throughput and packet loss-probability for \nameshort-Reno congestion control algorithm.
In 1997, Mathis et al.~\cite{tcp_tput_formula} established the relationship between throughput and packet loss for the conventional TCP Reno algorithm:
\begin{equation}
    Conventional~Reno~Throughput \approx \cfrac{1}{RTT \times \sqrt{p}}
\end{equation}
where $p$ is the packet loss probability. This formulation has been widely used to estimate TCP's throughput in various deployment scenarios~\cite{modeling_tcp, BBR, floyd_formula, vishal_aqm}. The above equation is derived by analyzing how Reno behaves under loss during the congestion avoidance phase. We do a similar analysis to 
show that the throughput of \nameshort-Reno flows is inversely proportional to the loss probability, $p$:

\begin{equation}
    \name~Reno~Throughput \approx \cfrac{1}{RTT \times p}
\end{equation}

\para{Derivation}.
Consider M infinitely long-lived flows with identical round-trip times RTT, competing on a single bottleneck link of capacity C. We further assume that the M flows are
synchronized; i.e. their window dynamics are in-phase and all the flows see packet drops at the same time. Let us assume that the packet drop for each flow happens when $cwnd = 2\omega$. Given that the multiplicative decrease factor is 0.5, the window was able to grow from $\omega$ to $2\omega$ before experiencing packet loss at steady state. If the network's loss probability is $p$, then every flow sends exactly $\frac{1}{p}$ packets between successive packet drop events.

During the additive increase phase, the cwnd increases by some value (i.e. additive increase for AIMD algorithms) every RTT. Let us assume that at the time instant of $t = 0$, the flow's cwnd is $W_0 = \omega$ and at an arbitrary $t = m \times RTT$, the cwnd grows to $W_m$. Also, assume that at time $t = N \times RTT$, the flows experience a packet drop, hence $W_N = 2\omega$. Note that cwnd is expressed in \textit{packets}.

If the flows use \name congestion control, then their additive increase factor at any time is given by : 
$$ \fblack = Slope \times \bytes + Intercept$$
$$ where, \ \ \ \  \bytes = \cfrac{bytes\_sent}{total\_bytes} = \cfrac{packets\_sent \times MTU}{total\_bytes}$$
where each packet is of size $MTU$ bytes. Note that $total\_bytes$ is a constant for each flow, 
$$Let, \ \ \textbf{a} = \frac{Slope \times MTU}{total\_bytes} \ \ and \ \ \textbf{b} = Intercept$$ Our aggressiveness functions for additive increase becomes: 
$$ \mathcal{F} = a \times packets\_sent + b$$
Hence, the congestion window recurrence for each flow between successive packet drops looks like the following: 
$$W_0 = \omega$$
$$W_1 = W_0 + a\times W_0 + b$$
$$W_2 = W_1 + a \times (W_0 + W_1) + b$$
$$.....$$
\begin{equation}
\label{eq:recurr_relation}
W_n = W_{n-1} + a \times \sum_{m=0}^{n-1} W_m + b  
\end{equation}
$$\implies W_n = a_1 W_{n-1} + a_2 W_{n-2} + .. + a_{n-2}W_2 + a_{n-1}W_1 + a W_0$$
$$where \ \ a_{n-1} = a_{n-2} = .. = a_3 = a_2 = a \ \ and \ \ a_1 = (1+a)$$
The above recurrence series can be solved using the method of generating functions of a variable $x$ with domain $(0,1)$. Consider two such generating functions : 
$$Z(x) = \sum_{m \geq 0} W_m x^m \ \ and  \ \ A(x) = \sum_{m > 0} a_m x^m$$
Equation~\ref{eq:recurr_relation} can be written in terms of generating functions: 
$$ Z(x) = Z(x) \times A(x) + \omega + b\sum_{m > 0} x^m$$
\begin{equation}
\label{eq:generating_fn}
    \implies Z(x) = \cfrac{\omega + b \sum_{m > 0} x^m}{1 - A(x)}
\end{equation}
The congestion window at $t = n \times RTT$, given by $W_n$ is actually the coefficient of $x^n$ in $Z(x)$ in Equation~\ref{eq:generating_fn}. We expand all the functions as power series to compute the coefficient of $x^n$.\\
The power series of the function $A(x)$ is nothing but an infinite geometric series : 
$$
\ \ \ A(x) = \sum_{m > 0} a_m x^m \ = x + a(x + x^2 + x^3 +...) \  = x + \cfrac{ax}{1-x}
$$
Similarly, the other geometric series of $x$ in the numerator on the right-hand side of Equation~\ref{eq:generating_fn}, can be converted to its functional form, giving the functional form of $Z(x)$ 
\begin{equation}
    \implies Z(x) = \cfrac{\omega + \cfrac{bx}{1-x}}{1 - (x + \cfrac{ax}{1-x})} = \cfrac{\omega - x (\omega-b)}{1 - x(a+2-x)}
\end{equation}\\
Note that $x(a+2-x) \in (0,1)$ for $x \in (0,1)$ and $a << 1$. Hence, 
\begin{equation}
\label{eq:power_series}
    Z(x) = (\omega - x (\omega-b)) \left( \sum_{m=0}^{\infty} x^m (a+2-x)^m\right)
\end{equation}
The congestion window at $t = n \times RTT$, given by $W_n$ is actually the coefficient of $x^n$ in $Z(x)$ in Equation~\ref{eq:power_series}. We have derived the expression for congestion window for $n$ being an even number and odd number in Equation~\ref{eq:cwnd_even} and \ref{eq:cwnd_odd} respectively.

\begin{equation}
\label{eq:cwnd_even}
   W_{2k} =  \omega \mathcal{P}_{k}(a) - (\omega-b) \mathcal{Q}_{k}(a) 
\end{equation}
\begin{equation}
\label{eq:cwnd_odd}
   W_{2k+1} = \omega \mathcal{Q}_{k+1}(a) - (\omega-b) \mathcal{P}_{k}(a) 
\end{equation}

where $\mathcal{P}$ and $\mathcal{Q}$ are polynomials of $a$
$$ \mathcal{P}_{k}(a) =  \sum_{i=0}^{k} \binom{k+i}{2i} (-1)^{k-i} (a+2)^{2i}$$
$$ \mathcal{Q}_{k}(a) =  \sum_{i=0}^{k-1} \binom{k+i}{2i+1} (-1)^{k-i-1} (a+2)^{2i+1}$$
There exists a natural number \textbf{$N$} such that at $t = N \times RTT$, the flow experiences a drop as the congestion window $W_{N}$ exceeds $2\omega$. Without losing generality, let us assume that $N = 2K$ is an even number so that the congestion window can be obtained from Equation~\ref{eq:cwnd_even}. We assume that $b << \omega$, i.e. the additive increase is much smaller than the actual cwnd.
$$\implies W_N = \omega (\mathcal{P}_K(a) - \mathcal{Q}_K(a)) + b\mathcal{Q}_K(a) \approx 2\omega$$
$$\implies (2 - \mathcal{P}_K(a) + \mathcal{Q}_K(a)) = \frac{b}{\omega}\mathcal{Q}_K(a)$$
$$\implies \cfrac{(\mathcal{P}_K(a) - 2)}{Q_K(a)} = 1-\frac{b}{\omega} \approx 1$$
$$\implies (\mathcal{P}_K(a) - \mathcal{Q}_K(a)) \approx 2$$
The value of $K$ and $N=2K$ can be obtained by finding the roots of the above equation. Note that $N$ is only a function of $a$ and is nearly independent of $\omega$.

We then compute the total number of packets sent between time $t = 0$ when cwnd is $W_0 = \omega$ and $t = N \times RTT = 2K \times RTT$ when the cwnd becomes $W_{2K} = 2\omega$. By using Equation~\ref{eq:recurr_relation},
$$
    \sum_{m=0}^{N-1} W_m = \cfrac{W_N - W_{N-1} - b}{a}
$$
\begin{equation}
    \implies \sum_{m=0}^{N} W_m = \cfrac{W_N(1+a) - W_{N-1} - b}{a}
\end{equation}
Note that by definition of $N$, $W_N = 2\omega$ and, $W_{N-1} = W_{2K-1}$ can be substituted using Equation~\ref{eq:cwnd_odd}
$$ \implies \sum_{m=0}^{N} W_m = \cfrac{2\omega(1+a) - \omega \mathcal{Q}_{K}(a) - (\omega-b) \mathcal{P}_{K-1}(a) - b}{a}
$$

From the above relation, since $K$ is only a function of $a$, it is clear that $\mathcal{Q}_{K}(a)$ and $\mathcal{P}_{K-1}(a)$ is only a function of $a$, hence
\begin{equation}
\label{eq:cwnd_vs_start_wind}
 \sum_{m=0}^{N} W_m \propto \omega
\end{equation}
Since the network drop probability is $p$, exactly $1/p$ packets are sent between two successive drops:
$$ \sum_{m=0}^{N} W_m = \frac{1}{p}$$
Using Equation~\ref{eq:cwnd_vs_start_wind},
$$ \implies \omega \propto \frac{1}{p}$$
Throughput for congestion window based protocol is given by $\cfrac{cwnd}{RTT}$, hence
\begin{equation}
    Throughput \approx \cfrac{\omega}{RTT} \propto \frac{1}{RTT \times p}
\end{equation}\\
Intuitively, this implies that given the same packet drop probability, an \nameshort-Reno flow claims more bandwidth share than a standard TCP-Reno flow. . 
However, \nameshort-Reno flows would not starve the other legacy flows because \name allocates non-zero bandwidth to the less favored flow as well instead of fully blocking it. 
To avoid delaying latency-sensitive legacy flows, service providers can assign them a higher additive increase value than the range of bandwidth aggressiveness function. This way, important short flows won't be affected by the DNN traffic.

\section{\namebf using multiplicative decrease}
\label{sec:aug_cca_detail}
This section explains how we augment Reno, CUBIC, DCTCP, and DCQCN algorithms with \name using their window (or rate) decrease phase. 

\textbf{Reno multiplicative decrease.} 
With every dropped packet, TCP Reno updates its cwnd as follows:
\begin{equation}
\setlength\abovedisplayskip{0pt}
cwnd \leftarrow 0.5 \times cwnd
\end{equation}

In \name, we update the above equation to: 
\begin{equation}
\setlength\abovedisplayskip{0pt}
cwnd \leftarrow \f \times 0.5 \times cwnd
\end{equation}

\textbf{DCTCP multiplicative decrease.} DCTCP uses ECN-marked packets to detect congestion. Upon the receipt of a Congestion Notification Packet (CNP) from the receiver, the sender reduces its current congestion window using a multiplicative decrease factor:
\begin{equation}
\setlength\abovedisplayskip{0pt}
cwnd \leftarrow (1-\cfrac{\alpha}{2}) \times cwnd
\end{equation}
where $\alpha$ is the fraction of the ack packets marked with ECN. In \name, the multiplicative decrease factor is updated to: 
\begin{equation}
\setlength\abovedisplayskip{0pt}
cwnd \leftarrow \f \times (1-\cfrac{\alpha}{2}) \times cwnd
\end{equation}

\textbf{CUBIC multiplicative decrease.} The multiplicative decrease step of CUBIC is similar to Reno, but instead of having a fixed value of 0.5, CUBIC specifies the multiplicative decrease factor as a parameter called $\beta$. The cwnd update during multiplicative decrease for CUBIC is given by:
\begin{equation}
\setlength\abovedisplayskip{0pt}
cwnd \leftarrow \beta \times cwnd
\end{equation}
In \name, we update the multiplicative decrease factor of CUBIC similar to Reno:
\begin{equation}
\setlength\abovedisplayskip{0pt}
cwnd \leftarrow \f \times \beta \times cwnd
\end{equation}

\textbf{DCQCN rate decrease.} DCQCN uses ECN-marked packets to detect congestion. Upon the receipt of a CNP from the receiver, the sender reduces its current rate using a multiplicative decrease factor:
\begin{equation}
\setlength\abovedisplayskip{0pt}
curr\_rate \leftarrow (1-\cfrac{\alpha}{2}) \times curr\_rate
\end{equation}
where $\alpha$ is the rate reduction factor of the DCQCN algorithm. We update the above equation to: 
\begin{equation}
\setlength\abovedisplayskip{0pt}
curr\_rate \leftarrow \f \times (1-\cfrac{\alpha}{2}) \times curr\_rate
\end{equation}

Table~\ref{tab:mltcp_cc_md} summarizes the implementation of \name using additive increase and multiplicative decrease for all congestion control algorithms.

\begin{table}[tbp]
\scriptsize
\centering
\def\arraystretch{1}
\begin{tabular}{|p{1.3cm}|p{1cm}|p{5.5cm}|} 

\hline 
\multirow{2}{2cm}{\nameshortbf\textbf{-Reno}}
\scriptsize
& Add. Inc. & $cwnd \,\,\,\, \xleftarrow{} \,\,\,\, cwnd \,\, + \,\, \textcolor{cyan}{\fblack} \, \left( \cfrac{num\_acks}{cwnd} \right) $ \vspace{0.15cm} \\ 
& Mul. Dec. & $cwnd \,\,\,\, \xleftarrow{} \,\,\,\, \textcolor{cyan}{\fblack}\, (0.5 \, cwnd) $  \vspace{0.15cm} \\
\hline
\multirow{2}{2cm}{\nameshortbf\textbf{-DCTCP}}
& Add. Inc. & $cwnd \,\,\,\, \xleftarrow{} \,\,\,\, cwnd \,\, + \,\, \textcolor{cyan}{\fblack} \, \left( \cfrac{num\_acks}{cwnd} \right) $ \vspace{0.15cm} \\
& Mul. Dec. & $ cwnd \,\,\,\, \xleftarrow{} \,\,\,\, \textcolor{cyan}{\fblack}\, \left( 1 - \cfrac{\alpha}{2} \right) \, cwnd $ \vspace{0.15cm} \\
\hline
\multirow{2}{2cm}{\nameshortbf\textbf{-Cubic}}
& Add. Inc. & $cwnd \,\,\,\, \xleftarrow{} \,\,\,\, Cubic ( \textcolor{cyan}{\fblack}\,. time)$ \textcolor{white}{$\cfrac{1}{1}$} \vspace{0.15cm} \\
& Mul. Dec. & $ cwnd \,\,\,\, \xleftarrow{} \,\,\,\, \textcolor{cyan}{\fblack}\, (0.8 \, cwnd) $ \vspace{0.15cm} \\
\hline
\multirow{2}{2cm}{\nameshortbf\textbf{-DCQCN}}
& Add. Inc. & $rate \,\,\,\, \xleftarrow{} \,\,\,\, rate \,\, + \,\, \textcolor{cyan}{\fblack} \, \Delta R$ \textcolor{white}{$\cfrac{1}{1}$} \vspace{0.15cm} \\
& Mul. Dec. & $rate \,\,\,\, \xleftarrow{} \,\,\,\, \textcolor{cyan}{\fblack} \, \left( 1 - \cfrac{\alpha}{2} \right) .\, rate$ \vspace{0.15cm} \\
\hline
\end{tabular}

\caption{Augmenting different congestion control algorithms with \namebf.}
\label{tab:mltcp_cc_md}

\end{table}

\begin{figure}[t]
    \centering
    \includegraphics[width=0.35\textwidth]{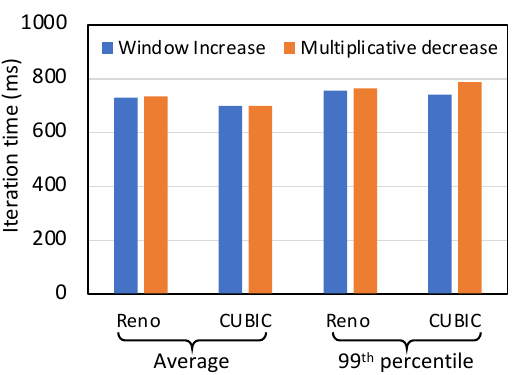}
    \caption{Window increase vs. multiplicative decrease.}
     \label{fig:reno_ai_vs_md}
\end{figure}

\textbf{Adjust the increase or decrease?} Thus far, our evaluations augment the congestion window (or rate) increase phases of Reno, CUBIC, and DCQCN. As mentioned in Section~\ref{sec:design}, the \name's modification can be augmented to both the increase and decrease phases. Figure~\ref{fig:reno_ai_vs_md} compares the training iteration time for two jobs running GPT-2 model and sharing a single bottleneck link in a dumbbell topology for \nameshort-Reno-WI, \nameshort-Reno-MD, \nameshort-CUBIC-WI and \nameshort-CUBIC-MD. We find that the WI or MD variants of \nameshort-Reno and \nameshort-CUBIC perform similarly for both average and tail iteration times.

\section{Testbed Benchmark for Reno and CUBIC}
\label{sec:app_benchmarks}
\begin{figure*}[t]
    \centering
    \includegraphics[width=0.89\textwidth]{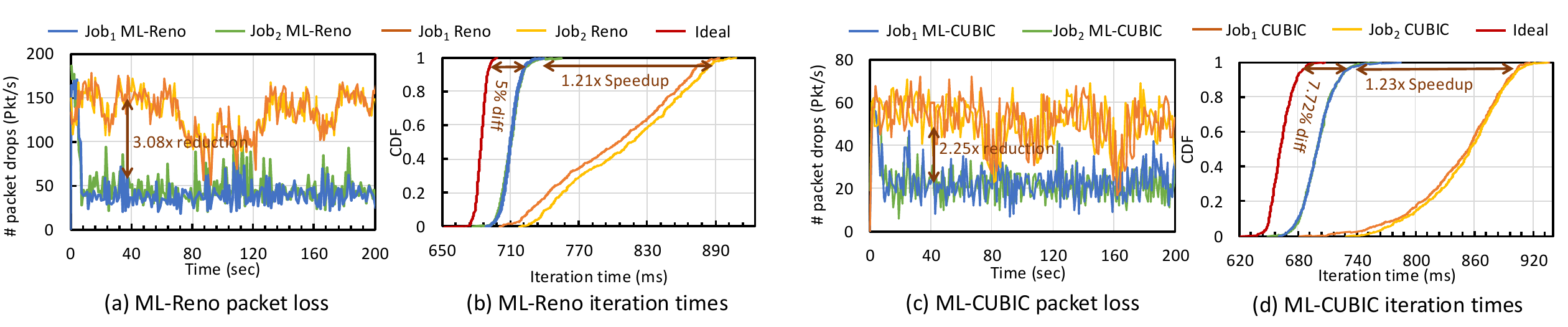}
    \vspace{-0.3cm}
    \caption{Performance of \nameshortbf-Reno and \nameshortbf-CUBIC.}
    \vspace{-0.3cm}
    \label{fig:gpt2_classic_cc}
\end{figure*}

\textbf{\nameshortbf-Reno.} Figure~\ref{fig:gpt2_classic_cc}(a) compares the number of packet drops per second for the first 200 seconds of the training and shows that, on average, \nameshort-Reno reduces the packet loss rate by 3.08$\times$. Consequently, \nameshort-Reno accelerates the average and 99$^{th}$ percentile training iteration time of both jobs by $1.1\times$ and $1.18\times$, respectively, as shown in Figure~\ref{fig:gpt2_classic_cc}(b). Note that the two \nameshort-Reno curves corresponding to the two jobs, overlapped on top of each other. Moreover, the figure shows that \nameshort-Reno's iteration times are within 5\% of the ideal scheme. \looseness=-1

\textbf{\nameshortbf-CUBIC.} 
Figure~\ref{fig:gpt2_classic_cc}(c) shows that \nameshort-CUBIC reduces the packet loss rate by a factor of 2.25$\times$ compared to default CUBIC. Figure~\ref{fig:gpt2_classic_cc}(d) shows that \nameshort-CUBIC accelerates the average and 99$^{th}$ percentile training iteration time of both jobs by $1.2\times$ and $1.23\times$, respectively. Similar to \nameshort-Reno, the two \nameshort-CUBIC curves overlap. We also observe a 7.72\% difference in tail iteration time of the jobs from the ideal scheme.

\section{Flow-Level Simulations}
\label{sec:fl_sim}
In this section, we evaluate \name's performance, convergence, and fairness properties using a Python-based flow-level simulator.

\label{sec:flsim_method}
\para{Topology.}
We use a dumbbell topology similar to Figure~\ref{fig:topology}(a) with a single bottleneck link and multiple servers on either side of the link. Each distributed DNN job communicates gradients using the bottleneck link. 

\para{Simulator.} Our flow-level simulator allocates bandwidth to the competing jobs based on their congestion control variant. The additive increase coefficients of flows using \name is given by the value of $\fblack$ which increases with time. Hence, the bandwidth allocation changes with time and is given by the instantaneous ratio of the bandwidth aggressiveness functions of the flows. In practice, the congestion control algorithm takes a few RTTs before the bandwidths settle to the ratio of additive increase coefficients. We ignore this bandwidth-settling time because the RTTs in a GPU cluster are much smaller than a training iteration time. \looseness=-1

\para{Periodic jobs.} In our simulations, we randomly sample ($i$) the number of competing jobs, ($ii$) the size and number of their communication \peaks per iteration, and ($iii$) the initial start time difference of the jobs. We experiment with different compatibility scores between the jobs. 
The simulator captures practical aspects of a DNN cluster by adding a Gaussian random noise to the compute phases of jobs.

\subsection{Convergence and Performance}
\label{sec:flsim_perf}

\para{Convergence.}

\begin{figure}
    \centering
    \includegraphics[width=0.99\linewidth]{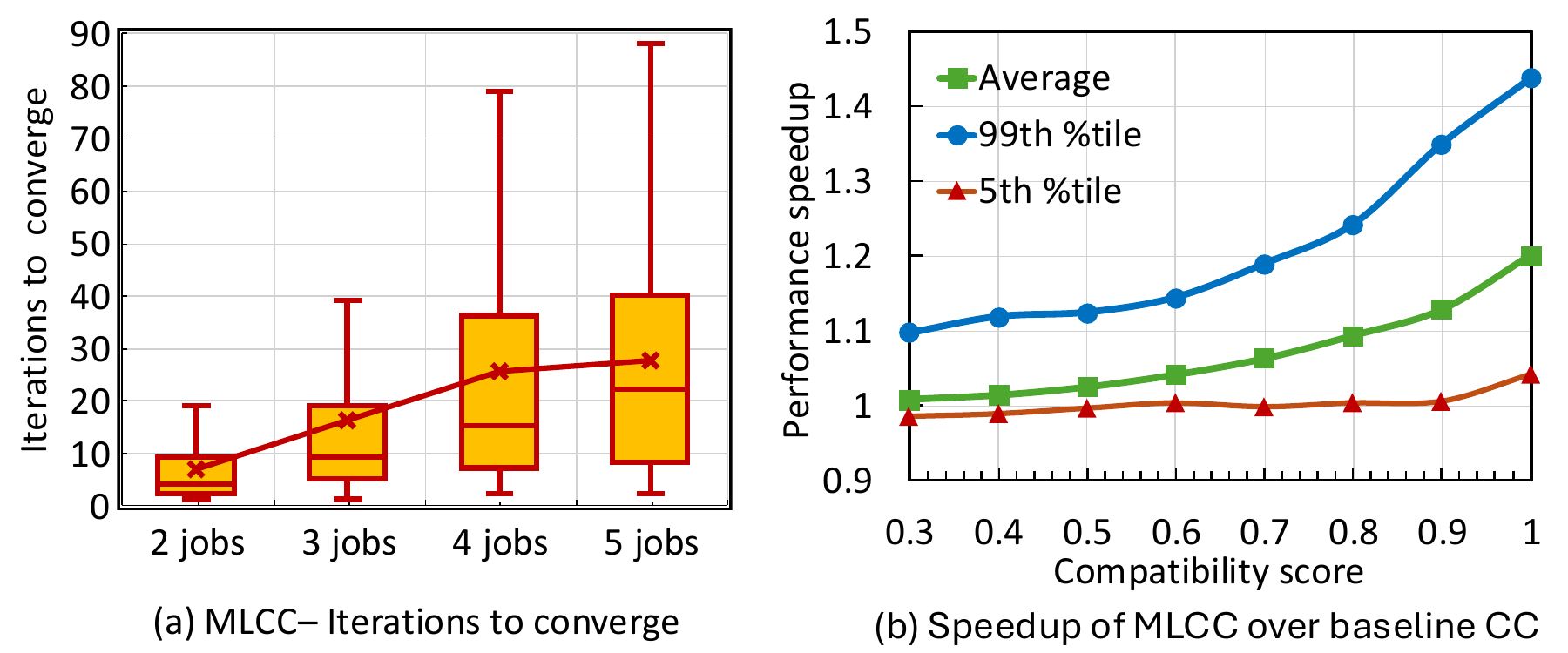}
    \vspace{-0.1cm}
    \caption{Convergence and performance speedup.}
    \label{fig:mltcp_iters_perf}
\end{figure}

We execute thousands of simulation runs, each spanning hundreds of training iterations of randomly sampled jobs. Each simulation run concludes when the jobs reach a stable interleaved state. Figure~\ref{fig:mltcp_iters_perf}(a) shows the distribution of the number of iterations of each job before convergence to an interleaved state. Although, the number of iterations to converge increases with the number of jobs, it is still a relatively small number (a few tens of iterations) compared to the training time with thousands of iterations. 

\para{Performance Speedup}.

Figure~\ref{fig:mltcp_iters_perf}(b) shows the average, $99^{th}$ percentile and $5^{th}$ percentile average iteration time speedup of \name over TCP as a function of compatibility score between the jobs. We simulate a new set of jobs in every simulation run for a thousand training iterations. Hence, the speedup varies across simulation runs depending on the characteristics of the sampled jobs. Note that \name is improving performance even for  partially compatible jobs. For jobs with compatibility scores as low as 0.3, \name's $5^{th}$ percentile speedup over TCP is greater than 1. For jobs with compatibility scores greater than 0.8, the $99^{th}$ percentile speedup increases to over $1.25\times$ over TCP. For fully compatible jobs, \name maintains the iteration times of the jobs within 5\% of their ideal values, achieving a $99^{th}$ percentile job speedup of $1.45 \times$.

\subsection{Fairness}
\label{sec:fairness}

\para{Fairness across ML flows}. 

In this experiment, we study MLCC's fairness in terms of training progress made by different DNN jobs and assess whether MLCC causes starvation in any job.
Following prior work~\cite{unfair1}, we use the average iteration time of the competing DNN jobs to compute the fairness of the congestion control algorithms. 
Since DNN jobs have different iteration times, we normalize all average iteration times by the ideal iteration time of the job when run in isolation~\cite{pfabric}. For a pair of DNN jobs competing on a bottleneck link, we define \textit{Average Iteration Time Fairness} as the ratio of the normalized average iteration time of the faster and the slower job throughout the experiment.
\begin{figure}
    \centering
\includegraphics[width=0.95\linewidth]{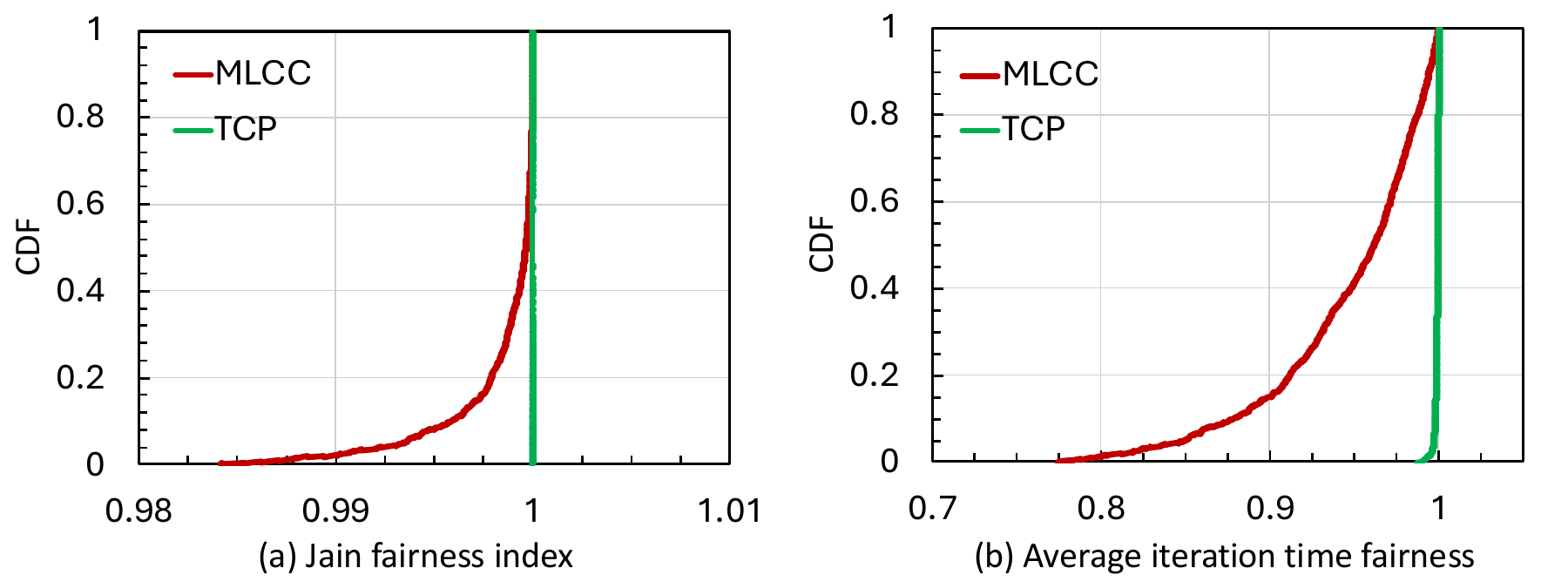}
    \vspace{-0.2cm}
    \caption{Comparison of TCP and \namebf fairness. }
    \vspace{-0.2cm}
    \label{fig:fairness}
\end{figure}

\begin{figure}
    \centering
\includegraphics[width=0.98\linewidth]{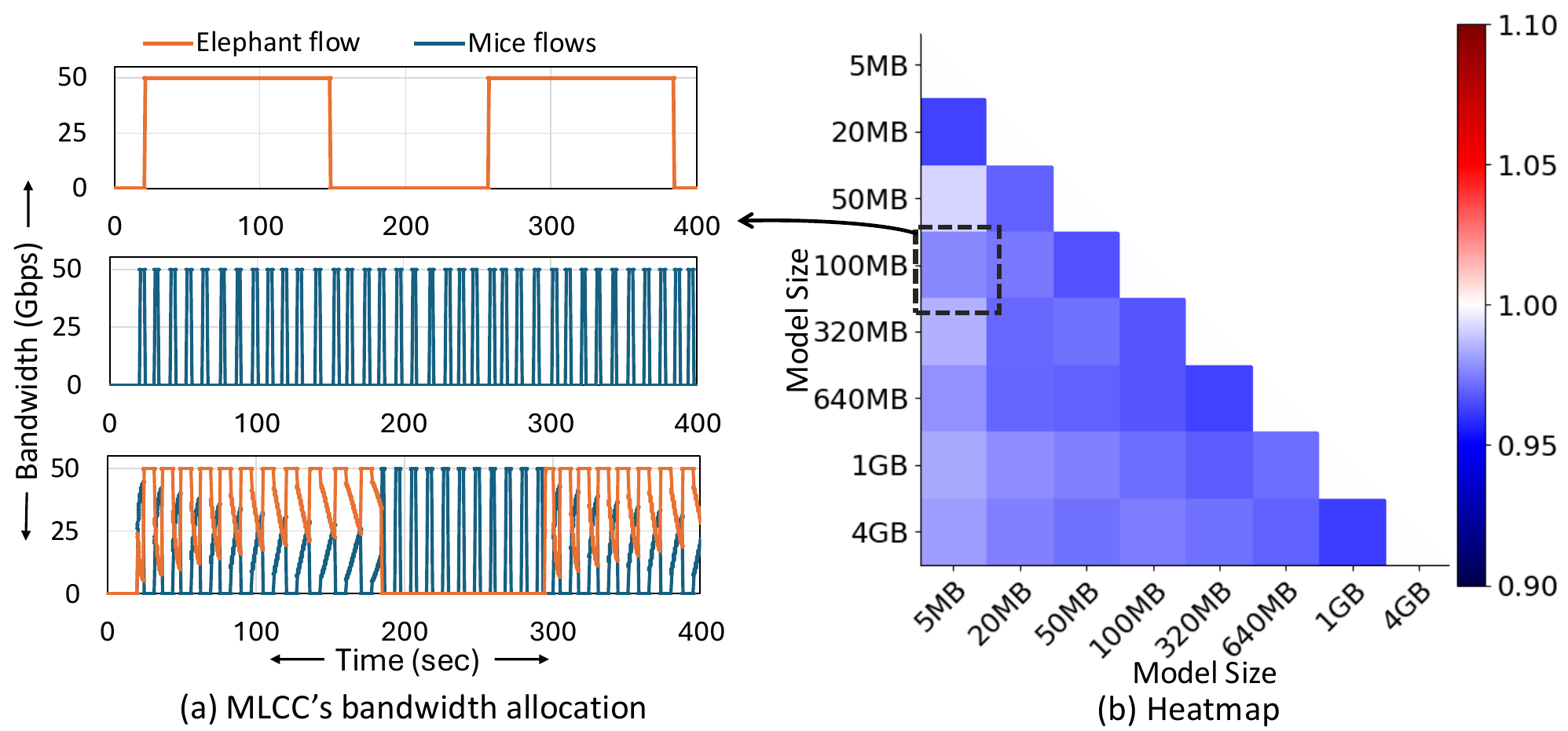}
    \vspace{-0.2cm}
    \caption{Heatmap of ratio of normalized average iteration times of elephant vs mice jobs.}
    \vspace{-0.1cm}
    \label{fig:fairness_heatmap}
\end{figure}

Figure~\ref{fig:fairness}(a) plots the cumulative distribution of Jain's fairness index using the jobs' normalized average iteration times in its expression. Each data point in the distribution corresponds to a single run of the simulator with a randomly sampled set of jobs. For TCP, Jain's fairness metric has a value of 1  because all jobs achieve the same normalized average iteration times in all simulation runs. For \name, the value of this metric is within $2\%$ of TCP, showing that the normalized average iteration times of the jobs are close despite the unequal instantaneous bandwidth allocation.

Figure~\ref{fig:fairness}(b) plots the distribution of the Average Iteration Time Fairness index for \name and TCP. \name's fairness under this metric is nearly 20\% far from TCP at the $99^{th}$ percentile. This shows that in the worst case scenario, the normalized average iteration time of the slower job is at most 20\% longer than the faster job across all our simulation runs with \name. We do not observe starvation of any job 
with \name irrespective of jobs' sizes and communication patterns. \looseness=-1

To study the impact of DNN model sizes on the fairness of \name, we plot the ratio of normalized average iteration time of the elephant job and the mice job, as we sweep the model sizes. We take an average from multiple simulations for each job combination with randomly sampled ideal iteration times. Figure ~\ref{fig:fairness_heatmap}(a) plots MLCC's bandwidth allocation between an elephant and a mice job, and figure ~\ref{fig:fairness_heatmap}(b) plots the heatmap of the ratio of normalized average iteration times between elephant vs mice jobs.
Figure~\ref{fig:fairness_heatmap}(b) shows that \name marginally favors the elephant job over the mice jobs in terms of job-completion times by at most $4\%$. 
This experiment also shows that \name does not starve either elephant jobs or mice jobs.

\label{lastpage}

\end{document}